\documentclass[11pt]{article}
\usepackage{amssymb,amsmath}
\usepackage{ascmac}
\usepackage[dvipdfmx]{hyperref}
\usepackage[dvips]{graphicx,psfrag}
\usepackage{epsfig}
\usepackage{array}
\usepackage{lscape}
\usepackage{accents}
\usepackage{color}
\usepackage{braket}
\usepackage{bm}

\numberwithin{equation}{section}

\makeatletter
\def\alt{\mathrel{\mathpalette\gl@align<}}
\def\agt{\mathrel{\mathpalette\gl@align>}}
\def\gl@align#1#2{\lower.6ex\vbox{\baselineskip\z@skip\lineskip\z@
\ialign{$\m@th#1\hfil##\hfil$\crcr#2\crcr\sim\crcr}}}
\makeatother

\newcommand{\doi}[1]{\href{https://doi.org/#1}{doi:#1}}

\newcommand{\mubar}{\overline{\text{Mu}}}
\newcommand{\dash}{\hspace{1pt}--\hspace{1pt}}

\begin{document}

\begin{flushright}
{\large
August, 2021
}
\end{flushright}
\vspace*{5mm}

\begin{center}

{\LARGE\bf
Models of the Muonium to Antimuonium Transition 
} 

\vspace{1.0cm}

{\Large 
Takeshi Fukuyama$\,{}^{a}$,
Yukihiro Mimura$\,{}^{b}$
and
Yuichi Uesaka$\,{}^{c}$
}

\vspace{8mm}

{\large
${}^a${\it 
Research Center for Nuclear Physics (RCNP),
Osaka University, \\Ibaraki, Osaka, 567-0047, Japan
}\\
\vspace{3mm}
${}^{b}${\it
Department of Physical Sciences, College of Science and Engineering, \\
Ritsumeikan University, Shiga 525-8577, Japan
}\\
\vspace{3mm}
${}^{c}${\it
Faculty of Science and Engineering, Kyushu Sangyo University, \\
2-3-1 Matsukadai, Higashi-ku, Fukuoka 813-8503, Japan
}\\
}

\vspace{1.5cm}

{\Large
{\bf Abstract}}\end{center}
\vspace{5mm}
\baselineskip 18pt
{\large
Muonium is a bound state composed of an antimuon and an electron,
and it constitutes a hydrogen-like atom.
Because of the absence of the hadronic matter in the bound state,
the muonium is a useful probe to explore new physics
being free from the hadronic uncertainties.
The process of the muonium-to-antimuonium transition is considered 
to be effective to identify fundamental interactions 
which relate to the lepton flavor and lepton number violation.
New experiments are being planned at J-PARC in Japan and CSNS in China,
and it is expected to attract more attention in the near future. 
In this paper, we will study what kind of model can be verified 
in the next generation of the muonium-to-antimuonium transition
search experiments while escaping the limitations from other experiments.
Though the transition probability is strongly suppressed by 
the lepton flavor conservation in the standard model,
it can be much larger by the exchanges of neutral and doubly charged bosons,
and by box loop diagrams in new physics beyond the standard model.
We study the neutrino models with heavy Majorana neutrinos at TeV scale, 
a type-II seesaw model, left--right models,
and models for radiative neutrino masses such as the Zee-Babu model in particular,
in addition to other possible models to induce the sizable transition probability, 
which can be tested in the forthcoming experiments.
}


\thispagestyle{empty}
\newpage
\addtocounter{page}{-1}

\section{Introduction}

\fontsize{13pt}{18pt}\selectfont

The muonium (Mu : $\mu^+e^-$) to antimuonium ($\mubar$ :   $\mu^-e^+$) transition 
is an interesting phenomenological possibility \cite{Pontecorvo:1957cp,Feinberg:1961zza,Lee:1977tib,Halprin:1982wm}.
There has been no new experimental result of the Mu-to-$\mubar$ transition 
since the Paul Scherrer Institute (PSI) experiment in the late 1990s \cite{Willmann:1998gd}.
The coefficient of the four-fermion transition operator is bounded to be less than $3\times 10^{-3}$ 
in the unit of the Fermi constant by the experiment.
A brand new idea of the transition experiment is now planning at 
Japan Proton Accelerator Research Complex (J-PARC) \cite{Kawamura:2021lqk}.
A new experiment is also planning at China Spallation Neutron Source (CSNS) \cite{MACE}.
It is expected that the bound will be updated for more than one digit 
by high-intensity $\mu^+$ beam lines.
From the theoretical point of view, 
the Mu-to-$\mubar$ transition is an important ingredient to accumulate our knowledge 
on lepton flavor violation and lepton number violation,
and to extract the fundamental interactions in the lepton sector.
For these twenty years, 
there are lots of new experimental results: measurements of the parameters in neutrino oscillations, 
updated results to bound the Mu-to-$\mubar$ transition indirectly,
and direct bounds on the new particles at the Large Hadron Collider (LHC).
We believe that
it is worth organizing the models of the Mu-to-$\mubar$ transition.

The Mu-to-$\mubar$ transition resembles $K^0$\dash$\overline{K^0}$ mixings in the quark sector.
The box diagrams via $W$ boson exchanges generate $K^0$\dash$\overline{K^0}$ mixings 
in the standard model (SM).
It is well-known that the  $K^0$\dash$\overline{K^0}$ mixings are suppressed 
due to the unitarity of the quark mixing matrix in the SM,
but they do not vanish completely
because the up-type quark masses are hierarchical, $m_u \ll m_c \ll m_t \sim M_W$.
Similarly to the $K^0$\dash$\overline{K^0}$ mixings, the
Mu-to-$\mubar$ transition operator can be potentially generated by the box diagram 
via $W$ boson exchanges in the SM.
However, the transition operator is strongly suppressed $\sim 10^{-30}$ 
due to the unitarity of the neutrino mixing matrix
and tiny neutrino masses, which is very different from the quark sector.
If there is a new particle around a TeV scale
and it couples to electrons and muons, the new interaction can induce the Mu-to-$\mubar$ transition
in an observable size.
%

%
The induced size of the Mu-to-$\mubar$ transition 
via the TeV-scale particles in the models beyond the SM
is indirectly restricted by the non-observation of
lepton flavor violating (LFV) processes \cite{Adam:2013mnn,Bellgardt:1987du}, 
such as $\mu \to e\gamma$ and $\mu^+ \to e^- e^+ e^+$ ($\mu \to 3e$).
%
We remark that 
the severely constrained LFV decays are $\Delta L_e = - \Delta L_\mu =  \pm 1$ processes,
while the change of the lepton flavor numbers in the Mu-to-$\mubar$ transition is $\Delta L_e = - \Delta L_\mu = -2$.
Therefore, if the lepton flavor numbers that new particles carry are odd,
one needs the multiple flavor violating interactions to induce the transition operator,
and therefore, it turns out that the size of the Mu-to-$\mubar$ transition will be much less than the one
which can be observed in near-future experiments.
However, 
if the lepton flavor numbers of the new particles are even,
the severe experimental constraints can be avoided and
an observable size of the Mu-to-$\mubar$ transition at near-future experiments can be induced at the tree level.

Those circumstances of the new physics contributions from flavor violation 
are similar to the meson mixings in the models beyond the SM.
In the case of the Mu-to-$\mubar$ transition, an additional contribution from the lepton number violation 
can be considered.
The transition operators can be generated by box diagrams
in which the mass terms of the fields in the internal lines violate the lepton numbers,
even if the interactions conserve the lepton flavor numbers.
In this case, the size of the Mu-to-$\mubar$ transition relates to the mechanism 
to generate the proper size of the tiny active neutrino masses.

The purpose of this paper is to scrutinize the models to induce the transition operators 
in eager anticipation of the new experiments.
In particular, we will study in detail those models with neutrino mass production.

We first review the model-independent issues of the Mu-to-$\mubar$ transition 
and Mu spectroscopy (Section \ref{sec2}).
We next classify the new particles and interactions
that causes the Mu-to-$\mubar$ transition in the models beyond the SM,
and make introductory remarks on 
how a sizable transition to be observed in the near-future experiments
can be induced avoiding the constraints such as LFV processes, which we have briefly mentioned above
 (Section \ref{sec3}).
After those traditional reviews, we start up the Mu-to-$\mubar$ transitions in the orthodox neutrino models:
TeV-scale Majorana neutrinos (Section \ref{sec4}), type-II seesaw  
(Section \ref{sec5}), and left--right models (Section \ref{sec6}).
We learn how the LFV processes restrict the transition operators in the respective models.
The box loop contribution is restricted by the $\mu \to e\gamma$,
and three-body LFV decays such as $\mu \to 3e$ restrict
the transition operators generated at the tree level.

We also learn 
the lepton number violation to induce the Mu-to-$\mubar$ transition in the orthodox models, 
and one finds that the Mu-to-$\mubar$ transition induced by the lepton number violation
is restricted by the natural realization of the sub-eV neutrino masses.
We thus study the models with the radiatively generated neutrino masses (Section \ref{sec7}),
which will be the main issue of this paper.
The radiative neutrino mass models fall into two broad categories: 
models with and without right-handed neutrinos.
In the models with right-handed neutrinos, the Dirac neutrino masses are forbidden by a discrete symmetry
and the tree-level neutrino mass is absent.
The Mu-to-$\mubar$ transitions induced by the lepton number violation in such situations are discussed.
The so-called Zee-Babu model \cite{Zee:1980ai,Zee:1985id,Babu:1988ki,Babu:2002uu} 
is one of the representative radiative neutrino models 
without the right-handed SM singlet fermions.
We show that the Zee-Babu model can produce the largest Mu-to-$\mubar$ transition 
among the radiative neutrino mass models, which can be tested in near-future experiments.

We also describe other models of the Mu-to-$\mubar$ transitions via the tree-level mediator exchanges,
including the ones that have been known for a long time:
neutral scalar exchange (Section \ref{sec8}), $R$-parity violating supersymmetry (SUSY) (Section \ref{sec9}), 
dilepton gauge bosons (Section \ref{sec10}),
and flavored neutral gauge bosons (Section \ref{sec11}).
We will also investigate the radiative neutrino masses as a version of a SUSY model with $R$-parity.
Other possible exotics can be considered (Section \ref{sec12}):
leptoquarks, vector-like fermions, and axion-like particles.

\section{Model independent description of Mu-to-$\mubar$ transition}
\label{sec2}

We review the model-independent issues on the Mu-to-$\mubar$ transitions.
We first describe the quantum mechanics of the Mu-to-$\mubar$ transition
and the probability of the transition.
We next introduce the four-fermion operators of the Mu-to-$\mubar$ transition,
and we obtain the transition amplitudes.
Since the experiments for the Mu-to-$\mubar$ transition have been done in a magnetic field,
we need to know the magnetic field dependence of the transition probability
in order to decode the experimental results.
We also comment on the corrections to the ground-state Mu hyperfine structure 
from the transition operators.

\subsection{Mu\dash$\mubar$ mixings}

The Schr\"odinger equation of the Mu\dash$\mubar$ system is
\begin{equation}
i \frac{\partial}{\partial t} 
\left( 
\begin{array}{c}
 \alpha \\ \beta 
\end{array} 
\right)
= 
\left( 
\begin{array}{cc}
 {\cal M}_{11} & {\cal M}_{12} \\
 {\cal M}_{21} & {\cal M}_{22} 
\end{array} 
\right)
\left( 
\begin{array}{c}
 \alpha \\ \beta 
\end{array} 
\right),
\end{equation}
for $|\psi(t) \rangle = \alpha(t) | {\rm Mu} \rangle + \beta(t) | \mubar \rangle$.
The matrix elements can be written as ${\cal M}_{ij} = M_{ij} - i \Gamma_{ij}/2$.
The CPT symmetry holds ${\cal M}_{11} = {\cal M}_{22}$, and
Mu and $\mubar$ can mix largely even if the off-diagonal element is tiny.
Solving the Schr\"odinger equation,
we obtain the time evolution of the Mu state, which is purely Mu at $t=0$,
as
\begin{eqnarray}
| {\rm Mu} (t) \rangle = 
f_+ (t) |{\rm Mu}  \rangle + (q/p) f_- (t) | \mubar  \rangle, 
\end{eqnarray}
where
\begin{equation}
q/p = \sqrt{\frac{{\cal M}_{21}}{{\cal M}_{12}}},
\qquad
f_\pm (t)= \frac12 ( e^{- i \lambda_+ t} \pm e^{-i \lambda_- t} ),
\end{equation}
\begin{equation}
\lambda_\pm = M - i \frac{\Gamma}{2} \pm \frac12 (\Delta M - i \frac{\Delta \Gamma}2),
\qquad 
\Delta M - i \frac{\Delta \Gamma}2 = 2\sqrt{{\cal M}_{12} {\cal M}_{21}}. 
\end{equation}
Here, $M$ and $\Gamma$ are the averages of the masses and widths, respectively,
and $\Delta M$ and $\Delta \Gamma$ are the differences of them.
The transition probability at a time $t$ can be written as
\begin{equation}
P({\rm Mu} \to \mubar ;t ) \sim |q/p|^2 |f_-(t)|^2,
\qquad
P({\rm Mu} \to {\rm Mu} ;t ) \sim  |f_+(t)|^2,
\end{equation}
and 
one can calculate 
\begin{equation}
|f_\pm (t)|^2 = \frac12 e^{-\Gamma t} (\cosh \frac{\Delta \Gamma }{2}t \pm \cos \Delta M t).
\end{equation}
If there is CP symmetry or $|\Gamma_{12}/M_{12} | \ll 1$, one obtains $|q/p| =1$.
We take a plausible assumption $|\Gamma_{12}/M_{12}| \ll 1$ to describe the following,
and ${\cal M} \equiv {\cal M}_{12} = (\Delta M)/2$.

The time-integrated probability of the Mu-to-$\mubar$ transition is obtained by
\begin{eqnarray}
P( {\rm Mu} \to \mubar) = \int_0^\infty dt \Gamma e^{-\Gamma t} \sin^2 {\cal M} t 
= \frac{2 {\cal M}^2}{4 {\cal M}^2 + \Gamma^2},
\end{eqnarray}
which corresponds to the probability that the decay of the Mu produced in the laboratory comes from 
the Mu$(\mu^+ e^-)$ $\to$ $\mubar (\mu^- e^+) \to (e^- + \bar \nu_e + \nu_\mu) + e^+$ mode.
For $|\Delta M| \ll \Gamma = 1/\tau$ ($\tau$ is the Mu lifetime, $2.20 \, \mu $s), 
one can write the time-integrated transition propability as
\begin{equation}
P \simeq 2 \tau^2 {\cal M}^2.
\end{equation}

\subsection{Operators}

The operators which can induce the Mu-to-$\mubar$ transition
are \cite{Conlin:2020veq}
\begin{eqnarray}
Q_1 &=& (\bar \mu \gamma_\mu (1- \gamma_5) e ) ( \bar \mu \gamma^\mu (1- \gamma_5) e ), \label{eq:Q1} \\
Q_2 &=& (\bar \mu \gamma_\mu (1+ \gamma_5) e ) ( \bar \mu \gamma^\mu (1+ \gamma_5) e ), \label{eq:Q2} \\
Q_3 &=& (\bar \mu \gamma_\mu (1+ \gamma_5) e ) ( \bar \mu \gamma^\mu(1- \gamma_5) e ), \label{eq:Q3} \\
Q_4 &=& (\bar \mu  (1- \gamma_5) e ) ( \bar \mu (1- \gamma_5) e ), \label{eq:Q4} \\
Q_5 &=& (\bar \mu  (1+ \gamma_5) e ) ( \bar \mu (1+ \gamma_5) e ). \label{eq:Q5}
\end{eqnarray}
Any dimension-six four-fermion operators for the Mu-to-$\mubar$ transition can be written by a linear combination 
of the above five by using Fierz identities.
For example, one can find
\begin{eqnarray}
&& (\bar \mu  (1+ \gamma_5) e ) ( \bar \mu (1- \gamma_5) e ) = - \frac12 Q_3, \\
&& (\bar\mu e) (\bar\mu e) = \frac14 (-Q_3 + Q_4 + Q_5), \\
&& (\bar\mu \gamma_5 e) (\bar\mu \gamma_5 e) = \frac14 (Q_3 + Q_4 + Q_5), \\
&& (\bar \mu  \sigma_{\mu\nu} e ) ( \bar \mu \sigma^{\mu\nu}  e ) = -3 (Q_4 + Q_5).
\end{eqnarray}

We denote the terms in the effective Lagrangian as
\begin{equation}
-{\cal L}_{{\rm Mu} \mbox{\dash} \mubar} = \sum_{i = 1, \cdots, 5} \frac{G_i}{\sqrt2} Q_i ,
\label{eq:Gi}
\end{equation}
where the normalization of the coefficients mimics the Fermi constant $G_F$.

In practice, the state of the produced Mu is a mixture of four states made by the hyperfine structure.
The four states $(F,m)=(0,0)$, $(1,0)$, and $(1,\pm 1)$ are indicated by the magnitude of the total angular momentum $F$ and the $z$ component of the total angular momentum $m$.
The $F=0$ state is called paramuonium, while the $F=1$ state is called orthomuonium.
The amplitudes of the $\textrm{Mu}\left(F,m\right)\to\overline{\textrm{Mu}}\left(F,m\right)$ transition\footnote{If $F\ne F'$ or $m\ne m'$, $\braket{\overline{\textrm{Mu}};F,m|Q_i|\textrm{Mu};F',m'}=0$ for any $i$.} are written as
\begin{align}
\mathcal{M}_{F,m}= \sum_{i=1, \cdots, 5} \frac{G_i}{\sqrt{2}}\braket{\overline{\textrm{Mu}};F,m|Q_i|\textrm{Mu};F,m}.
\end{align}
Treating the bound leptons nonrelativistically, we obtain
\begin{equation}
{\cal M}_{0,0} =- \frac{8 \left|\varphi(0)\right|^2}{\sqrt2} \left( G_1 + G_2 - \frac32 G_3 - \frac 14 G_4 - \frac14 G_5\right) ,
\label{eq:amplitude00}
\end{equation}
for the spin-singlet paramuonium,
and
\begin{equation}
{\cal M}_{1,0} ={\cal M}_{1,\pm 1} =- \frac{8 \left|\varphi(0)\right|^2}{\sqrt2} 
\left( G_1 + G_2 + \frac12 G_3 - \frac 14 G_4 - \frac14 G_5\right) ,
\label{eq:amplitude10}
\end{equation}
for the spin-triplet orthomuonium.
The derivation of Eqs.~\eqref{eq:amplitude00}-\eqref{eq:amplitude10} is given in Appendix~\ref{app:transition_amplitude}.

Solving the Schr\"{o}dinger equation for the hydrogen-like atom, we find that the wave function $\varphi(\bm{r})$ is
\begin{align}
\varphi(\bm{r})=\,&\sqrt{\frac{\left(m_\textrm{red}\alpha\right)^3}{\pi}}\exp\left(-m_\textrm{red}\alpha r\right),
\end{align}
where $m_\textrm{red}=m_em_\mu/\left(m_\mu+m_e\right)\simeq m_e$ is the reduced mass between a muon and an electron and $\alpha\simeq 1/137$ is the QED fine structure constant.
The value of $\varphi(\bm{r})$ at the origin is given by
\begin{align}
\left|\varphi\left(0\right)\right|^2=\frac{\left(m_\textrm{red}\alpha\right)^3}{\pi}.
\end{align}

\subsection{Magnetic field dependence}

The transition probability of $\textrm{Mu}$ to $\overline{\textrm{Mu}}$ is changed in a finite magnetic field $B$.
Since we have to care about the effects of the external magnetic field
to describe the experimental constraints given by the PSI experiment,
let us review the magnetic field dependence \cite{Horikawa:1995ae,Hou:1995np}.
For the Mu spectroscopy, see Appendix \ref{app:spectroscopy}.

The external magnetic field splits the $(1,\pm 1)$ states
and makes $\Delta E = M_{11} - M_{22}$ to be non-zero.
As a consequence, the Mu\dash$\mubar$ mixing becomes small
and the transition probability becomes
\begin{equation}
P ({\rm Mu} \to \mubar, t) \simeq e^{-\Gamma t} \frac{{\cal M }^2_{1,\pm 1}}{{\cal M }^2_{1,\pm 1} + (\Delta E/2)^2} \sin^2 \sqrt{{\cal M}^2_{1,\pm 1}  + (\Delta E/2)^2}\, t,
\end{equation}
and the time-integrated probability is
\begin{equation}
\int_0^\infty dt\, \Gamma P ({\rm Mu}\to \mubar, t)   \simeq \frac{2 \tau^2 |{\cal M}_{1,\pm 1}|^2}{1+(\tau \Delta E)^2  }.
\end{equation}
The energy splitting $\Delta E$ of the $(1,\pm 1)$ states by the magnetic field 
can be obtained by Eq.(\ref{E_B_pm1}),
%
and
one obtains 
\begin{equation}
\tau \Delta E = 3.85 \times 10^5 \times \frac{B}{\rm Tesla}.
\end{equation}
Therefore, in the magnetic flux $B$ to be more than $1\, \mu T$ (micro Tesla),
the transition probability is suppressed for $(1,\pm 1)$ states.
For one's information, the geomagnetic flux density is $\sim 30$\dash$60$~$\mu T$.

We note that the oscillation time without a magnetic field 
is $O(1)$ second or longer under 
the current experimental bound.
Therefore, the Mu-to-$\mubar$ ``oscillations'' do not start before Mu decays.
The behavior of the transition probability for the $m = \pm 1$ states near $t=0$
is the same as the one without a magnetic field.
However, if the external magnetic field is $\sim 1\, \mu T$, the oscillation time is the same as the
Mu lifetime, and therefore, the transition probability for $m = \pm 1$ is suppressed for $B \agt O(1) \, \mu T$.

The Mu\dash$ \mubar$ mixing for $m=0$ states is (nearly) maximal even in the magnetic field,
contrary to the $m=\pm 1$ states.
The $(F,m) = (1,0)$ and $(0,0)$ states are mixed due to the magnetic filed,
and 
the transition amplitudes (halves of the mass differences) are modified as
\begin{eqnarray}
{\cal M}_{0,0}^B &=& \frac12 \left( {\cal M}_{0,0} - {\cal M}_{1,0} + \frac{{\cal M}_{0,0} + {\cal M}_{1,0}}{\sqrt{1+X^2}} \right), \\
{\cal M}_{1,0}^B &=& \frac12 \left( -{\cal M}_{0,0} + {\cal M}_{1,0} + \frac{{\cal M}_{0,0} + {\cal M}_{1,0}}{\sqrt{1+X^2}} \right),
\end{eqnarray}
where $X = 6.31 \times B/{\rm Tesla}$ 
is defined in Eq.(\ref{eq:X}).

The time-integrated transition probability is totally
\begin{equation}
P = 2 \tau^2 \left( |c_{0,0}|^2 | {\cal M}_{0,0}^B |^2 +  |c_{1,0}|^2 | {\cal M}_{1,0}^B |^2
+ \sum_{m= \pm 1} |c_{1,m}|^2 \frac{|M_{1,m}|^2}{1+ (\tau \Delta E)^2} 
\right),
\label{transition-probability}
\end{equation}
where $|c_{F,m}|^2$ gives the population of the Mu states.
The experimental result by the PSI experiment at the magnetic flux density $B = 0.1$ Tesla is \cite{Willmann:1998gd}
\begin{equation}
P < 8.3 \times 10^{-11}.
\end{equation}
The oscillations of the $(1,\pm 1)$ states are dropped in the magnetic flux density.
If $G_3 =0$, we obtain
\begin{eqnarray}
P &=& \frac{64 m_{\rm red}^6 \alpha^6 \tau^2 G_F^2}{\pi^2} \left(\frac{G_1 + G_2 - \frac14 (G_4+G_5)}{G_F}\right)^2 \frac{|c_{0,0}|^2+|c_{1,0}|^2}{1+X^2} \nonumber \\
&=& 2.57 \times 10^{-5} \left(\frac{G_1 + G_2 - \frac14 (G_4+G_5)}{G_F}\right)^2 \frac{|c_{0,0}|^2+|c_{1,0}|^2}{1+X^2} ,
\end{eqnarray}
and the experimental result 
 is decoded as
\begin{equation}
\left|G_1 + G_2 - \frac14 G_4 - \frac14 G_5\right| < 3.0 \times 10^{-3} G_F .  
\label{PSI-bound}
\end{equation}
If $G_3 \neq 0$ and the others are zero,
we find 
\begin{equation}
|G_3 | < 2.1 \times 10^{-3} G_F.
\label{G3_bound}
\end{equation}
We use the population of Mu states, $|c_{0,0}|^2 = 0.32$, $|c_{1,0}|^2  = 0.18$.
If the operators are turned on containing $Q_3$, the expression is a little complicated to write down here,
but one can easily calculate the bound from the expressions above.

The PSI experiment tried to observe the decay product (electron) from an expected $\mu^-$ in $\mubar$ ($\mu^- e^+$) 
after Mu ($\mu^+ e^-$) is produced in the laboratory.
The external magnetic field is thus needed.
The MACE group in China will also use this method \cite{MACE}.
The time is not specified, and the time-integrated transition probability is applied.
The intrinsic beam-related and accidental backgrounds disturb the detection 
of the electrons emitted from the $\mubar$ decays, which determines
the experimental bound of the Mu-to-$\mubar$ transition.

A new method of the Mu-to-$\mubar$ transition search is proposed 
using a high-intensity pulsed muon source in J-PARC and an intense laser \cite{Kawamura:2021lqk}.
The high-intensity beam (H-line) will work this summer.
In their method, an expected $\mubar$ is ionized by a laser shot at a time
and the dissolved $\mu^-$ is directly analyzed by a spectrometer.
The transition probability is not time-integrated and is given at a time $t$
when the laser is shot.
Therefore, the number of the possible transition events will be less 
(by a factor $(t/\tau)^2 \exp(-t/\tau)/2$ up to the other experimental lacks in the laboratory)
than the time-integrated 
transitions in the preceding method.
This method, however, is free from the background noises from the accelerator 
and messy positrons' scatterings from $\mu^+$ decays in the preceding experiment.
Their method does not need an external magnetic flux to 
detect the decay products from $\mubar$ and Mu decays,
and the Mu-to-$\mubar$ transition can be observed with the geomagnetic flux or with shielding it.
The controllability of the external magnetic field has further advantages to confirm new physics 
and to investigate the operator dominance by the characteristic magnetic-field dependence.

\subsection{Muonium hyperfine structure}

The MuSEUM group is planning measurements of the $1S$ hyperfine structure (HFS)
of Mu using the H-line at J-PARC \cite{MuSEUM:2020mzm,Tanaka:2021jtf}.
The current most accurate experimental value of the Mu HFS interval is~\cite{Liu:1999iz}
\begin{equation}
\nu_{\rm HFS}^{\rm exp} = 4 \ 463 \ 302 \ 765 \pm 53 \ {\rm Hz} ,
\end{equation}
which has been measured in a strong magnetic field.
%
%
The MuSEUM group will reduce the systematic errors of the measurements to a few Hz
in both zero and strong magnetic fields.
The theoretical expression of the HFS interval can be written in Heaviside-Lorentz units as
\begin{equation}
\nu_{\rm HFS}^{\rm th} =  
\frac1{4\pi} \frac{16}3 \mu_\mu \mu_e |\varphi(0)|^2
( 1 + \delta_{\rm QED} + \delta_{\rm weak} + \delta_{\rm hadronic} ),
\end{equation}
where $\mu_\mu$ and $\mu_e$ are the magnetic moments of the muon and electron, respectively.
The theoretical calculation with electroweak and intermediate hadronic corrections contains the uncertainty 
$\sim 300$\dash$500$~Hz~\cite{Mohr:2015ccw,Eides:2018rph,Karshenboim:2021pit}.
Precise measurement by MuSEUM also reduces the uncertainty
in the muon-proton magnetic moment and muon-electron mass ratios,
which can reduce the uncertainty in the theoretical calculation of the HFS interval.
In this subsection, 
we describe the corrections from the Mu-to-$\mubar$ transition operators 
to the HFS interval.
For the ground-state Mu spectroscopy, see Appendix \ref{app:spectroscopy}.

When the external magnetic field is zero $B=0$,
the HFS interval is defined as
\begin{equation}
h\nu_{\rm HFS} (B=0) \equiv E_{F=1}- E_{F=0},
\end{equation}
where
\begin{equation}
E_{F=1} = E_0+ \frac14 h\nu_{\rm HFS}, \qquad
E_{F=0} = E_0 -\frac34 h\nu_{\rm HFS},
\end{equation}
and the Planck constant is $h=2\pi$ since we are working in the natural unit 
$\hbar = 1\, (= 6.582 
 \times 10^{-25}$ GeV$\cdot$ s).
When there is a transition operator, 
the Mu and $\mubar$ is maximally mixed
(even if the coefficient of the operator is small).
The energy eigenstates correspond to CP eigenstates.
The HFS interval is measured by using the resonance of microwave frequency in the cavity.
The transitions to the different CP states are suppressed.
As a result,
the correction from the transition operators
is given as 
\begin{equation}
\Delta \nu_{HFS}(B=0) =\pm  \frac{M_{F=1} - M_{F=0}}{2\pi}=
\pm \frac{4m_{\rm red}^3 \alpha^3}{\sqrt2 \pi^2 } 2 |G_3|.
\end{equation}
If there is $Q_3$, the mass difference between Mu and $\mubar$
is different for spin-singlet and triplet,
and then, it can modify the Mu HFS interval between the spin-singlet and triplet \cite{Fujii:1993su}.
If there is only a $Q_3$ operator, the current bound of the Mu-to-$\mubar$ transition implies
\begin{equation}
|\Delta \nu_{\rm HFS}(B=0)| < 1.1 \, {\rm Hz}.
\end{equation}

When there is an external magnetic field, the states split
and thus the definition of the HFS interval should be modified:
\begin{equation}
h\nu_{\rm HFS} (B\neq 0) =h \nu_{12} + h\nu_{34} \equiv E_{(1,1)}  - E_{(1,0)} + E_{(1,-1)}- E_{(0,0)}.
\end{equation}
The energy eigenstates are not CP eigenstates in the magnetic field,
and thus
there are two resonant frequencies if there is a transition operator
and the measurements of the frequency are very accurate.
The corrections are given as
\begin{eqnarray}
\Delta \nu_{12} &\simeq& \pm \frac{M^B_{1,0}}{2\pi} = \pm \frac{4 m_{\rm red}^3 \alpha^3}{\sqrt2 \pi^2} \left|G_3 + \frac{G_1 + G_2 -\frac12 G_3 -\frac14 G_4 - \frac14 G_5}{\sqrt{1+X^2}} \right|, \\
\Delta \nu_{34} &\simeq& \pm \frac{M^B_{0,0}}{2\pi}
= \pm \frac{4 m_{\rm red}^3 \alpha^3}{\sqrt2 \pi^2} \left|-G_3 + \frac{G_1 + G_2 -\frac12 G_3 -\frac14 G_4 - \frac14 G_5}{\sqrt{1+X^2}} \right|.
\end{eqnarray}
The splitting of the resonant frequency is less than about 1 Hz
for the current experimental bound of the Mu-to-$\mubar$ transition.

The theoretical calculation of the HFS interval contains the uncertainty of the fine structure constant,
and it will be hard to reduce the uncertainty of the theoretical prediction to be less than 1 Hz.
Therefore, we cannot say anything about new physics
even if the HFS interval is accurately measured only in the case of $B=0$.
The accurate measurements of the HFS intervals for both $B=0$ and $B\neq 0$ will be important.
According to Ref.\cite{Tanaka:2021jtf},
the HFS intervals for both $B\simeq 0$ and $B\neq 0$ will be measured
with systematic errors of $2$\dash$3$ Hz.
The accurate measurement of the HFS intervals will give us an interesting cross-check, though
the accuracy is not enough to say something, and the Mu-to-$\mubar$ transition bound will be updated
when the HFS interval is accurately measured at J-PARC.

\section{Classification of the mediators}
\label{sec3}

The purpose of this paper is to study models
to induce the transition operators, $Q_i$.
Before moving to the concrete description of the individual models,
we classify them by the LFV couplings to generate the operators to learn how the Mu-to-$\mubar$ transition can be sizable avoiding the LFV decay constraints.
Though the assignments of the lepton (flavor) numbers may have ambiguity in respective models,
this classification can specify the mediator in the model.
This classification is useful to make clear what is needed to obtain the sizable Mu-to-$\mubar$ transition 
in a model-independent way.

\begin{enumerate}

\item
$\Delta L_e = \Delta L_\mu =0$

The interactions do not violate the flavor numbers,
but the mass terms of SM singlet fields have $L_e = \pm2$ and $L_\mu = \pm 2$.
Total lepton number conservation is violated in this case.
Therefore, this case is friendly to the models to generate neutrino masses.

The models with right-handed Majorana neutrinos are considered 
to induce the Mu-to-$\mubar$ transition, which will be studied mainly in Sections \ref{sec4}, \ref{sec7.2} and \ref{sec7.3}.
The transition operators can be generated by box loop diagrams (e.g., Fig.\ref{fig:box1} (right) and Fig.\ref{fig:charged-Higgs} (right)).

\item
$(\Delta L_e, \Delta L_\mu) = (\pm 2,0)$ and $(\Delta L_e, \Delta L_\mu) = (0,\pm 2)$

Interaction terms violate the flavor numbers separately.

The mediators have the lepton number to be 2, and they are called dilepton bosons.
The mediators have doubly electric charges.
The dilepton doubly charged scalars will be studied in Sections \ref{sec5}, \ref{sec6}, and \ref{sec7.1},
and dilepton gauge boson will be considered in Section \ref{sec10}.
The transition operators can be generated by tree diagrams (e.g., Fig.\ref{fig:doubly-charge-scalar} and Fig.\ref{fig:dilepton}).

 \item
$ \Delta L_e = - \Delta L_\mu = \pm 1$

Interaction terms violate both flavor numbers.

The mediators are neutral bosons.
The neutral scalars are considered in Sections \ref{sec8} and \ref{sec9.1},
and neutral gauge bosons are studied in Section \ref{sec11}.
The transition operators can be generated by tree diagram (e.g., Fig.\ref{fig:neutral-scalar} and Fig.\ref{fig:flavor-gauge-boson}).

We remark that this mediator should not couple with quarks
to generate the Mu-to-$\mubar$ transition
to avoid the $\mu$\dash $e$ conversion in nuclei induced at the tree level.

\item
$(\Delta L_e, \Delta L_\mu) = (\pm 1,0)$ and $(\Delta L_e, \Delta L_\mu) = (0,\pm 1)$

The transition operators can be generated by box loop diagrams (e.g., Fig.\ref{fig:box1} (left) and Fig.\ref{fig:charged-Higgs} (left)).
However, the interaction can induce $\mu \to e\gamma$ and/or $\mu \to 3e$,
which restricts the
size of coupling constants.
Then, the magnitudes of the coefficients of the transition operators
become much less than
the achievement of the planned experiments,
as we will see in many of the models.

\end{enumerate}

In the cases 1, 2, and 3, 
there are interactions or mass terms with
$\Delta L_e - \Delta L_\mu = \pm 2$, which are even numbers.
Even in those cases, the interactions with $\Delta L_e - \Delta L_\mu = \pm 1$ 
may be intermingled in the respective models, and then,
the magnitudes of the Mu-to-$\mubar$ transition is bounded by 
 $\mu \to e\gamma$ and/or $\mu \to 3e$ \cite{Adam:2013mnn,Bellgardt:1987du}:
\begin{eqnarray}
&&{\rm Br} ( \mu \to e\gamma ) < 4.2 \times 10^{-13},  \\
&&{\rm Br} ( \mu \to 3e ) < 1.0 \times 10^{-12}.
\end{eqnarray}
If $Z_n$ discrete symmetry can be imposed
to suppress
the $\Delta L_e - \Delta L_\mu = \pm 1$ interactions 
(while the $\Delta L_e - \Delta L_\mu = \pm 2$ interactions are allowed),
the Mu-to-$\mubar$ transition can be as large as the current experimental bound.
In other words, the observation of the Mu-to-$\mubar$ transition
in the near-future experiments 
implies the existence of such discrete symmetry in the lepton sector.

The new $\Delta L_e - \Delta L_\mu = \pm 2$ interactions can induce a ``wrong muon decay'':
\begin{equation}
\mu^+ \to \nu_\mu + e^+ + \bar\nu_e,
\end{equation}
and thus the couplings and masses of the mediators are restricted by the universality of the Fermi decay constant \cite{Fujii:1993su}.
They are also constrained by $e^+ e^- \to e^+ e^-$
Bhabha scattering data.
The couplings can also induce muon and electron anomalous magnetic moments.
The couplings and the mediator masses to induce the Mu-to-$\mubar$ transition 
which is allowed by the PSI experiment,
do not conflict with those low energy data at present.
Rather, the experimental results of the Mu-to-$\mubar$ transitions restrict them.
The data from the high-luminosity LHC, ILC, and Belle II will cooperate with the 
near-future Mu-to-$\mubar$ transition experiments.

\section{Heavy Majorana neutrinos}
\label{sec4}

\begin{figure}
\center
\includegraphics[width=7cm]{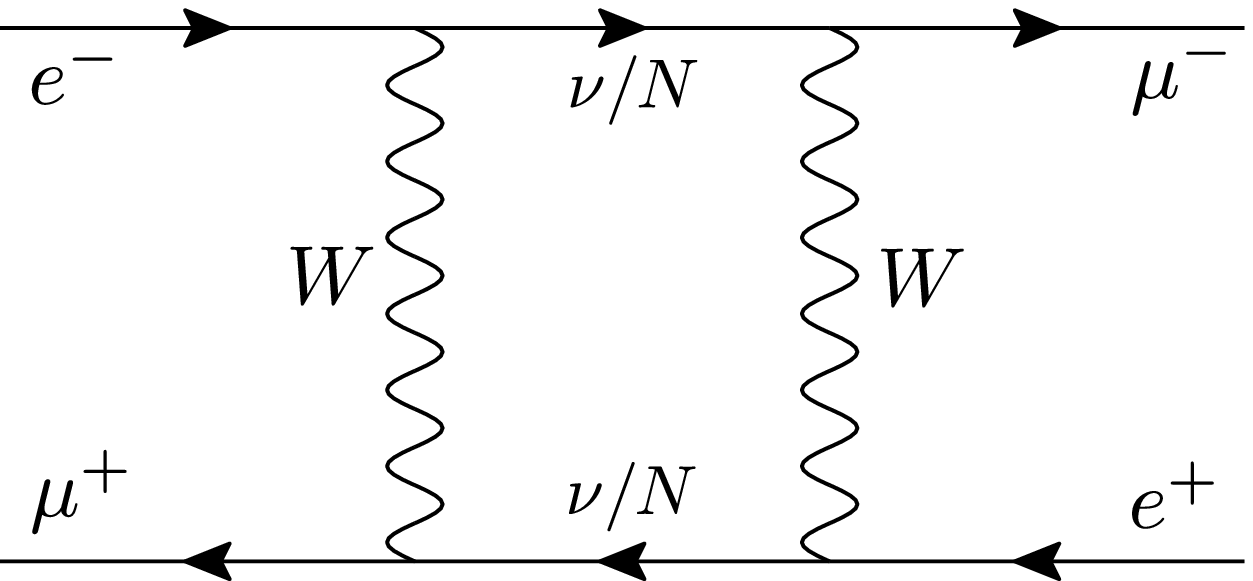} \hspace{1cm}
\includegraphics[width=7cm]{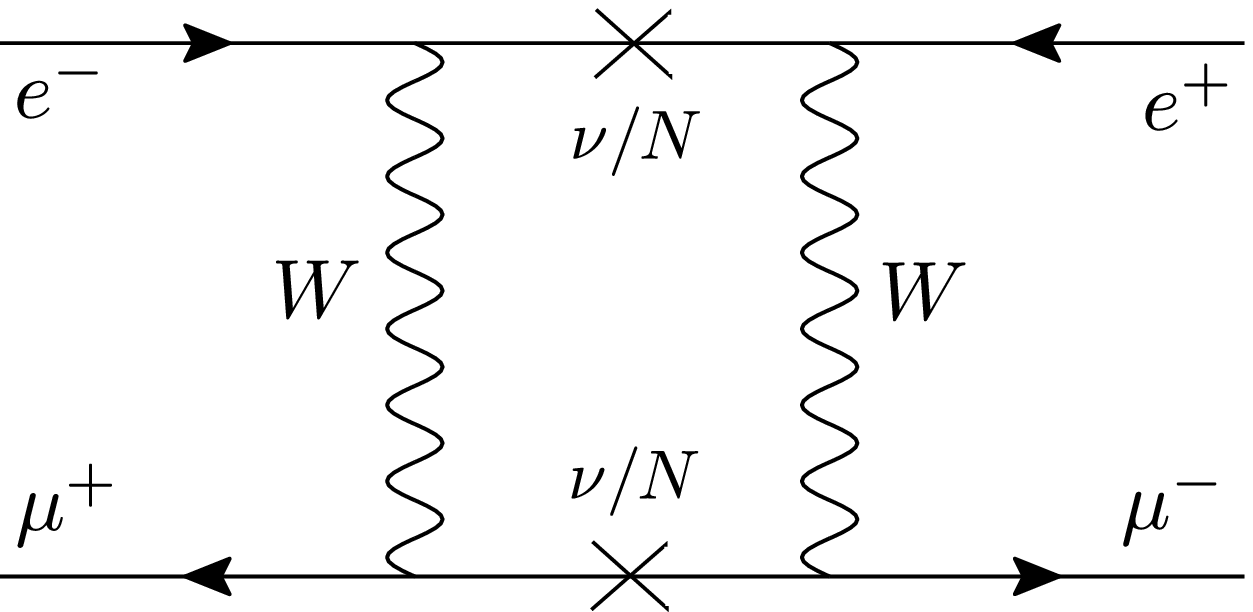}
\caption{
The box loop diagrams to generate the transition operator $Q_1$.
There are two types of contribution:
the momentum parts ($p \!\!\! / $) of the numerators of neutrino propagators 
are picked (left),
and 
the mass parts of the propagators are picked (right).
The lepton flavors are changed at the vertices in the left diagram,
while the lepton numbers are violated in the neutrino masses in the right diagram.
}
\label{fig:box1}
\end{figure}

The simplest neutrino model to acquire the Mu-to-$\mubar$ transition may be 
the models with TeV scale Majorana neutrinos, which are SM singlets.
The coefficient of the transition operator from the neutrino box loop contribution is written as \cite{Clark:2003tv,Cvetic:2005gx,Liu:2008it,Abada:2015oba}
\begin{eqnarray}
\frac{G_1}{\sqrt2} &=&  \frac{G_F^2 M_W^2}{16 \pi^2}  \sum_{\cal I,J}
\left[
{\cal U}_{\mu_L {\cal I}} {\cal U}^*_{e_L {\cal I}}
{\cal U}_{\mu_L {\cal J}} {\cal U}^*_{e_L {\cal J}}
E_0(x_{\cal I} ,x_{\cal J})+
({\cal U}_{\mu_L {\cal I}})^2 
({\cal U}^*_{e_L {\cal J}})^2 
E_1(x_{\cal I} ,x_{\cal J})\right] \nonumber \\
&\simeq&
\frac{G_F^2 M_W^2}{16 \pi^2}  \sum_{I,J}
\left[
X_{\mu {I}} X^*_{e { I}}
X_{\mu{J}} {X}^*_{e { J}}
\tilde E_0(x_{I} ,x_{J})+
(X_{\mu {I}})^2 
(X^*_{e {J}})^2 
E_1(x_{I} ,x_{ J})\right],
\label{eq:MutoMubar-neutrino}
\end{eqnarray}
where
\begin{equation}
x_{\cal I} = \frac{M_{\cal I}^2}{M_W^2},
\end{equation}
and 
see Appendix \ref{app:diagonalization} and \ref{app:loop_function}
for the neutrino mixing matrix $\cal U$ and the loop functions $E_0$ and $E_1$.
The $E_0$ term comes from box diagrams such as Fig.\ref{fig:box1} (left),
and $E_1$ term comes from diagrams such as Fig.\ref{fig:box1} (right). 
We ignore the light neutrino masses $x_i = m_i^2/M_W^2 \simeq 0$,
and one can rewrite the first line into the second line by using the unitary relation of the mixing matrix,
\begin{equation}
\sum_{\cal I} {\cal U}_{\mu_L {\cal I}} {\cal U}^*_{e_L {\cal I}} = \sum_{i} U_{\mu i} U^*_{e i} + \sum_{I} X_{\mu I} X^*_{e I} =0,
\end{equation}
and
\begin{equation}
\tilde E_0(x,y) = E_0 (x,y) - E_0(x,0) - E_0 (0,y) + E_0 (0,0) = E_0(x,y).
\end{equation}

We enumerate the necessary facts to evaluate the loop contribution.
\begin{enumerate}
\item
The mixings are bounded by electroweak precision data model-independently
\cite{Atre:2009rg,Deppisch:2015qwa}
\begin{equation}
|X_{eI}|^2, |X_{\mu I} |^2 \alt 0.003,
\end{equation}
individually (supposing only one of $M_I$ is in the TeV scale).

\item
The product of $|X_{eI} X_{\mu I}|$ is bounded by LFV processes,
especially $\mu \to e \gamma$.

The $\mu_R \to e_L \gamma$ decay amplitude
is given as
\begin{eqnarray}
A_R &=& \frac{e m_\mu}{16\pi^2} \frac{G_F}{\sqrt2} \left( \sum_i U_{\mu i} U_{e i}^* F(x_i) + \sum_I X_{\mu I} X_{eI}^* F(x_I) \right) \nonumber \\
&=& 
\frac{e m_\mu}{16\pi^2} \frac{G_F}{\sqrt2}  \sum_I X_{\mu I} X_{eI}^* \tilde F(x_I), 
\end{eqnarray}
where
\begin{equation}
\tilde F(x) = F(x) - F(0) = - \frac{x(1-6x+3x^2 + 2x^3 - 6x^2 \ln x)}{(1-x)^4}.
\label{tildeF}
\end{equation}
One finds\footnote{
The decay width in our convention of the amplitude is 
\begin{equation}
\Gamma(\mu \to e\gamma) = \frac{m_\mu^3}{16 \pi^2} (|A_L|^2 + |A_R|^2).
\end{equation}
The branching ratio can be written as 
\begin{equation}
{\rm Br} (\mu \to e\gamma) = \frac{3 \alpha}{16 \pi} ( | \tilde A_L |^2 + |\tilde A_R|^2 ),
\end{equation}
where the dimensionless amplitude $\tilde A_{L,R}$ is defined by
\begin{equation}
A_{L,R} = \frac{e m_\mu}{16\pi^2} G_F \tilde A_{L,R}.
\end{equation}
} 
\begin{equation}
{\rm Br} ( \mu \to e \gamma) = \frac{3\alpha}{32 \pi} 
\left| \sum_I X_{\mu I} X^*_{e I}  \tilde F\left( 
x_I
 \right)\right|^2,
\end{equation}
and the bound of Br$(\mu \to e \gamma) < 4.2 \times 10^{-13}$
requires 
\begin{equation}
\left| \sum_I X^*_{\mu I} X_{e I}  \tilde F\left( x_I 
 \right)\right| \alt 4.4 \times 10^{-5}.
\end{equation}

\item
In the box loop contribution of the Mu-to-$\mubar$ transition in Eq.(\ref{eq:MutoMubar-neutrino}), 
the $E_0$ term is generated by flavor violation, and therefore, its magnitude is bounded by 
$\mu \to e\gamma$.
The $E_1$ term, on the other hand, is generated by the Majorana property of heavy neutrinos
 even if there is no flavor violation in principle,
i.e., the $E_1$ term can be enlarged without a constraint from $\mu \to e \gamma$,
if $X_{eI}$ and $X_{\mu J}$ ($I \neq J$) can become large.

\item
One finds 
\begin{equation}
E_0 (x,y) \sim \frac14 \frac{xy}{x-y} \ln \frac{x}{y},
\qquad
E_1 (x,y) \sim \frac12 \sqrt{xy} \frac{y \ln x - x \ln y}{x-y},
\end{equation}
and
\begin{equation}
E_0 (x,x) \sim \frac14 x,
\qquad
E_1 (x,x) \sim -\frac12 x \ln x,
\end{equation}
for large $x,y$.
Therefore, ``if the neutrino mixings can be kept the same,'' the Mu-to-$\mubar$ transition can be larger for heavier neutrinos.
This is due to the longitudinal modes of gauge bosons in the unitary gauge (or Nambu-Goldstone bosons in 
the 't Hooft-Feynman gauge).

\item
For one generation ($2\times 2$ neutrino mass matrix),
the light neutrino mass in type-I seesaw is $m_\nu = m_D^2/M_N$,
and the mixing is $(X_{\alpha I})^2 = m_D^2/M_N^2 = m_\nu/M_N$, and therefore, the Mu-to-$\mubar$ transition is tiny.
For a three-generation case, there is freedom to enlarge the mixings, $X_{e I}$ and $X_{\mu I}$,
while keeping the active neutrino masses tiny.
Therefore, the $E_0$ term can be larger than the naive expectation from the size of the light--heavy neutrino mixing
in one generation.
If two $X_{\alpha I}$'s (say $X_{\alpha 1}$, $X_{\alpha 2}$) are large,
the heavy neutrino masses need to be degenerate, $M_1 = -M_2$ and $X_{\alpha 1} = X_{\alpha 2}$
(or conventionally, $M_1 = M_2$ and $X_{\alpha 1} = i X_{\alpha 2}$),
due to the freedom of the mass matrix.
Such degeneracy can eliminate the $E_1$ contribution.
If there are more than three singlet neutrinos, such degeneracy can be released.

\item
If the light--heavy neutrino mixing is enlarged,
a sizable active neutrino mass can be generated by $Z$ boson loop \cite{Pilaftsis:1991ug},
\begin{equation}
(M_\nu)_{\alpha\beta} \simeq \frac{\alpha_2}{4\pi \cos^2 \theta_W} \sum_I X_{\alpha I} X_{\beta I} \frac{M_I^3}{M_I^2 - M_Z^2} \ln \frac{M_Z^2}{M_I^2}.
\end{equation}
The loop-induced neutrino mass can be canceled if the heavy neutrino masses are degenerate ($M_1 = - M_2$).
Therefore, the $E_1$ contribution cannot be enlarged, unless the tree-level and loop-induced active neutrino masses
are miraculously canceled.
(Even if one allows such unnatural cancellation, the size of the coefficient is $|G_1| \alt O(10^{-6}) G_F$ due to the
constraints of light--heavy neutrino mixings from precision electroweak data, and near-future experiments cannot reach it.)

\end{enumerate}

In total, the $E_0$ contribution is bounded by $\mu \to e\gamma$ constraint,
and the $E_1$ contribution is bounded by the natural neutrino mass hierarchy.

Since the cancellation between the tree-level and loop-induced neutrino masses cannot be controlled by
symmetry, an elaborated construction of the neutrino mass model is needed (e.g.
the Dirac neutrino mass is forbidden) to enlarge the Mu-to-$\mubar$ transition naturally from the neutrino Majorana property,
which we will see in Section \ref{sec7}.
Here, we exhibit the Mu-to-$\mubar$ transition assuming the heavy neutrino mass degeneracy,
which can be controlled by a flavor symmetry. 

\begin{figure}
\center
\includegraphics[width=8cm]{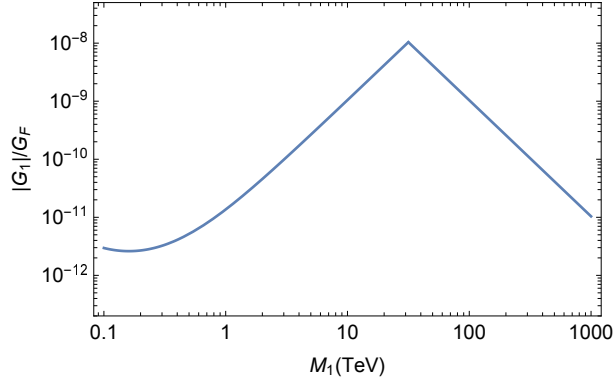}
\caption{
The upper bound of $|G_1|/G_F$ as a function of the heavy neutrino mass.
For the neutrino mass to be less than $\sim 30$ TeV, the Mu-to-$\mubar$ transition is bounded by
the $\mu \to e\gamma$ constraint.
}
\label{fig1}
\end{figure}

We assume that the right-handed neutrino mass matrix $M_N$ is given as
\begin{equation}
M_N = \left(
 \begin{array}{ccc}
  0 & 0 & M_1 \\
  0 & M_3 & 0 \\
  M_1 & 0 & 0
 \end{array}
\right).
\end{equation}
Then, the light neutrino mass after seesaw is 
\begin{equation}
-(M_\nu)_{\alpha\beta} = \frac{ (m_D)_{\alpha 1} (m_D)_{\beta 3}+ (m_D)_{\alpha 3} (m_D)_{\beta 1}}{M_1} + \frac{(m_D)_{\alpha 2} (m_D)_{\beta 2}}{M_3}.
\end{equation}
The heavy neutrino masses are $M_1$, $M_2 \, ( = -M_1)$, and $M_3$.
The light--heavy neutrino mixings are approximately
\begin{equation}
X_{\alpha 1} = X_{\alpha 2} \simeq \frac{ (m_D)_{\alpha 3} }{M_1} ,  \qquad
X_{\alpha 3} \simeq 0.
\end{equation}
To obtain the proper size of light neutrino masses with sizable light--heavy neutrino mixings, $(m_D)_{\alpha 1}$ needs to be small.

We plot the upper bound of $ |G_1|/G_F $ in the above setup in Fig.\ref{fig1}.
For $M_1 \alt 30$ TeV,
the Dirac mass $(m_D)_{\alpha 3}$ is chosen just to satisfy the $\mu \to e\gamma$ bound.
For $M_1 \agt 1$ TeV, the loop function $\tilde F$ for $\mu \to e\gamma$ does not depend on $M_1$
very much (due to the longitudinal modes of the gauge bosons in the unitary gauge),
and thus the maximal value of $|X_{e1} X_{\mu 1}|$ does not depend on $M_1$. 
Therefore, 
in the region of $1\ {\rm TeV} \alt M_1 \alt 30$ TeV, the upper bound of the Mu-to-$\mubar$ transition 
behaves as $|G_1| \propto M_1^2$
because of
$4E_0 \sim M_1^2/M_W^2$ for $M_1 \gg M_W$.
For $M_1 \agt 30$ TeV, the $\mu \to e \gamma$ bound can satisfy for $(m_D)_{\alpha 3} < 100$ GeV,
and then, the upper bound behaves as $|G_1| \propto 1/M_1^2$.

\section{Type-II seesaw model}
\label{sec5}

\begin{figure}
\center
\includegraphics[width=7cm]{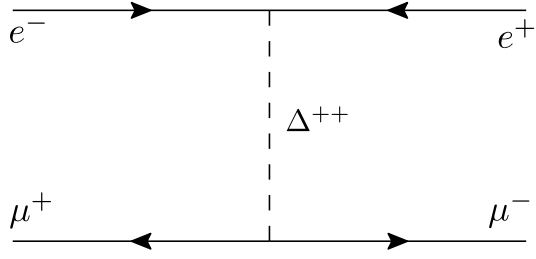}
\caption{The tree-level exchange of a doubly charged scalar boson $\Delta^{++}$ to induce the Mu-to-$\mubar$ transition.
}
\label{fig:doubly-charge-scalar}
\end{figure}

The doubly charged scalar can couple with two charged leptons,
and the Mu-to-$\mubar$ transition can be induced by the exchange of it at the tree level \cite{Chang:1989uk,Swartz:1989qz}
as shown in Fig.\ref{fig:doubly-charge-scalar}.
The doubly charged scalar which can couple to right-handed charged leptons
is a $SU(2)_L$ singlet with hypercharge $Y= 2$,
while a $SU(2)_L$ triplet scalar with hypercharge $Y= 1$ can couple to the left-handed lepton doublets $\ell$.

In the type-II seesaw model \cite{Schechter:1980gr,Cheng:1980qt,Lazarides:1980nt,Mohapatra:1980yp},
the neutrino masses are generated by a vacuum expectation value (vev) of the neutral component of the $SU(2)_L$ triplet scalar:
\begin{equation}
(\Delta_L)_{ab} = 
\left(
 \begin{array}{cc}
  \Delta_L^{++} & \Delta_L^+/\sqrt2 \\
  \Delta_L^+/\sqrt2 & \Delta_L^{0}
 \end{array}
\right).
\end{equation}
Therefore, the neutrino mass generation in the type-II seesaw model
can be related to the Mu-to-$\mubar$ transition \cite{Han:2021nod}.

The Lagrangian of the type-II seesaw can be written as
\begin{equation}
-{\cal L} = \left(\frac12 \kappa^L_{ij} \overline{(\ell_{iL})^c} \ell_{jL} \Delta_L + \mu_\Delta H H  \Delta_L^* +  h.c.\right) + M_\Delta^2 |\Delta_L|^2,
\end{equation}
where $H$ is a Higgs doublet with hypercharge $Y=1/2$,
and $\mu_\Delta$ is a dimensionful scalar trilinear coupling.
%
The ``$\ell \ell \Delta_L$'' term is written down as
\begin{equation}
\overline{(\ell_{iL})^c} \ell_{jL} \Delta_L
=
\overline{\nu^c_{iL}} \nu_{jL} \Delta_L^0
- \frac1{\sqrt2} \overline{\nu^c_{iL}} e_{jL} \Delta_L^+ - \frac1{\sqrt2} \overline{e^c_{iL}} \nu_{jL} \Delta_L^+
+ \overline{e^c_{iL}} e_{jL} \Delta_L^{++}.
\end{equation}
By integrating out $\Delta_L$, the dimension-five neutrino mass operator (so-called Weinberg operator, ``$\ell \ell H H$'') can be generated,
which can be also interpreted to mean that the vev of $\Delta_L^0$ is $v_L \equiv \langle \Delta_L^0 \rangle
= - \mu_\Delta \langle H^0 \rangle^2/M_\Delta^2$,
and the type-II neutrino mass is
\begin{equation}
M_\nu^{\rm II} = \kappa^L v_L.
\end{equation}

As mentioned, the type-II seesaw Lagrangian contains the doubly charged scalar couplings to the left-handed 
charged leptons.
We here describe it using two-component spinor convention to avoid the complication of the expressions\footnote{
\begin{equation}
{\sf e}_i {\sf e}_j = \overline{e^c_i} P_L e_{j} = e_i^T C P_L e_j.
\end{equation}

In the chiral representation, the four-component spinor can be expressed as
\begin{equation}
e = 
\left( 
\begin{array}{c} 
{\sf e} \\ 
\overline {\sf e^c}
\end{array} 
\right),
\quad
e^c = 
\left( 
\begin{array}{c} 
{\sf e^c} \\ 
\overline {\sf e}
\end{array} 
\right),
\quad
\bar e = 
\left( 
\begin{array}{cc} 
{\sf e}^c &
\overline {\sf e}
\end{array} 
\right),
\quad
\overline{ e^c} = 
\left( 
\begin{array}{cc} 
{\sf e} &
\overline {\sf e^c}
\end{array} 
\right).
\end{equation}
%
},
\begin{equation}
- {\cal L} \supset \frac12 \kappa^L_{ij}\, {\sf e}_i {\sf e}_j \Delta_L^{++} + \frac12 \kappa^{L*}_{ij} \,\overline{\sf e}_i \overline{\sf e}_j \Delta_L^{--}
 + M_{\Delta}^2 \Delta_L^{--} \Delta_L^{++},
\end{equation}
where ${\sf e}$ denotes the two-component spinor.
Integrating out $\Delta_L^{++}  = (\Delta_L^{--} )^*$
by equation of motion, $1/2 \kappa^L_{ij} {\sf e}_i {\sf e}_j + M_\Delta^2 \Delta^{--} = 0$,
 one obtains
\begin{equation}
- {\cal L} \supset - \frac14  \frac{1}{M^2_{\Delta}} \kappa^L_{ij} \kappa^{L*}_{kl} ( {\sf e}_i {\sf e}_j  ) ( \overline{\sf e}_k \overline{\sf e}_l) .
\end{equation}
Using
\begin{equation}
\delta_{\beta}^{\alpha} \delta^{\dot \beta}_{\dot \alpha} 
= \frac12 \sigma^\mu_{\beta \dot\alpha} \bar\sigma_\mu^{\dot\beta \alpha},
\end{equation}
one finds
\begin{equation}
( {\sf e}_i {\sf e}_j ) (  \overline{\sf e}_k \overline{\sf e}_l )
 = \frac12 (\overline{\sf e}_l \bar\sigma^\mu {\sf e}_i ) (\overline{\sf e}_k \bar\sigma_\mu {\sf e}_j).
\end{equation}
Expressing it in four-component fermion convention, we obtain 
\begin{equation}
- {\cal L} \supset - \frac18 \frac{1}{M_\Delta^2} \kappa^L_{ij} \kappa^{L*}_{kl} (\overline{e_l} \gamma^\mu P_L e_i) (\overline{e_k} \gamma_\mu P_L e_j) ,
\label{eeee}
\end{equation}
and the coefficient of the transition operator can be written as 
\begin{equation}
\frac{G_1}{\sqrt2} = - \frac{\kappa^L_{ee} \kappa^{L*}_{\mu\mu}}{32 M_{\Delta}^2}=
- \frac{1}{32 v_L^2 M_\Delta^2} (M_\nu^{\rm II})_{ee} (M_\nu^{\rm II})^*_{\mu\mu}.
\end{equation}
%
%
The four-Fermi operator Eq.(\ref{eeee}) can generate LFV decays
\begin{equation}
{\rm Br} (l_a^- \to l_b^+ l_c^- l_d^-) 
= \frac{1}{2(1+\delta_{cd})}  \left| 
\frac{\kappa^L_{ab} \kappa^{L*}_{cd}}{4G_F M_\Delta^2}
 \right|^2 \times {\rm Br} (l_a^- \to l_b^-  \nu\bar\nu) .
\end{equation}
The $\mu \to 3e$ decay process gives the most stringent constraint to the Mu-to-$\mubar$ transition in the model:
\begin{equation}
 {\rm Br}(\mu \to 3 e) <1.0 \times 10^{-12}.
\end{equation}
We find 
\begin{equation}
\frac{|G_1|}{G_F} = \sqrt{{\rm Br} (\mu \to 3e)}
\frac1{2\sqrt2} \left| \frac{\kappa^L_{\mu \mu}}{\kappa^L_{e\mu}} \right| \alt 3.5 \times 10^{-7} \left| \frac{\kappa^L_{\mu \mu}}{\kappa^L_{e\mu}} \right| .
\end{equation}

We suppose that the type-II term dominates the active neutrino mass
(e.g., there is no right-handed neutrino, or 
the type-I contribution is negligible for the right-handed neutrinos to be very heavy),
and the type-II neutrino mass matrix is 
\begin{equation}
M_\nu^{\rm II} = U_{\rm PMNS}^* {\rm diag} (m_1  e^{i \alpha_1}, m_2 e^{i\alpha_2} , m_3) U_{\rm PMNS}^\dagger,
\label{PMNS-para}
\end{equation}
where $U_{\rm PMNS}$ is the Pontecorvo-Maki-Nakagawa-Sakata (PMNS) neutrino mixing matrix 
given by Particle Data Group convention \cite{Zyla:2020zbs},
and $m_i$'s are the active neutrino masses.
Naively,
one obtains 
\begin{equation}
\frac{\kappa^L_{e\mu}}{\kappa^L_{\mu\mu} } 
\sim O\left(\theta_{13},\sqrt{{\Delta m^2_{\rm sol}}/{\Delta m^2_{\rm atm}}}\right),
\end{equation}
%
and $|\kappa^L_{e\mu}/\kappa^L_{\mu\mu}|$ is roughly $ 10$\dash$20$\%.
Then, $|G_1|/G_F$ is smaller than $ O(10^{-6})$, which cannot be observed in near-future experiments. 
However, 
the observed neutrino mixings can be realized even if $\kappa^L_{e\mu} \to 0$.
From the viewpoints of the masses and mixings,
$\kappa^L_{e\mu} \to 0$ can happen if
\begin{enumerate}
\item
The neutrino masses are degenerate, $m_1 e^{i\alpha_1} \simeq m_2 e^{i\alpha_2}( \simeq m_3)$.

\item
$\sum U_{ei}^* U_{\mu i}^* m_i e^{i \alpha_i}$ is accidentally canceled,
which can happen since $\theta_{13}\sim \sqrt{{\Delta m^2_{\rm sol}}/{\Delta m^2_{\rm atm}}}$
and $\theta_{23}, \theta_{12} \sim O(1)$.
\end{enumerate}

%
In the case of the degenerate solution,
the transition probability is maximized (for fixed $v_L$ and $M_\Delta$),
and then, the half-life of neutrinoless double beta decay can be just above the current bound
(if other experimental data allow the solution).

\begin{figure}
\center
\includegraphics[width=8cm]{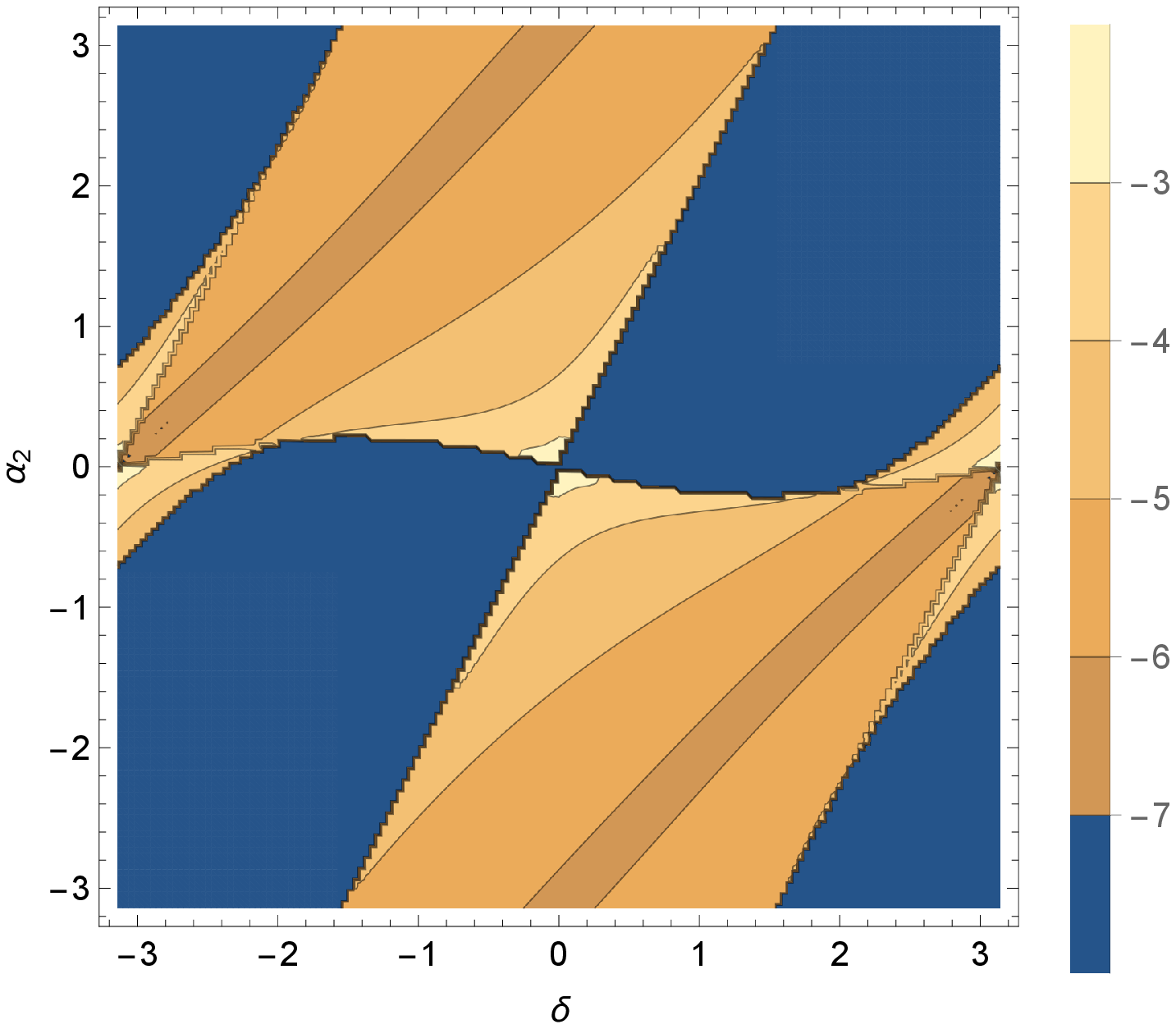} \hspace{3mm}
\includegraphics[width=8cm]{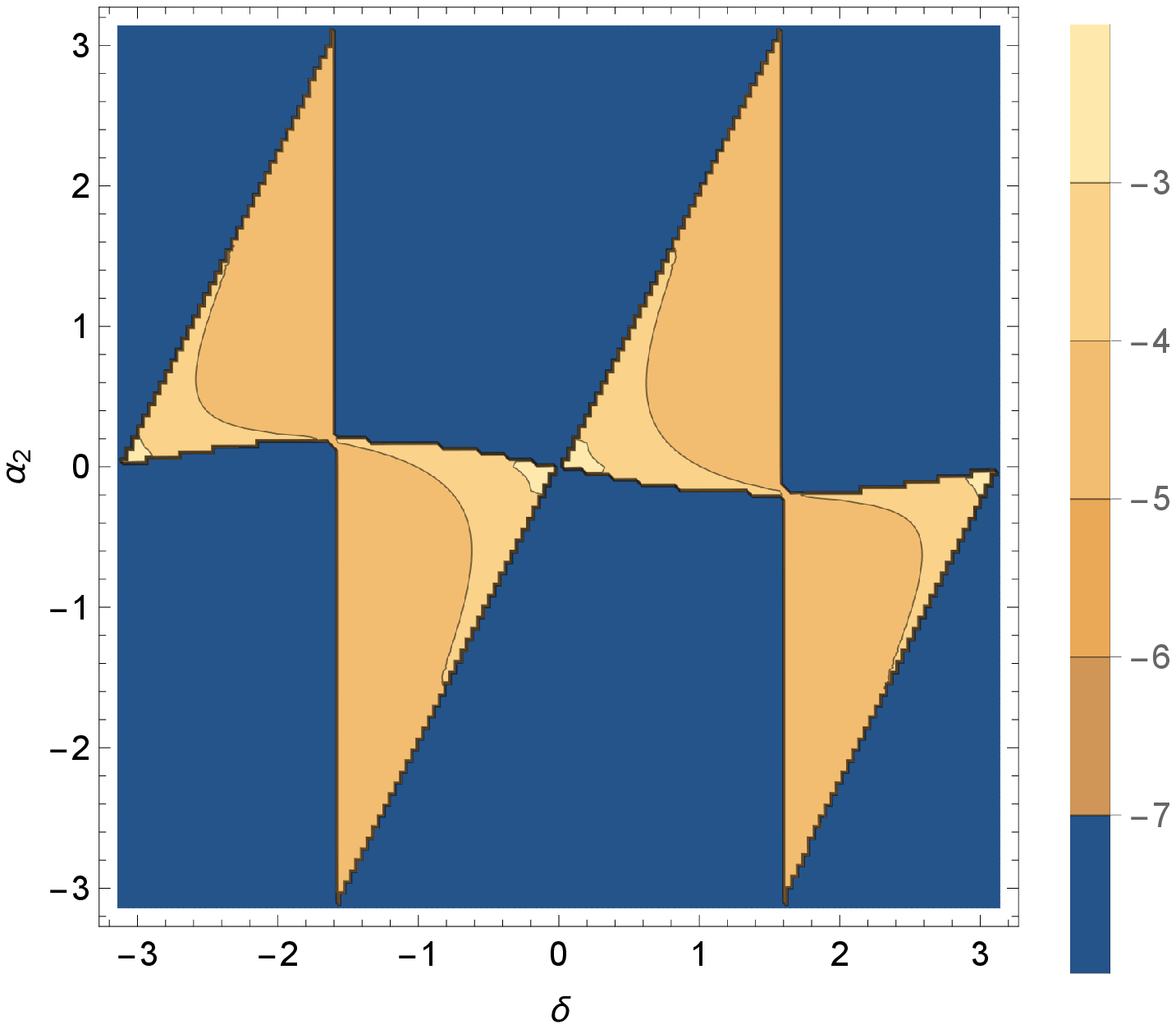}
\caption{
Contour plots of upper bound of $\log_{10} (|G_1|/G_F)$
from LFV decays 
%
%
as a function of $\delta$ and $\alpha_2$ (in radian)
for normal mass ordering (left) and inverted mass ordering (right).
In the dark blue region, there is no solution to make $\kappa^L_{e\mu} \to 0$.
}
\label{fig-type2-ratio}
\end{figure}

Let us calculate the numerical upper bounds of $|G_1|/G_F$ 
allowed by the constraints from LFV decays \cite{Zyla:2020zbs}:
\begin{eqnarray}
&&\{{\rm Br}(\tau \to 3e), {\rm Br}(\tau \to 3\mu) 
\}
< \{2.7,2.1 
 \}\times 10^{-8} , \\
%
&&\{
{\rm Br}(\tau^- \to e^+ \mu^- \mu^-),
{\rm Br} (\tau^- \to \mu^+ e^- e^-) \}
< \{
1.7,    
1.5  
\}\times 10^{-8} .
\end{eqnarray}
The LFV decay bounds can give the upper bounds of the Mu-to-$\mubar$ transition as
\begin{eqnarray}
\frac{|G_1|}{G_F}& \simeq& \sqrt{   \frac{{\rm Br} (\tau \to 3e)}{{\rm Br} (\tau \to e \nu\bar\nu)}     }
\frac1{2\sqrt2} \left| \frac{\kappa^L_{\mu \mu}}{\kappa^L_{e\tau}} \right| \alt 1.4 \times 10^{-4} \left| \frac{\kappa^L_{\mu \mu}}{\kappa^L_{e\tau}} \right| , \\
\frac{|G_1|}{G_F}& \simeq& \sqrt{   \frac{{\rm Br} (\tau^- \to \mu^+ e^- e^-)}{{\rm Br} (\tau \to \mu \nu\bar\nu)}     }
\frac1{2\sqrt2} \left| \frac{\kappa^L_{\mu \mu}}{\kappa^L_{\mu\tau}} \right| \alt 1.0 \times 10^{-4} \left| \frac{\kappa^L_{\mu \mu}}{\kappa^L_{\mu\tau}} \right| , \\
\frac{|G_1|}{G_F}& \simeq& \sqrt{   \frac{{\rm Br} (\tau^- \to e^+ \mu^-\mu^-)}{{\rm Br} (\tau \to e \nu\bar\nu)}     }
\frac1{2\sqrt2} \left| \frac{\kappa^L_{e e}}{\kappa^L_{e\tau}} \right| \alt 1.1 \times 10^{-4} \left| \frac{\kappa^L_{ee}}{\kappa^L_{e\tau}} \right| , \\
\frac{|G_1|}{G_F}& \simeq& \sqrt{   \frac{{\rm Br} (\tau \to 3\mu)}{{\rm Br} (\tau \to \mu \nu\bar\nu)}     }
\frac1{2\sqrt2} \left| \frac{\kappa^L_{ee}}{\kappa^L_{e\tau}} \right| \alt 1.2 \times 10^{-4} \left| \frac{\kappa^L_{ee}}{\kappa^L_{e\tau}} \right| .
\end{eqnarray}
The $\mu \to e \gamma$ (and $\mu$\dash$e $ conversion \cite{Raidal:1997hq}) constraint
can be written as\footnote{
\begin{equation}
{\rm Br} (\mu \to e\gamma) = \frac{3\alpha}{16\pi} \left| 
\frac{\kappa^L_{ei} \kappa^{L*}_{\mu i} }{3 G_F} \left( \frac1{M_{\Delta^{++}}^2} + \frac1{8M_{\Delta^+}^2 } \right) 
\right|^2.
\end{equation}
The $SU(2)_L$ triplet contains a single charged scalar, which contributes $\mu \to e\gamma$.
}
\begin{equation}
\frac{|\kappa^L_{e\tau} \kappa^{L*}_{\mu \tau}|}{M_\Delta^2} \alt 1 \times 10^{-3} \times \frac{1}{(1\, {\rm TeV})^2},
\end{equation}
which is interpreted as
\begin{equation}
\frac{|G_1|}{G_F} < 3.8 \times 10^{-6} \left| \frac{\kappa^L_{ee} \kappa^L_{\mu\mu} }{ \kappa^L_{e\tau} \kappa^L_{\mu\tau} } \right|.
\label{mueg-doubly-charge}
\end{equation}
%
%
We solve the equation $\kappa^L_{e\mu} = 0$ by $m_1 e^{i\alpha_1}$ to satisfy the severe $\mu \to 3e$ bound.
Eliminating $m_1 e^{i \alpha_1}$ from the equation,
we obtain the neutrino mass matrix 
as a function of a Dirac phase $\delta$ in the PMNS matrix and
a Majorana phase $\alpha_2$.
%
%
%
Therefore, the upper bounds of the Mu-to-$\mubar$ transition are obtained as shown in Fig.\ref{fig-type2-ratio}.
Surely, the plot
in Fig.\ref{fig-type2-ratio}
is symmetric under $\delta \to -\delta$ and $\alpha_2 \to - \alpha_2$,
because of $G_1 \to G_1^*$.
Near $\delta \sim 0, \pi$ (and $\alpha_2 \sim 0$),
the degenerate solution can be obtained, and thus, 
the Mu-to-$\mubar$ transition can be largest there.
In the inverted hierarchy case, 
$m_1$ and $m_2$ are degenerate by themselves,
and the Mu-to-$\mubar$ transition can be large at all the points
(if there is a solution to make $\kappa^L_{e\mu} \to 0$).
In the normal hierarchy case,
$\kappa^L_{e\mu}$ can be canceled even without mass degeneracy.
Actually, both $\kappa^L_{ee}$ and $\kappa^L_{e\mu}$ can be small 
to reproduce the neutrino oscillation data.
Therefore, there is a band where the Mu-to-$\mubar$ transition is small in the plot.

\begin{figure}
\center
\includegraphics[width=8.1cm]{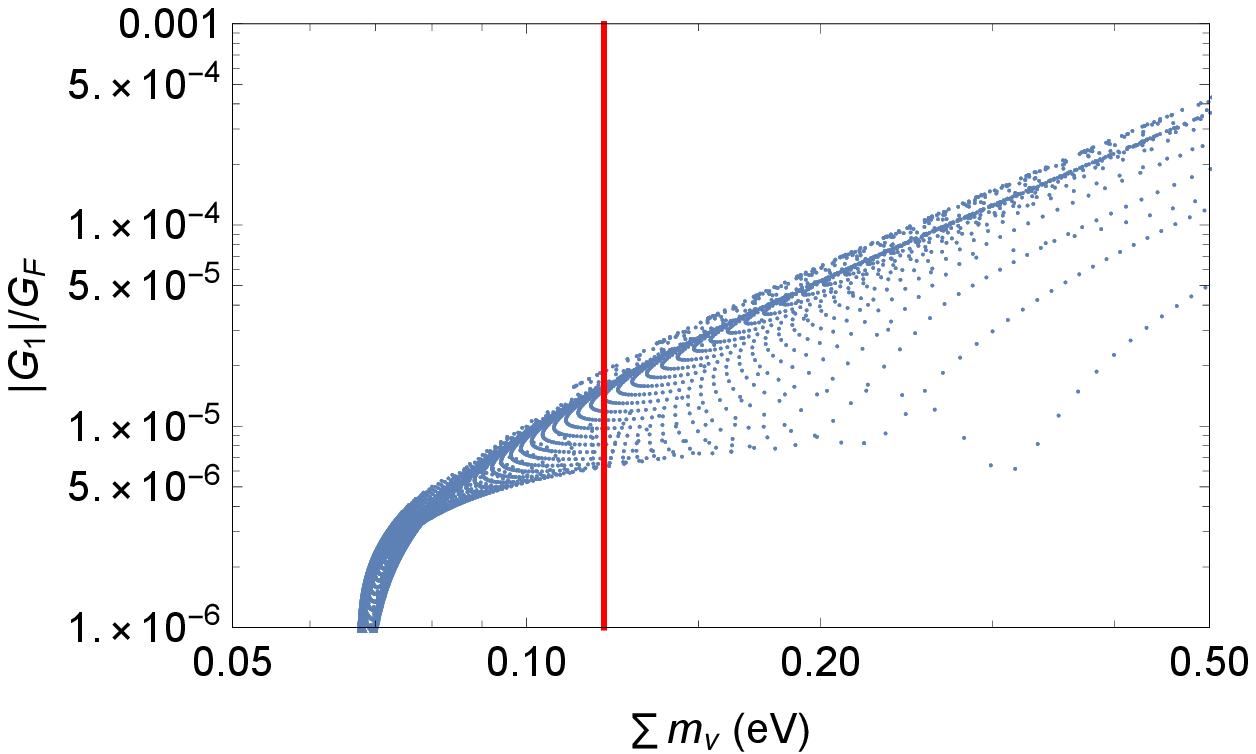} \hspace{3mm}
\includegraphics[width=8cm]{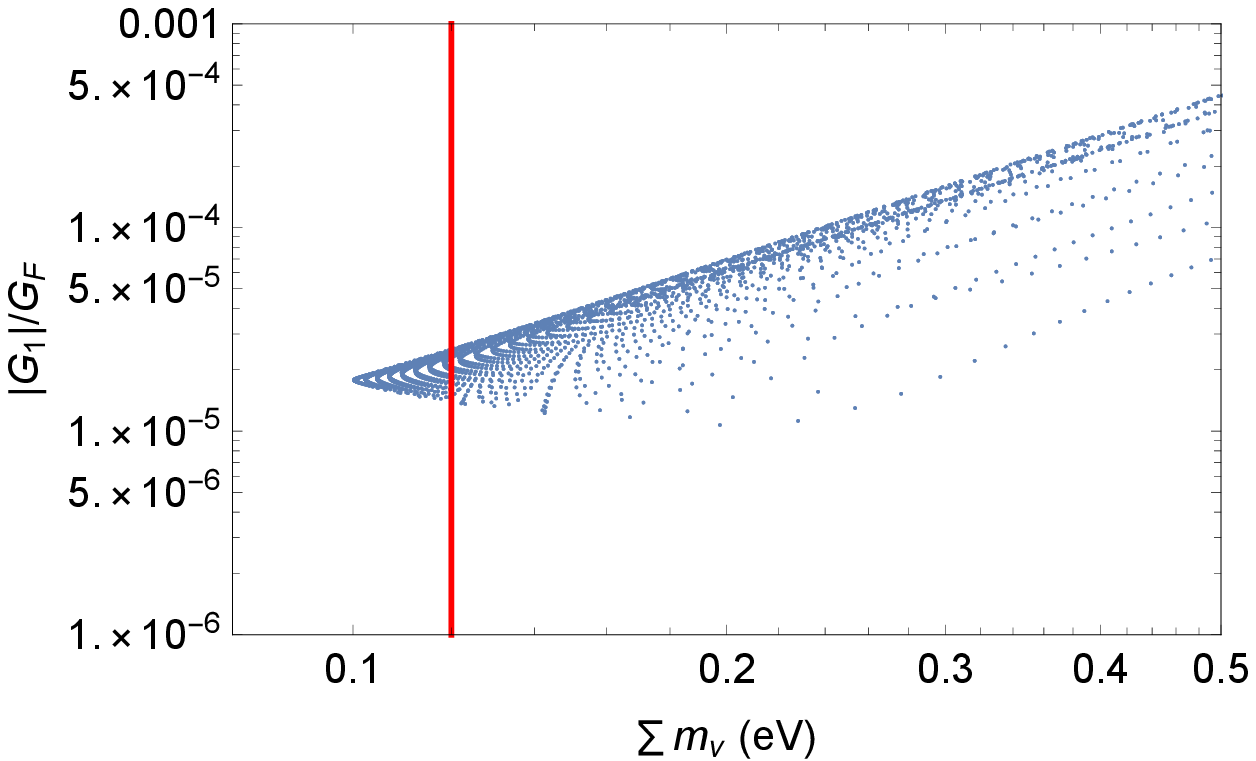}
\caption{
The plot of $|G_1|/G_F$ allowed by LFV constraints versus the total neutrino mass
for normal mass ordering (left) and inverted mass ordering (right).
The vertical red line shows $\sum m_\nu = 0.12$ eV.
}
\label{fig-type2-sumnu}
\end{figure}

The current strongest bound of the absolute neutrino mass is
from cosmological measurement:
the total neutrino mass
$\Sigma\, m_\nu < 0.12$ eV \cite{Aghanim:2018eyx}.
Therefore, unless there is a loophole (e.g. the neutrinos are not stable in the cosmological 
time scale \cite{Chacko:2019nej}),
the solution of the large degree of degeneracy is excluded and 
the Mu-to-$\mubar$ transition is bounded.
In Fig.\ref{fig-type2-sumnu},
we show the plot of the coefficient of the Mu-to-$\mubar$ transition operator
versus the total neutrino mass.
The shown Mu-to-$\mubar$ transition in the plot is the upper bound from LFV as described above,
generated by a mesh of $\delta$ and $\alpha_2$.
The cosmological measurements bound the Mu-to-$\mubar$ transition
as $|G_1|/G_F \alt O(10^{-5})$.

If we adopt the type-I seesaw contributions in addition to type-II,
one can tune $\kappa^L_{e\mu} $ to be zero irrespective of the neutrino masses and mixings,
and because of the many parameters,
any size of the Mu-to-$\mubar$ transition can be fit satisfying experimental constraints in principle~\cite{Han:2021nod}.
Supposing $\kappa^L_{ee} \sim \kappa^L_{\mu \mu} \sim 0.3$,
and $M_\Delta = 600 $ GeV 
(the experimental bound of the doubly charged scalar mass can be found in \cite{Aaboud:2017qph,Dev:2018sel}),
one finds
\begin{equation}
|G_1| \sim 1 \times 10^{-3}\, G_F,
\end{equation}
which can be soon tested by the near-future experiments.

\section{Left--right model}
\label{sec6}

The $SU(3)_c \times SU(2)_L \times SU(2)_R \times U(1)_{B-L} (\equiv G_{3221})$ gauge theory (left--right model)
to induce the Mu-to-$\mubar$ transition \cite{Pal:1983bf,Herczeg:1992pt}
is one of the representative models
 where the new experimental results in this quarter-century have brought about changes drastically.
In the early 1990s, there was still room that the active neutrinos can lie around $10$\dash$100$ keV.
Surely, the neutrino oscillations exclude the room, and a large mixing between left-handed neutrinos
and SM singlet right-handed neutrinos at such mass scale is not allowed.
The meson mixing data pushes up the $W_R$ gauge boson mass  to be more than 3 TeV \cite{Blanke:2011ry,Bertolini:2014sua}.
The direct LHC data from 
$W_R^+ \to N_R \ell_R^+ \to jj \ell_R^\pm \ell_R^+ $ processes
gives the lower bound of the $W_R$ mass to be more than 4 TeV \cite{Nemevsek:2018bbt,Sirunyan:2018pom,Aaboud:2019wfg}.
Therefore, the resume for the Mu-to-$\mubar$ transition in the quarter-century ago is not valid anymore.

Various experimental constraints in the left--right model, especially on the flavor physics, can be found in
\cite{Barry:2013xxa,Dev:2014xea}.
We note that the same sign and opposite sign of the two-lepton signals 
from the $W_R^+ \to jj \ell_R^\pm \ell_R^+ $ processes
can be a probe of the structure in the neutrino mass matrix,
which is related to the degeneracy of the heavy neutrino masses
\cite{Das:2017hmg}.

The Dirac mass of tau neutrino is supposed to be (at least) of the order of GeV
due to left--right symmetry,
and thus, one needs fine-tuning to obtain the sub-eV active neutrino mass 
in the TeV-scale left--right model.
Therefore, 
an extended seesaw model to generate sub-eV active neutrino mass is often considered
in the TeV-scale left--right model.
We employ three SM singlet fermions $S_i$,
and consider the neutrino mass as
\begin{equation}
-{\cal L} = \frac12 \left(
 \begin{array}{ccc}
  \overline{(\nu^c)_R} & \overline{N_R} & \overline{(S^c)_R}
 \end{array}
\right)
{\cal M}
\left(
 \begin{array}{c}
  \nu_L \\
  (N^c)_L \\
  S_L
 \end{array}
\right) + h.c.,
\end{equation}
where ${\cal M}$ is a $9\times 9$ mass matrix (in the basis where the charged-lepton mass matrix is diagonal),
\begin{equation}
{\cal M} = 
\left(
 \begin{array}{ccc}
  0 & m_D & 0 \\
  m_D^T & \mu_N & M_S \\
  0 & M_S^T & \mu_S 
 \end{array}
\right).
\end{equation}
The light neutrino mass matrix is
\begin{equation}
M_\nu^{\rm light} = m_D ( M_S \mu_S^{-1} M_S^T - \mu_N)^{-1} m_D^T \simeq m_D (M_S^T)^{-1} \mu_S M_S^{-1} m_D^T.
\end{equation}
We suppose that the Majorana mass $\mu_S$ of the singlet $S$ is small,
and then the active neutrino mass can be sub-eV easily even in the TeV-scale left--right model.
This is sometimes called an inverse seesaw.

The Dirac mass $m_D$ comes from the usual Dirac Yukawa coupling to Higgs bi-doublet : 
$({\bf 1},{\bf 2},{\bf 2}, 0)$ under $G_{3221}$,
and $M_S$ comes from the $\Phi \overline{\ell_R} S_L$ coupling
with $\Phi : ({\bf 1},{\bf 1},{\bf 2}, -1)$ under $G_{3221}$.
The vev of $\Phi$ breaks $G_{3221}$ down to SM gauge symmetry.
The Majorana mass $\mu_N$ is generated if 
there is a $SU(2)_R$ triplet $\Delta_R : ({\bf 1},{\bf 1},{\bf 3}, 2)$ 
and it acquires a vev to break $G_{3221}$.
How the Mu-to-$\mubar$ transition is induced in the left--right model depends
on
with or without the $SU(2)_R$ triplet.
In the case without the triplet, the Mu-to-$\mubar$ transition is generated at the loop level,
while in the case with the triplet,
it can be generated at the tree level since the triplet contains the doubly charged scalar.

We parameterize\footnote{
Since the singlet $S$ does not have a reference current basis,
 one can parameterize
the matrix $M_S$ to be given in Eq.(\ref{MS}) without loss of generality
and $\mu_S$ to be a general $3\times 3$ matrix.
}
\begin{eqnarray}
&&m_D = U_0^* \, {\rm diag} (m_D^1,m_D^2,m_D^3)\, V_0^\dagger, \\
&&\mu_N = V_1^* \, {\rm diag} (\mu_N^1,\mu_N^2,\mu_N^3)\,  V_1^\dagger, \\
&&M_S = V_2^*  \, {\rm diag} (M_S^1,M_S^2,M_S^3). \label{MS}
\end{eqnarray}
The convention in the diagonalization of the neutrino mass matrix is given in Appendix \ref{app:diagonalization}.
The matrix $U$ in the $9\times 9$ diagonalization matrix in Eq.(\ref{9x9}),
 which corresponds to nearly the $3\times 3$ PMNS matrix, is
%
a diagonalization matrix of 
\begin{equation}
U_0^* m_D^{\rm diag} V_0^\dagger V_2 (M_S^{\rm diag})^{-1} \mu_S (M_S^{\rm diag})^{-1} V_2^T V_0^* m_D^{\rm diag} U_0^T.
\end{equation}
In the left--right model,
the mixings in $U_0$ and $V_0$ are expected to be small as CKM mixings,
but the structure of $M_S$ and $\mu_S$ can have freedom to generate large neutrino mixings.
Surely, one can also employ a $SU(2)_L$ triplet and consider the type-II seesaw contribution
for the active neutrino mass.

\subsection{Case 1: Without $SU(2)_R$ triplet}
\label{sec6.1}

If there is no $SU(2)_R$ triplet, the Mu-to-$\mubar$ transition
is generated by a box loop diagram.
In addition to the $W_L$\dash $W_L$ loop diagram in Fig.\ref{fig:box1},
we have $W_R$\dash $W_R$ box loop contributions:
\begin{eqnarray}
\frac{G_2}{\sqrt2} =
 \frac{G_F^2 M_{W_L}^2}{16 \pi^2}\frac{g_R^4}{g_L^4}
  \frac1{z} 
  \sum_{I,J}
\left[
Y^*_{\mu {I}} Y_{e { I}}
Y^*_{\mu{J}} {Y}_{e { J}}
 E_0(\tilde x_{I} , \tilde x_{J})+
(Y^*_{\mu {I}})^2 
(Y_{e {J}})^2 
E_1(\tilde x_{I} ,\tilde x_{ J})\right] ,
\end{eqnarray}
%
%
%
%
and $W_L$\dash $W_R$ box loop contributions:
\begin{eqnarray}
\frac{G_3}{\sqrt2} =
- \frac{G_F^2 M_{W_L}^2}{8 \pi^2} \frac{g_R^2}{g_L^2}   \sum_{I,J}
\left[
X_{\mu {I}} Y^*_{\mu { I}}
X^*_{e{J}} {Y}_{e { J}}
E_0( x_{I} ,  x_{J},z)+
X_{\mu I} Y_{e {I}}
X^*_{e J} Y^*_{\mu {J}} 
E_1( x_{I} , x_{ J},z)\right] ,
\end{eqnarray}
where
\begin{equation}
\tilde x_I = \frac{M_{N_I}^2}{M_{W_R}^2},\qquad
 x_I =  \frac{M_{N_I}^2}{M_{W_L}^2}, \qquad z = \frac{M_{W_R}^2}{M_{W_L}^2}.
\end{equation}
The loop functions $E_0$ and $E_1$ are given in Appendix \ref{app:loop_function}.
Strictly speaking, 
since there is a $W_L$\dash$W_R$ mixing due to vevs of Higgs bi-doublets,
the mass eigenstates should be quoted as their mixed states.
We here neglect their mixing in the box contributions.

One can find that 
$E_1$ term in $G_2$ and $E_0$ term in $G_3$ correspond to
the Mu-to-$\mubar$ transition utilized by the Majorana property of the heavy neutrinos.
If there is no $SU(2)_R$ triplet, the Majorana mass of the right-handed neutrino
$\mu_N$ is absent, and the heavy neutrino masses are degenerate
in the setup of the inverse seesaw.
Then, their contributions are canceled.
Therefore, our concerns are $E_0$ term in $G_2$ and $E_1$ term in $G_3$,
which are bounded by $\mu \to e\gamma$.
The $\mu_R \to e_L \gamma$ amplitude via $W_L$ loop is
\begin{equation}
A_R (W_L) =
\frac{e m_\mu}{16 \pi^2} \frac{G_F}{\sqrt2}  
\sum_I X^*_{\mu I} X_{e I}  \tilde F\left( x_I \right),
\end{equation}
where $\tilde F$ is given in Eq.(\ref{tildeF}),
and the $\mu_L \to e_R \gamma$ amplitude via $W_R$ loop is
\begin{equation}
A_L (W_R)= \frac{e m_\mu}{16 \pi^2} \frac{G_F}{\sqrt2} \frac{g_R^2}{g_L^2} \frac1{z} 
\sum_I Y_{\mu I} Y_{e I}^*  \tilde F\left( \tilde x_I \right).
\end{equation}
Because there is a $W_L$\dash$W_R$ mixing $\xi_{LR}$, the chirality can flip at the internal line in the loop and
the decay amplitudes are
\begin{eqnarray}
 A_R (W_L{\mbox \dash}W_R)&=& 
 \frac{e}{16 \pi^2} \xi_{LR} \frac{g_R}{g_L}\frac{G_F}{\sqrt2} 
  \sum_I Y_{\mu I} X_{e I} M_{N_I} \left(G\left( x_I \right) - \frac1{z} G( \tilde x_I ) \right),\\
A_L (W_L{\mbox \dash}W_R)&=&
 \frac{e }{16 \pi^2} \xi_{LR} \frac{g_R}{g_L} \frac{G_F}{\sqrt2} 
  \sum_I X_{\mu I}^* Y_{e I}^* M_{N_I} \left( G\left( x_I \right)- \frac1{z} G( \tilde x_I )\right),
\end{eqnarray}
where
\begin{equation}
G(x) = \frac{2(4-15x+12x^2-x^3-6x^2 \ln x)}{(1-x)^3}.
\end{equation}
The $\mu \to e\gamma$ experimental result implies
\begin{equation}
\left|\sum_I X^*_{\mu I} X_{e I}  \tilde F\left( x_I \right) \right| ,\quad
\frac{g_R^2}{g_L^2} \frac1{z} 
\left| \sum_I Y_{\mu I} Y_{e I}^*  \tilde F\left( \tilde x_I \right) \right| \alt 4 \times 10^{-5},
\end{equation}
if we assume that there is no cancellation in each $A_L$ and $A_R$.
These two constraints restrict the Mu-to-$\mubar$ transition operators from $W_L$\dash $W_L$ box and 
$W_R$\dash $W_R$ box diagrams, respectively.
The $\mu\to e\gamma$ bound via the $W_L$\dash$W_R$ mixing is written as
\begin{equation}
\left| \sum_I Y_{\mu I}^* X_{e I}^* M_{N_I} G(x_I) \right|, \quad
\left| \sum_I Y_{e I}^* X_{\mu I}^* M_{N_I} G(x_I) \right| \alt 40 \ {\rm MeV} \times \frac{g_L}{g_R} \times \frac{10^{-4}}{\xi_{LR}},
\end{equation}
which restricts the Mu-to-$\mubar$ transition from the $W_L$\dash $W_R$ box diagram.
Because of 
\begin{equation}
\sum_I X_{\alpha I}^* Y_{\beta I}^* M_{N_I} \simeq (m_D)_{\alpha\beta},
\end{equation}
those roughly correspond to the bounds of the $e\mu$ and $\mu e$ elements of the Dirac neutrino mass matrix.
We remark that the restriction to the Mu-to-$\mubar$ transition via $W_L$\dash $W_R$ box ($G_3$) is severer 
due to the internal chirality flipping in the $\mu \to e\gamma$ diagram.

\begin{figure}
\center
\includegraphics[width=8cm]{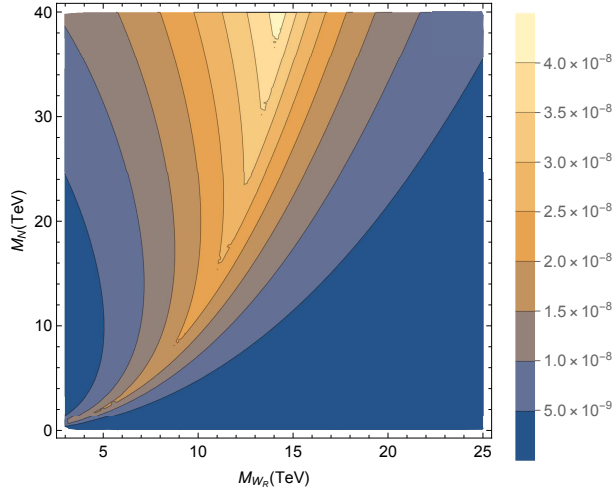}
\caption{
The upper bound of $|G_2|/G_F$ as a function of $M_{W_R}$ and the heavy neutrino mass $M_N$.
}
\label{fig-lr}
\end{figure}

In the case without a $SU(2)_R$ triplet, the Majorana mass $\mu_N =0$ at the tree level.
The flavor violation of the right-handed neutrino is characterized by
$V_2$ in Eq.(\ref{MS}).
To show the evaluation of the size of the Mu-to-$\mubar$ transition,
we assume $M_S^1 \simeq M_S^2$ and $(V_2)_{13} = (V_2)_{23}=0 $ (so that $M_S^3$ does not contribute).
In Fig.\ref{fig-lr},
we show the upper bound of the Mu-to-$\mubar$ transition by the $W_R$ box loop
allowed by $\mu \to e\gamma$ constraint.
We suppose $g_L= g_R$ in the plot.
When $M_N (= M_S^1)$ is fixed,
the $\mu \to e\gamma$ bound to the mixing $(V_2)_{12}$ is relaxed for a heavier $W_R$.
The upper bound of the Mu-to-$\mubar$ transition becomes the largest for a mass of $W_R$
just when the maximal mixing is allowed.
The largest upper bound (for fixed $M_N$) becomes larger 
for larger $M_N$ because of the behavior of the box loop function.
The mass $M_N$ comes from
$\Phi \bar\ell_R S_L$ coupling
and the vev of $\Phi$ gives $W_R$ mass.
Therefore, the mass $M_N$ should not be much larger than $M_{W_R}$,
and the
bound of the Mu-to-$\mubar$ transition is estimated as $|G_2|/G_F \alt O(10^{-8})$.

As we have remarked, the $\mu \to e\gamma$ bound
is stronger in the case of internal chirality flipping.
The Mu-to-$\mubar$ transition 
via $W_L$\dash $W_R$ diagram is bounded to be $|G_3|/G_F \alt O(10^{-10})$
as long as $\xi_{LR} \agt 0.01 M_{W_L}^2/M_{W_R}^2$.
We note that 
the $W_L$\dash $W_R$ mixing $\xi_{LR}$
is proportional to $M_{W_L}^2/M_{W_R}^2$,
and the proportionality coefficient is determined by the ratio of the vevs of Higgs bidoublet,
though we do not describe it in detail in this paper.

\subsection{Case 2: With $SU(2)_R$ triplet}
\label{sec6.2}

If there is $SU(2)_R$ triplet Higgs to break 
$SU(2)_R \times U(1)_{B-L}$,
Majorana masses of the right-handed neutrinos can be generated:
\begin{equation}
-{\cal L} \supset \frac12 \kappa^R_{ij}\, \overline{\ell_{iR}} (\ell^c_j)_L \Delta_R + h.c.,
\end{equation}
\begin{equation}
\mu_N = \kappa^R \langle \Delta_R^0 \rangle.
\end{equation}
The box loop contribution can be larger than in case 1.
In this case, however,
the coupling to induce the Majorana mass
can generate the transition operator at the tree level,
which can be surely larger than the box loop:
\begin{equation}
- {\cal L} \supset - \frac18 \frac{1}{M_\Delta^2} \kappa^R_{ij} \kappa_{kl}^{R*} (\overline{e_i} \gamma^\mu P_R e_l) (\overline{e_j} \gamma_\mu P_R e_k), 
\end{equation}
and
\begin{equation}
\frac{G_2}{\sqrt2} = - \frac{\kappa^{R*}_{ee} \kappa^R_{\mu\mu}}{32 M_{\Delta}^2}.
\end{equation}
When we parameterize the Majorana mass matrix as
\begin{equation}
\mu_N = \tilde V {\rm diag} (\mu_{N1}, \mu_{N2}, \mu_{N3}) \tilde V^T,
\end{equation}
the $\kappa^R$ matrix is written as
\begin{equation}
\kappa^R_{\alpha \beta} = \tilde V_{\alpha I} \tilde V_{\beta I} \mu_{NI}.
\end{equation}


The $\mu \to 3e$ bound, Br($\mu \to 3e) <1.0 \times 10^{-12}$, 
restricts the Mu-to-$\mubar$ transition similarly to the previous,
\begin{equation}
\frac{|G_2|}{G_F} \alt 10^{-6} \times \frac1{2\sqrt2} \left| \frac{\kappa^R_{\mu \mu}}{\kappa^R_{e\mu}} \right|.
\end{equation}

There are three ways to suppress $\kappa^R_{e\mu} = \tilde V_{e1} \tilde V_{\mu 1} \mu_{N1} +  \tilde V_{e2} \tilde V_{\mu 2} \mu_{N2}
+ \tilde V_{e3} \tilde V_{\mu 3} \mu_{N3}$.
\begin{enumerate}
\item
The mixings are small: $\tilde V \simeq {\bf 1}$.

\item
The right-handed neutrino masses are degenerate: $\mu_{N1} \simeq \mu_{N2}$.

\item
The mixings are not small, and the masses are not degenerate,
but the
$\kappa^R_{e\mu}$ is accidentally canceled by the $\mu_{N3}$ contribution.

\end{enumerate}


In models with ``left--right parity" (exchange symmetry $\ell_L \leftrightarrow (\ell_R)^c$),
one obtains
\begin{equation}
\kappa^L = \kappa^R,
\end{equation}
and Yukawa matrices are symmetric.
Therefore, in the case of the type-II dominance ($\mu_S \to 0$),
$\kappa^R$ is also related to the neutrino masses and mixings:
\begin{equation}
\kappa^R \propto M_\nu = U_{\rm PMNS}^* {\rm diag} (m_1  e^{i \alpha_1}, m_2 e^{i\alpha_2} , m_3) U_{\rm PMNS}^\dagger,
\end{equation}
and the Mu-to-$\mubar$ transition is estimated in parallel to the analysis in the type-II seesaw.

In general, there is no reason that $\tilde V_{e2}$ and $\tilde V_{\mu 1}$ are small in the model construction in the left--right model.
Rather, the mixing is not small in the unification scenarios,
and the Mu-to-$\mubar$ transition is much smaller than the near-future experimental reach.
If we do not go beyond the left--right symmetry,
a global discrete flavor symmetry to suppress LFV can be assigned in the lepton sector 
and $\tilde V \simeq {\bf 1}$.
(Large neutrino mixings can originate from a hidden sector with singlet fermions,
where the discrete symmetry is broken.)
The right-handed neutrinos
 (more precisely, mass eigenstates of the heavy neutrinos from $N_R$ and $S$ for $\mu_N \sim M_S$), as well as $W_R$ gauge boson, should be heavier than 
$4$\hspace{1pt}--\hspace{1pt}$5$ TeV to satisfy 
the bound from the
$W_R^+ \to N_R \ell_R^+ \to jj \ell_R^\pm \ell_R^+ $ processes at the LHC
\cite{Nemevsek:2018bbt,Sirunyan:2018pom,Aaboud:2019wfg}.
The doubly charged scalar mass, on the other hand, can be around 1 TeV \cite{Aaboud:2017qph,Dev:2018sel},
%
and thus, the Mu-to-$\mubar$ transition with $|G_2|/G_F \sim 10^{-3}$
can be obtained,
which can be tested by the near-future Mu-to-$\mubar$ transition experiments
along with the direct search at high-luminosity LHC experiment.


\section{Radiative neutrino mass}
\label{sec7}

There are plenty of models in which the neutrino masses are
induced radiatively.
The models can be roughly classified into two groups.

(1) There is no SM singlet fermion.

(2) There are SM singlet fermions, but the Dirac neutrino Yukawa coupling, $N H \ell$, is forbidden by a discrete symmetry. 

The representative model for (1) is called Zee-Babu model
\cite{Zee:1980ai,Zee:1985id,Babu:1988ki,Babu:2002uu,McDonald:2003zj,Nebot:2007bc,Herrero-Garcia:2014hfa}.
The improved version of the model has a hypercharge $\pm 2$, $SU(2)_L$ singlet scalar, 
which is a doubly charged scalar and can be a mediator to induce the transition operator.
Neutrino masses are generated at the two-loop level.
The model where the neutrino masses are induced at a three-loop level is also considered
(so-called Cocktail model) \cite{Gustafsson:2012vj,Cepedello:2020lul}.

In the models for (2), the tree-level active neutrino masses are forbidden by discrete symmetries.
Because the discrete symmetries can be exploited,
the models are often discussed together with dark matter candidates~\cite{Ma:2006km}.
As we have studied in Section \ref{sec4},
the enlargement of the Mu-to-$\mubar$ transition from the Majorana property suffers from 
the natural neutrino mass hierarchy due to 
the light--heavy neutrino mixings induced by the Dirac neutrino masses. 
Because of the absence of the Dirac neutrino mass, the models are also suitable to discuss 
the Mu-to-$\mubar$ transition from the Majorana property.
The model for (2) has
a Yukawa coupling $N \eta \ell\, (= N \eta^+ e_L - N \eta^0 \nu_L)$ 
to generate the neutrino mass at the one-loop level,
where the neutral component of the $SU(2)_L$ doublet $\eta$
does not acquire a vev ($\eta$ is often called an inert Higgs doublet).
Alternatively,
the model has a $N S^+ e_R$ type coupling ($S^+$ is a hypercharge $+1$ $SU(2)_L$ singlet scalar),
and 
the neutrino masses are generated at the three-loop level \cite{Cepedello:2020lul,Krauss:2002px,Aoki:2008av}.
The  $N \eta^+ e_L$ and $N S^+ e_R$ couplings can induce the transition operators
via box diagrams.

\subsection{Models with doubly charged scalar}
\label{sec7.1}

\subsubsection{Zee-Babu model}
\label{sec7.1.1}

In the Zee-Babu model, 
there are $SU(2)_L$ singlet scalars, $h^+$ and $k^{++}$, with hypercharge $Y=1$ and $Y=2$, respectively.
The couplings to the leptons and the masses of the scalars are given as 
%
\begin{equation}
-{\cal L} \supset ( f_{ij} \overline{\ell_{i}^c} \cdot \ell_j h^+ + g_{ij} \overline{e_i} e_j^c k^{--} 
+ \mu_{hhk}\, h^+ h^+ k^{--}
+ h.c.) + m_h^2 h^- h^+ + m_k^2 k^{--} k^{++}, 
\end{equation}
where `$\cdot$' stands for the contraction of the $SU(2)_L$ doublet:
$A\cdot B \equiv \epsilon_{ab} A_a B_b = A_1 B_2 - A_2 B_1$.
The coupling matrix $f$ is antisymmetric under the flavor index, and
$g$ is symmetric.
The scalar trilinear coupling $\mu_{hhk}$ violates the lepton number symmetry.

\begin{figure}
\center
\includegraphics[width=7cm]{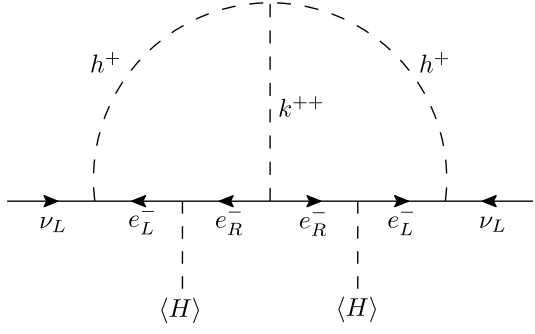}
\caption{
The diagram to induce the neutrino masses in the Zee-Babu model.
The symbol $\langle H \rangle$ stands for the vev of the SM Higgs boson.
}
\label{figZB-diagram}
\end{figure}

The neutrino mass can be induced by two-loop diagram in Fig.\ref{figZB-diagram}, and
the mass matrix is given as
\begin{equation}
M_\nu = \frac{1}{M_0} f M_e g M_e f^T,
\label{ZB}
\end{equation}
where $M_e = {\rm diag} (m_e, m_\mu, m_\tau)$, 
\begin{equation}
\frac1{M_0} = \frac{\mu_{hhk}}{48 \pi^2\, {\rm max} (m_h^2,m_k^2)} \tilde I,
\end{equation}
and 
the loop function is approximately given as \cite{McDonald:2003zj,Nebot:2007bc,Herrero-Garcia:2014hfa}
\begin{equation}
\tilde I \simeq 
\left\{
\begin{array}{cc}
 1 & \mbox{for }\  m_k \ll m_h, \\
 1+ \frac{3}{\pi^2}\left( \ln^2 \frac{m_k^2}{m_h^2} -1\right) & \mbox{for } \ m_k \gg m_h.
\end{array}
\right.
\end{equation}

Because $f$ is anti-symmetric, the neutrino mass matrix is rank 2 (i.e., $m_1 = 0$),
and the neutrino mass matrix in the normal mass hierarchy is given by the PMNS matrix $U$ as
\begin{equation}
M_\nu = U^* {\rm diag} (0, m_2, m_3) U^\dagger = m_2 u_2^* u_2^\dagger + m_3 u_3^* u_3^\dagger,
\end{equation}
where 
$m_i$'s are the active neutrino masses as used throughout this paper,
and $u_2$ and $u_3$ are column vectors in $U = (u_1, u_2, u_3)$:
\begin{equation}
u_1 = \left(
\begin{array}{c}
 c_{12} c_{13} \\
 - s_{12} c_{23} - e^{i \delta} c_{12} s_{13} s_{23} \\
 s_{12} s_{23} - e^{i \delta} c_{12} s_{13} c_{23}
\end{array}
\right), \ 
u_2 = \left(
\begin{array}{c}
 s_{12} c_{13} \\
  c_{12} c_{23} - e^{i \delta} s_{12} s_{13} s_{23} \\
 -c_{12} s_{23} - e^{i \delta} s_{12} s_{13} c_{23}
\end{array}
\right),
\ 
u_3 = \left(
\begin{array}{c}
 e^{-i \delta} s_{13} \\
  c_{13} s_{23} \\
 c_{13} c_{23} 
\end{array}
\right).
\end{equation}

We parameterize the anti-symmetric matrix $f$ as
\begin{equation}
f = \left(
 \begin{array}{ccc}
  0 & f_3 & - f_2 \\
  -f_3 & 0 & f_1 \\
  f_2 & -f_1 & 0
 \end{array}
\right),
\end{equation}
and then,
\begin{equation}
f f^\dagger = |v_f|^2 I - v_f^* v_f^T,
\end{equation}
where $v_f$ is a column vector, $v_f = (f_1, f_2,f_3)^T$, and $I$ is an identity matrix.
Suppose that $v_f$ is orthogonal to $u_2$ and $u_3$, i.e., $v_f^T u_2^* = v_f^T u_3^* = 0$,
which means that $v_f = f_0 u_1$ ($f_0$ is a coefficient).
Then, one finds
\begin{equation}
f f^\dagger M_\nu f^* f^T =|f_0|^4 M_\nu,
\end{equation}
and 
\begin{equation}
f u_1 = 0.
\end{equation}
One can also obtain
\begin{equation}
f^\dagger u_2^* =- f_0^* u_3, \quad f^\dagger u_3^* =  f_0^* u_2.
\end{equation}
Therefore, we find that the solution of Eq.(\ref{ZB}) is
\begin{equation}
f = f_0 \left(
 \begin{array}{ccc}
  0 & U_{\tau 1} & - U_{\mu 1} \\
  -U_{\tau 1} & 0 & U_{e1} \\
  U_{\mu 1} & - U_{e1} & 0
 \end{array}
\right),
\end{equation}
and
\begin{equation}
\frac{f_0^2}{M_0} M_e g M_e =  m_2\, u_3 u_3^T + m_3\, u_2 u_2^T + a_{1} u_1 u_1^T
+ a_{2} (u_1 u_2^T + u_2 u_1^T) + a_3  (u_1 u_3^T + u_3 u_1^T) ,
\end{equation}
where $a_i$'s are arbitrary coefficients with mass dimension.
Because any vectors can be given by a linear combination of $u_i$,
there are three free complex parameters $a_i$ (and one parameter $f_0$ in $f$) in the solution.

Roughly, we obtain (supposing $a_i = 0$)
\begin{equation}
\frac{ f_0^2 m_\mu^2 g_{\mu\mu}}{M_0} \sim (U_{\mu 2})^2 \sqrt{\Delta m^2_{\rm atm}},
\end{equation}
and the size of the scalar mass is estimated as 
\begin{equation}
\frac{M_0}{48\pi^2} = \frac{{\rm max} (m_h^2, m_k^2)}{\mu_{hhk} \tilde I} \sim 420\  {\rm GeV} \times \frac{f_0^2 g_{\mu\mu}/(U_{\mu 2})^2}{10^{-3}}.
\label{eq:gmumu-constraint}
\end{equation}

The coupling $f$ can generate 
$\mu \to e \gamma$ process by $f_{e\tau} f_{\mu\tau}^*$ product:
\begin{equation}
{\rm Br} (\mu \to e\gamma) = \frac{3 \alpha}{16 \pi} \left| \frac{f_{e\tau} f_{\mu\tau}^*}{3 G_F m_h^2} \right|^2,
\end{equation}
and therefore, the magnitude of $f_0$ is bounded to satisfy the $\mu \to e\gamma$ experimental constraint:
roughly, $|f_0|^2 \alt 0.002 \times (m_h/1 \ {\rm TeV})^2$.

Since there are three free parameters $a_i$, one can eliminate
all off-diagonal elements of the coupling matrix $g$ to suppress the LFV three-body decays of charged leptons.
In that case, however, $g_{ee}$ becomes larger than 1 since $g_{ee} m_e^2 \sim g_{\mu \mu} m_\mu^2$.
Therefore,
using one degree of freedom, we need to adjust the $ee$ element of $M_e g M_e$.
Then, one of the three off-diagonal elements of $g$ cannot be eliminated.
Because it is expected that $g_{\mu\tau}$ is small ($g_{\mu \tau} \simeq g_{\mu\mu} m_\mu/m_\tau$),
$\tau \to 3 \mu$ bound can be satisfied and the
other bounds of LFV processes can be satisfied by eliminating $e\mu$ and $e\tau$ elements of $M_e g M_e$
using the remaining two degrees of freedom.

\begin{figure}
\center
\includegraphics[width=7.7cm]{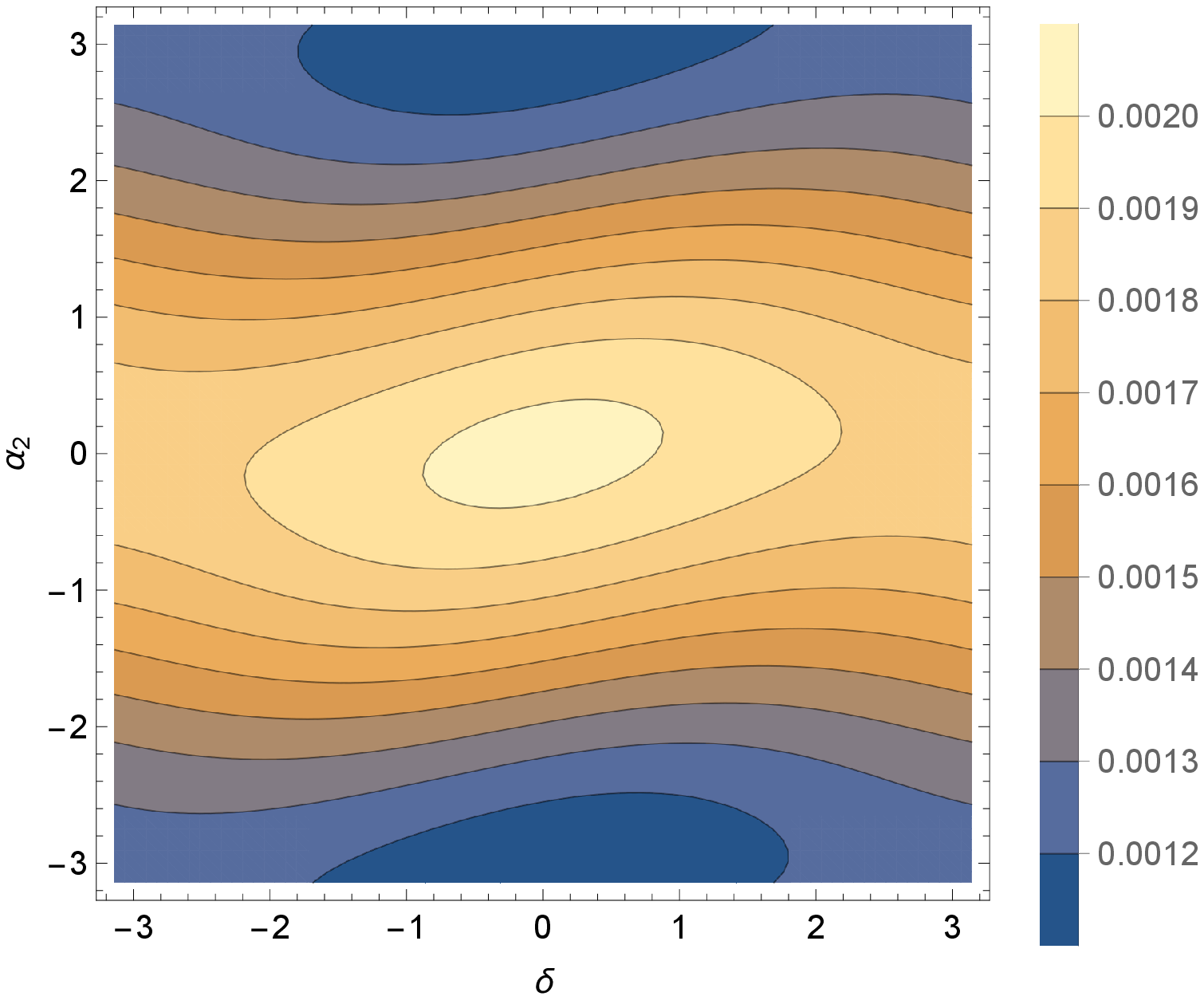} \hspace{3mm}
\includegraphics[width=8cm]{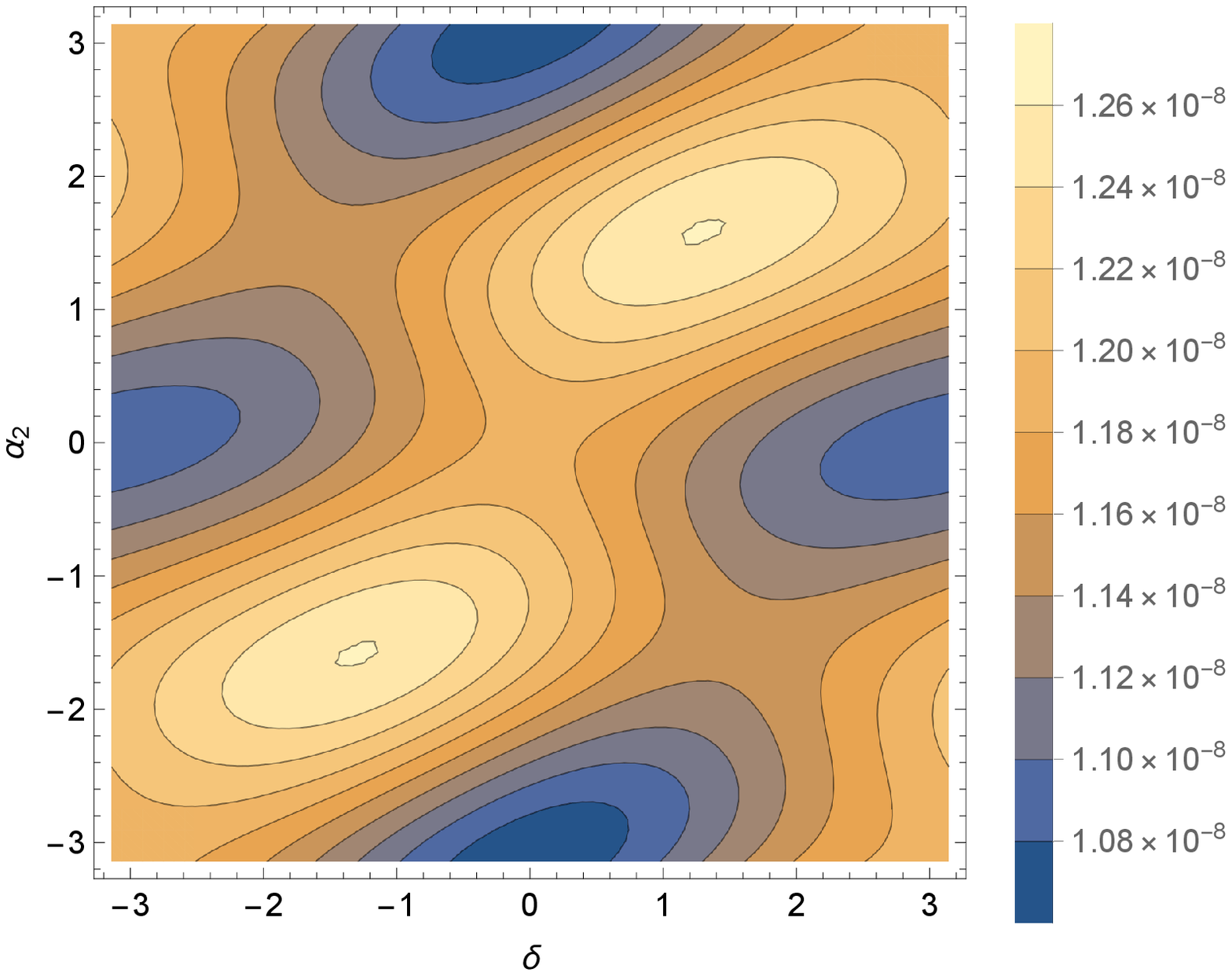}
\caption{
The contour plots of $|G_2|/G_F$ (left), Br($\tau \to 3\mu$) (right) 
as functions of the Dirac phase $\delta$ and the Majorana phase $\alpha_2$.
The choices of parameters in the Zee-Babu model are given in the text.
}
\label{figZB}
\end{figure}

In Fig.\ref{figZB} (left), we show the contour plot of $|G_2|/G_F$ as a function of the Dirac phase $\delta$ and 
the Majorana phase $\alpha_2$,
by adjusting $g_{e\mu} = g_{e\tau} = 0$ and $g_{ee} = g_{\mu\mu}$.
The Majorana phase $\alpha_2$ defined in the convention by Particle Data Group \cite{Zyla:2020zbs}
is $\alpha_2 = {\rm arg} (m_2/m_3)$ here.
%
%
We choose $f_0^2 = 0.002$, $m_k = 1.2$ TeV and $M_0/(48\pi^2) = 500$ GeV.
In Fig.\ref{figZB} (right), we show Br($\tau \to 3\mu$).
Because $g_{ee} = g_{\mu\mu}$ is chosen, we obtain
${\rm Br} (\tau^- \to \mu^+ e^- e^-) = {\rm Br} (\tau \to 3\mu)$.
As can be seen in Fig.\ref{figZB} (left), 
$|G_2|/G_F$ can be as large as the current experimental bound shown in Eq.(\ref{PSI-bound}).

We comment on the model-parameter dependence of the Mu-to-$\mubar$ transition. 
The coefficient of the transition operator is roughly proportional to the model parameters as
\begin{equation}
\frac{|G_2|}{\sqrt2} = \frac{|g_{ee} g_{\mu\mu}|}{8m_k^2} \propto
\frac{1}{f_0^4} \frac{{\rm max} (m_k^4, m_h^4)}{\mu_{hhk}^2 m_k^2} \frac{g_{ee}}{g_{\mu\mu}}.
\end{equation}
Here we use Eq.(\ref{eq:gmumu-constraint}) to include the constraint to reproduce the neutrino mass.
The non-observation of $\mu \to e\gamma$ gives the lower bound of $m_h^4/f_0^4$.
If $f_0$ (namely $f$) becomes smaller, the coupling $g_{\mu\mu}$ needs to be larger 
to reproduce the size of neutrino mass $m_3$,
and thus, the Mu-to-$\mubar$ transition becomes larger.
The scalar trilinear coupling $\mu_{hhk}$ should not be much larger than $m_h$ and $m_k$
to avoid a charge breaking global minimum.
Therefore, the search of the Mu-to-$\mubar$ transition gives a good test of the Zee-Babu model 
in the range of $g_{ee} \sim g_{\mu\mu}$.

Since $g_{\mu\tau}$ cannot be eliminated, the $\tau \to 3\mu$ and $\tau^- \to \mu^+ e^- e^-$ processes are generated:
\begin{equation}
\{ {\rm Br} (\tau \to 3\mu), {\rm Br} (\tau^- \to \mu^+ e^- e^-)  \}= 
8 \left| \frac{G_2}{G_F} \right|^2 {\rm Br} (\tau \to \mu \nu\bar\nu)  \times 
\left\{ \left| \frac{g_{\mu\tau}}{g_{ee} } \right|^2, \left| \frac{g_{\mu\tau}}{g_{\mu\mu} } \right|^2\right\}.
\end{equation}
Because of $g_{\mu\tau} \simeq g_{\mu\mu} m_\mu/m_\tau$,
if the Mu-to-$\mubar$ transition is observed at $G_2/G_F \sim 10^{-3}$, those two LFV tau decays will be observed.

We note on the case of inverted mass hierarchy.
Similarly, the rank-2 neutrino mass matrix ($m_3=0$ in this case) is given as
\begin{equation}
M_\nu = m_1 u_1^* u_1^\dagger + m_2 u_2^* u_2^\dagger,
\end{equation}
and the solution of Eq.(\ref{ZB}) is
\begin{equation}
f = f_0 \left(
 \begin{array}{ccc}
  0 & U_{\tau 3} & - U_{\mu 3} \\
  -U_{\tau 3} & 0 & U_{e3} \\
  U_{\mu 3} & - U_{e3} & 0
 \end{array}
\right),
\end{equation}
and
\begin{equation}
\frac{f_0^2}{M_0} M_e g M_e = m_1\, u_2 u_2^T + m_2\, u_1 u_1^T
+ a_{1} (u_3 u_1^T + u_1 u_3^T) + a_2  (u_3 u_2^T + u_2 u_3^T) + a_{3} u_3 u_3^T.
\end{equation}
The size of $g_{\mu\mu}$ becomes larger than the one in the normal hierarchy
to make $ee$, $e\mu$, and $e\tau$ elements of $M_e g M_e$ to be small, under the same model parameters above.
This is because $U_{e3}$ is small compared to the other elements, and $a_3$ needs to be larger.
Consequently, the coefficient $G_2$ becomes larger than the current bound unless $g_{ee}$ is made to be much smaller than
$g_{\mu\mu}$.

\subsubsection{Cocktail model}
\label{sec7.1.2}
\begin{figure}
\center
\includegraphics[width=7cm]{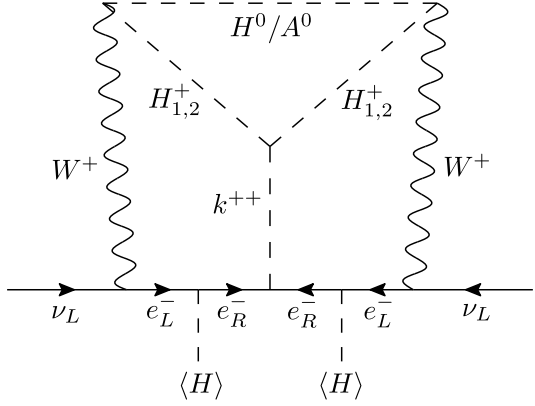}
\caption{
A diagram to induce the neutrino mass in the cocktail model.
The charged scalar in the inert doublet $\eta$ and the $SU(2)_L$ singlet scalar $S^+$ are mixed to be $H_{1,2}^+$.
By splitting the masses of the real and imaginary parts ($H^0, A^0$) of the neutral scalar in $\eta$,
 neutrino masses are generated.
The symbol $\langle H \rangle$ stands for the vev of the SM Higgs boson. 
}
\label{fig:cocktail}
\end{figure}

In the Cocktail model \cite{Gustafsson:2012vj},
an inert Higgs doublet $\eta$ (which does not acquire a vev)
 and a hypercharge $Y=1$ $SU(2)_L$ singlet $S^+$ are introduced
in addition to the doubly charged scalar $k^{++}$.
Contrary to the Zee-Babu model, the $S^+$ scalar does not couple to leptons directly,
and the $\mu \to e \gamma$ induced by $S^+$ loop does not bother us.
The neutrino masses are generated by three-loop diagrams,
which look like cocktail glass as shown in Fig.\ref{fig:cocktail}. The mass matrix is given as
\begin{equation}
(M_\nu)_{\alpha\beta} = \frac{1}{(16\pi^2)^3} m_\alpha g_{\alpha \beta} m_\beta \frac{F_{\rm cocktail}}{m_k},
\end{equation}
where $g$ is a doubly charged scalar coupling to right-handed charged leptons (same as in Zee-Babu model),
$m_\alpha$ ($\alpha = e, \mu, \tau$) is the charged lepton mass,
$m_k$ is a doubly charged scalar mass,
and $F_{\rm cocktail}$ stands for a loop function containing couplings in the model.

\begin{figure}
\center
\includegraphics[width=8cm]{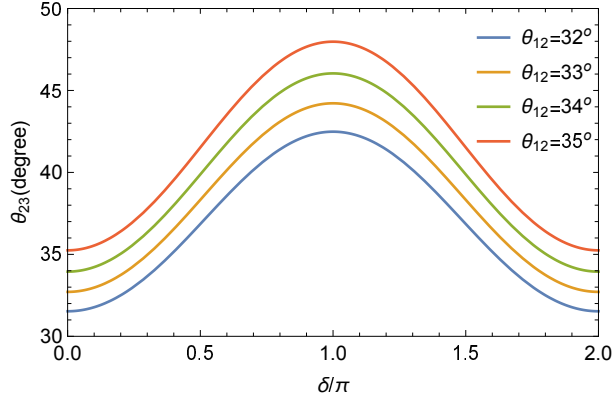}
\caption{
The relation between $\delta$ (radian) and $\theta_{23}$ for various $\theta_{12}$ in the case of $(M_\nu)_{ee,e\mu} \to 0$.
We use central values for the 1-3 neutrino mixing and mass squared differences:
$\theta_{13} = 8.57^{\rm o}$, 
$\Delta m^2_{\rm sol} = 7.42 \times 10^{-5}$ eV$^2$,
and 
$\Delta m^2_{\rm atm} = 2.52 \times 10^{-3}$ eV$^2$ \cite{Esteban:2020cvm}.
}
\label{fig-tex2}
\end{figure}

The mass scale to generate the neutrino mass matrix, $M_\nu = U^* {\rm diag}(m_1,m_2,m_3) U^\dagger$, in the normal mass ordering is estimated as
\begin{equation}
\frac{m_k}{F_{\rm cocktail}} \sim \frac1{(16\pi^2)^3} \frac{m_\mu^2 g_{\mu\mu} }{m_3 U_{\mu 3}^2}
\sim 100 \ {\rm GeV} \times g_{\mu\mu}.
\end{equation}
A large value of $F_{\rm cocktail}$ is needed for a realistic model to satisfy the experimental constraints.
More numerical works to obtain the scale by model parameters can be found in
\cite{Cepedello:2020lul}.
One can immediately notice that 
the magnitude of $ee$ and $e\mu$ elements of the neutrino mass matrix $M_\nu$
should be much smaller than the magnitude of $\mu\mu$ element
unless $g_{ee}$ and $g_{e\mu}$ are much larger than 1.
Indeed, we want to make $g_{e\mu} \to 0$ to obtain a reachable Mu-to-$\mubar$ transition
while suppressing the $\mu \to 3 e$ process.
If the $ee$ and $e\mu$ elements of $M_\nu$ are much smaller than the $\mu\mu$ element, 
neutrino mixings and mass ratio, and phases are constrained.
The analytic relation of the neutrino mixings and mass ratio is given in Ref.\cite{Fukuyama:2016vgi}.
In Fig.\ref{fig-tex2}, we show the relation
between the PMNS phase $\delta$ and $\theta_{23}$.
We vary $\theta_{12}$ since the relation is sensitive to it.
The $3\sigma$ range of $\theta_{12}$ is $31.3^{\rm o}$\dash$35.9^{\rm o}$ by NuFIT 5.0 \cite{Esteban:2020cvm}.
Because $\theta_{23}$ is in the $3\sigma$ range of $40^{\rm o}$\dash$52^{\rm o}$,
it predicts that $\delta$ is preferred to be in the second or third quadrant roughly.

Let us choose $\theta_{12} = 33.4^{\rm o}$, $\theta_{13} = 8.57^{\rm o}$, 
$\Delta m^2_{\rm sol} = 7.42 \times 10^{-5}$ eV$^2$,
and 
$\Delta m^2_{\rm atm} = 2.52 \times 10^{-3}$ eV$^2$.
Then, we can choose $\theta_{23} = 45^{\rm o}$ and $\delta \simeq \pi$
as a benchmark point.
For $g_{ee}= g_{\mu\mu}$ and $g_{e\mu} = 0$, we obtain
\begin{equation}
g_{\alpha\beta} = \left(
 \begin{array}{ccc}
  1 & 0 & -4.57 \\
  0 & 1 & 0.0454 \\
  -4.57 & 0.0454 & 0.00321 
 \end{array}
\right) g_{\mu\mu} .
\end{equation}
Because of $m_e \ll m_\mu$, the numerical values of the elements are insensitive 
to $g_{ee}$ to reproduce the neutrino mass matrix.
It is important to notice that $g_{e\tau}$ is large, and the Mu-to-$\mubar$ transition is bounded
by the $\tau \to 3e$ process:
\begin{equation}
\frac{|G_2|}{G_F} = \frac{{\rm Br} ( \tau \to 3 e )}{{\rm Br} (\tau \to e \nu\bar\nu) } \frac{1}{2\sqrt2} 
\left| \frac{g_{\mu\mu} }{g_{e\tau} } \right| < 3\times 10^{-5}.
\end{equation}
The $\mu \to e\gamma$ process also bounds the Mu-to-$\mubar$ transition similarly to Eq.(\ref{mueg-doubly-charge}) as
\begin{equation}
\frac{|G_2|}{G_F} < 3.8 \times 10^{-6} \left| \frac{g_{ee} g_{\mu\mu} }{ g_{e\tau} g_{\mu\tau} } \right|
= 1.8 \times 10^{-5} \times  \left| \frac{g_{ee}}{g_{\mu\mu} } \right|
\end{equation}
in this benchmark point.
We note that the Zee-Babu model has freedom to suppress $g_{e\tau}$,
while in the cocktail model, $g_{e\tau}$ is needed to generate $\theta_{12}$ and $\theta_{13}$ neutrino mixings.
Consequently, the Mu-to-$\mubar$ transition is bounded in the cocktail model rather than the Zee-Babu model.

\subsection{Charged Higgs contribution}
\label{sec7.2}

\begin{figure}
\center
\includegraphics[width=8cm]{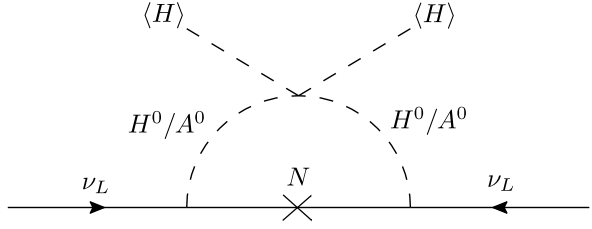} 
\caption{
The diagram to generate the neutrino mass radiatively in the model without the Dirac neutrino masses.
The symbol $\langle H \rangle$ stands for the vev of the SM Higgs boson.
}
\label{fig:radiative-seesaw}
\end{figure}

The Dirac mass is supposed to be forbidden by $Z_2$ symmetry.
Namely, 
the Dirac Yukawa coupling with the SM Higgs doublet $\Phi$ is absent,
but couplings with an additional inert doublet $\eta$ 
are allowed:
\begin{equation}
-{\cal L} \supset y_{\alpha i} \overline{\ell_\alpha} P_R N_i \eta + \frac12 M_i \overline{N_i^c} N_i + h.c.
\end{equation}
If the scalar potential contains the $\lambda_5$ term,
\begin{equation}
V \supset \frac{\lambda_5}{4} (\eta \Phi^\dagger)^2 + h.c.,
\end{equation}
the masses ($m_H$ and $m_A$) of the real and imaginary parts ($H^0$ and $A^0$) of the neutral Higgs boson in the inert doublet $\eta$ are split:
\begin{equation}
m_H^2 - m_A^2 = \lambda_5 v^2,
\end{equation}
where $v$ is the vev of the SM Higgs boson.
Then, the active neutrino masses are generated radiatively 
by the diagram given in Fig.\ref{fig:radiative-seesaw}
as \cite{Ma:2006km}
\begin{eqnarray}
(M_\nu)_{\alpha\beta} &=&
\frac{1}{16\pi^2} y_{\alpha i} y_{\beta i} M_i 
\left( 
\frac{m_H^2}{m_H^2 - M_i^2} \ln \frac{m_H^2}{M_i^2}
-
\frac{m_A^2}{m_A^2 - M_i^2} \ln \frac{m_A^2}{M_i^2}
\right) \nonumber \\
&\simeq&
\frac{\lambda_5 v^2}{16\pi^2} \frac{y_{\alpha i} y_{\beta i}}{M_i}
\left(
\frac{M_i^2}{m_H^2- M_i^2} - \frac{M_i^4}{(m_H^2 - M_i^2)^2} \ln \frac{m_H^2}{M_i^2}
\right).
\end{eqnarray}

\begin{figure}
\center
\includegraphics[width=7cm]{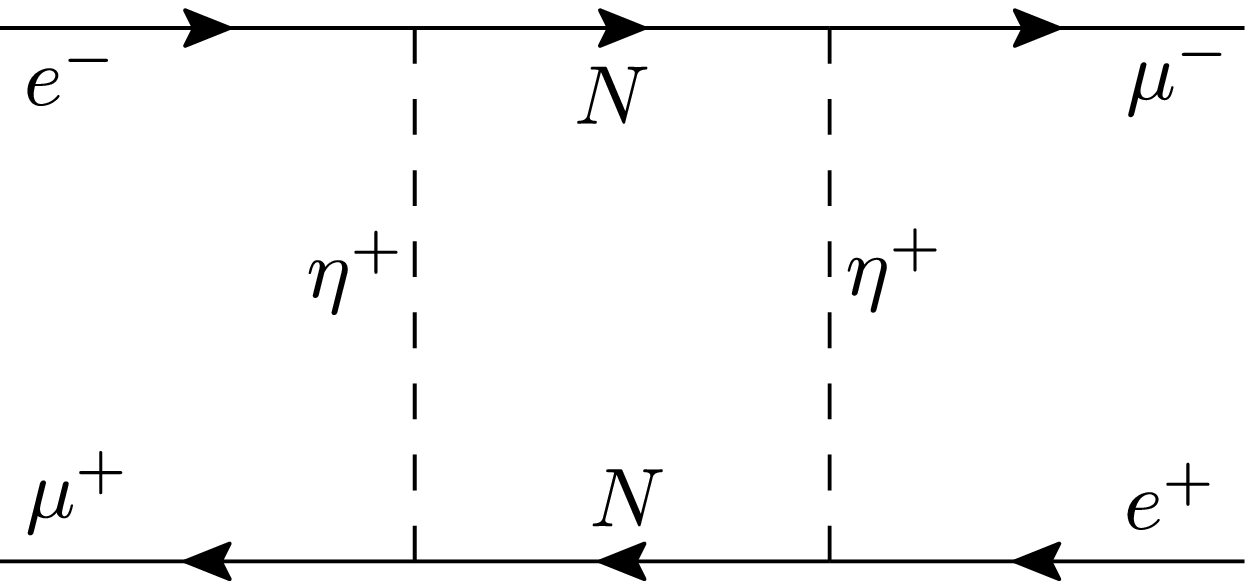}  \hspace{1cm}
\includegraphics[width=6.9cm]{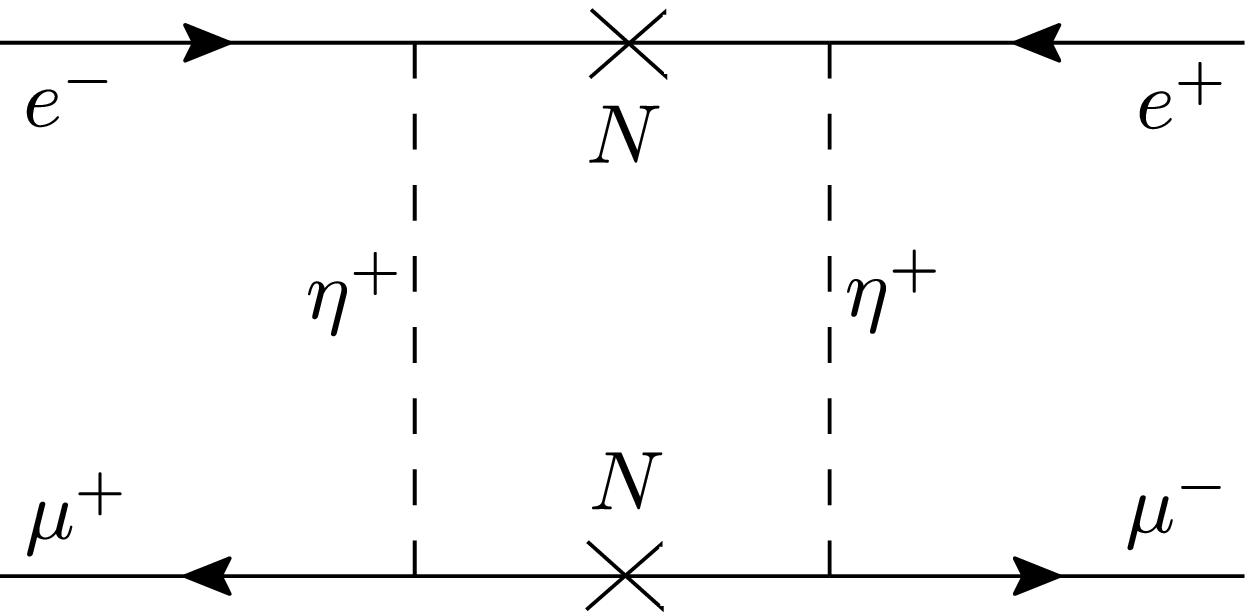}
\caption{
The box loop diagrams to generate the Mu-to-$\mubar$ transition via the charged Higgs bosons.
}
\label{fig:charged-Higgs}
\end{figure}

The charged Higgs boson in $\eta$ can generate the Mu-to-$\mubar$ transition by the box diagrams in Fig.\ref{fig:charged-Higgs}:
\begin{equation}
\frac{G_1}{\sqrt2} = \frac{y^*_{ei} y_{\mu i} y_{e j}^* y_{\mu j} }{512\pi^2 m_\eta^2} I_2( x_i, x_j )
+ \frac{(y^*_{ei})^2  (y_{\mu j})^2  }{256\pi^2 m_\eta^2} \sqrt{x_i x_j}\, I_1( x_i, x_j ),
\end{equation}
where $x_i = M_{i}^2/m_\eta^2$, 
and $I_n (x_i, x_j)$'s are the box loop functions given in Appendix \ref{app:loop_function},
and $m_\eta$ stands for the charged Higgs boson mass.
The first term is bounded by $\mu \to e\gamma$.
Let us consider whether the Mu-to-$\mubar$ transition from the second term can be generated
avoiding the $\mu \to e\gamma$ constraint.
To do that, we consider
\begin{equation}
y_{\alpha i} = \left(
 \begin{array}{ccc}
  y_{e 1} & 0 & 0 \\
  0 & y_{\mu 2} & 0 \\
  y_{\tau 1} & y_{\tau 2} & y_{\tau 3} 
 \end{array}
\right)
\label{DiracY-me}
\end{equation}
to eliminate the one-loop $\mu \to e \gamma$ amplitude via $\eta$ loop.
(For the purpose to eliminate $\mu \to e\gamma$, one of $h_{e3}$ and $h_{\mu 3}$ can be non-zero.
We here suppose that both are zero to reduce the number of parameters.)
Then the neutrino mass matrix is
\begin{equation}
M_\nu = \frac{\lambda_5 v^2}{16\pi^2}
 \left(
 \begin{array}{ccc}
  \frac{y_{e 1}^2}{\tilde M_1} & 0 &  \frac{ y_{e 1} y_{\tau 1}}{\tilde M_1} \\
  0 &  \frac{ y_{\mu 2}^2}{\tilde M_2} & \frac{y_{\mu 2} y_{\tau 2}}{\tilde M_2}  \\
\frac{ y_{e 1} y_{\tau 1}}{\tilde M_1} & \frac{ y_{\mu 2} y_{\tau 2}}{\tilde M_2}  & 
 \frac{y_{\tau 1}^2}{\tilde M_1} + \frac{y_{\tau 2}^2}{\tilde M_2} + \frac{y_{\tau 3}^2}{\tilde M_3}
 \end{array}
\right),
\end{equation}
where $\tilde M_i$ is defined to be $M_i$ divided by a loop function so that the
neutrino mass matrix is $\propto y_{\alpha i} y_{\beta i}/{\hat M_i}$.
The size of the coupling is estimated as
\begin{equation}
\lambda_5 y_{\mu 2}^2 \sim 10^{-10} \times \frac{\tilde M_2}{1\,{\rm TeV}}.
\end{equation}
Because there are four complex parameters (up to normalization),
one can fit three  
neutrino mixings, one mass ratio, and phases, in principle.
The lightest neutrino mass $m_1$ (with its phase) is a function of the other parameters
since the $e\mu$ element is chosen to be zero.

Though the couplings have been chosen to eliminate the $\mu \to e\gamma$ process,
one needs to care about $\tau \to l_\alpha \gamma$ ($l_\alpha = e, \mu$) processes
since $y_{\tau 1}$ and $y_{\tau 2}$ are needed to reproduce the neutrino mixings:
\begin{equation}
{\rm Br} (\tau \to l_\alpha \gamma) = \frac{3 \alpha}{16 \pi} \left| \frac{y_{\alpha i} y_{\tau i}^* F_N (x_i)}{ G_F m_\eta^2} \right|^2
{\rm Br} (\tau \to l_\alpha \nu\bar\nu), 
\end{equation}
where
\begin{equation}
F_N (x) = \frac{1-6x + 3x^2 +2x^3 - 6 x^2 \ln x}{12(1-x)^4}.
\label{FN_function}
\end{equation}
We also need to care about muon $g-2$ since the loop contribution gives a negative contribution to it: 
\begin{equation}
\Delta a_\mu = - \frac{1}{8\pi^2} \frac{m_\mu^2}{m_\eta^2} |y_{\mu i}|^2 F_N (x_i).
\end{equation}
We can check that the experimental bound can be satisfied even if 
the non-zero elements of
$y_{\alpha i}$ are $O(1)$ for $m_\eta = 500$ GeV.
The transition amplitude can be maximal for $M_i \simeq m_\eta$,
and we find $|G_1|/G_F \alt O(10^{-5})$ 
for $m_\eta = 500$ GeV and $|y_{\alpha i}| < 1$.

\subsection{$N S^+ e_R$ coupling}
\label{sec7.3}

We consider models with right-handed charged lepton couplings to SM singlet fermion $N$:
\begin{equation}
- {\cal L} = h_{\alpha i} \overline{e^c_\alpha} P_R N_i S^+ + \frac12 M_{i} \overline{N_i^c} N_i + M_S^2 S^+ S^-,
\end{equation}
where $S^+$ is a $SU(2)_L$ singlet with hypercharge $Y=1$.
The transition operator $Q_2$ is generated and the coefficient is
\begin{equation}
\frac{G_2}{\sqrt2} = \frac{h_{ei} h_{\mu i}^* h_{ej} h_{\mu j}^* }{512\pi^2 M_S^2} I_2( x_i, x_j )
+ \frac{(h_{ei})^2  (h_{\mu j}^*)^2  }{256\pi^2 M_S^2} \sqrt{x_i x_j}\, I_1( x_i, x_j ),
\label{G2-eR}
\end{equation}
where $x_i = M_{i}^2/M_S^2$.

\subsubsection{KNT model}
\label{sec7.3.1}

\begin{figure}
\center
\includegraphics[width=7cm]{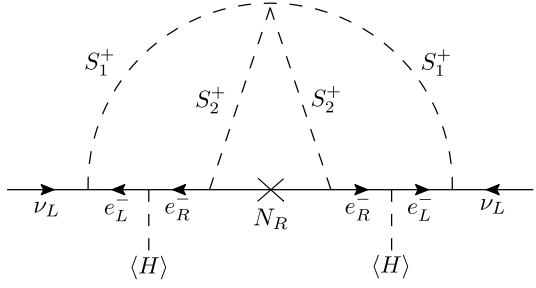}
\caption{
A diagram to induce the neutrino mass in the KNT model.
The symbol $\langle H \rangle$ stands for the vev of the SM Higgs boson.
}
\label{fig:KNT}
\end{figure}

In the model by Krauss-Nasri-Trodden (KNT) \cite{Krauss:2002px},
two $Y=1$ $SU(2)_L$ singlet scalars (we call them $S_1$ and $S_2$) are introduced:
\begin{equation}
-{\cal L} \supset f_{\alpha\beta} \overline{\ell^c_\alpha} \cdot \ell_\beta S_1^+ +  h_{\alpha j} \overline{e^c_\alpha} P_R N_j S^+_2 +  \lambda_S (S_1^+ S_2^-)^2 + h.c.
\end{equation}
The neutrino mass is generated by a three-loop diagram shown in Fig.\ref{fig:KNT}:
\begin{equation}
(M_\nu)_{\alpha\beta} = \frac{\lambda_S}{(16\pi^2)^3 M_{S_2}^2} f_{\alpha \alpha'} m_{\alpha'} h_{\alpha' i} 
F_{\rm KNT}(M_{Ni}/M_{S_2}, M_{S_1}/M_{S_2}) h_{ \beta' i} m_{\beta'} f_{ \beta \beta'},
\end{equation}
where $m_\alpha = (m_e, m_\mu, m_\tau)$ and $F_{\rm KNT}$ is a loop function.
The equation can be solved just similarly to the Zee-Babu model.
However, the coupling $h$ becomes large to be in a non-perturbative region
if one assumes that the observed neutrino masses are all covered by this contribution.
Actually, in the Zee-Babu model in which the neutrino masses are generated by the two-loop diagram,
the coupling can be $O(1)$ for several hundred GeV scalar masses, and thus, 
one can imagine that the coupling needs to be large to generate the neutrino mass by three-loop. 
This is because the anti-symmetric coupling $f$ needs to be small to avoid $\mu \to e\gamma$ constraint.
Of course, 
it is possible to give up on explaining the entire neutrino mass matrix with this loop-induced contribution
and assume that the neutrino masses come primarily from somewhere else.
In such a situation,
the $h$ couplings can be $O(1)$ 
without a contradiction with observables, 
and they 
can induce the Mu-to-$\mubar$ transition, potentially as large as the current bound,
from the second term in Eq.(\ref{G2-eR}).

\subsubsection{AKS model}
\label{sec7.3.2}

\begin{figure}
\center
\includegraphics[width=6cm]{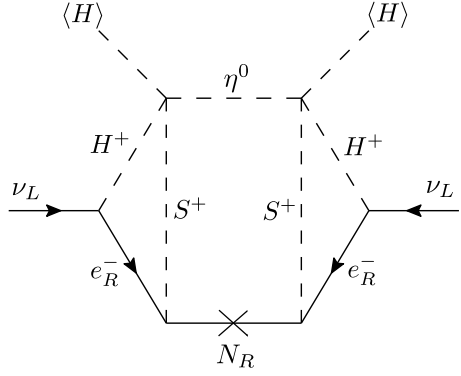}
\caption{
A diagram to induce the neutrino mass in the AKS model.
The symbol $\langle H \rangle$ stands for the vev of the SM Higgs boson.
}
\label{fig:AKS}
\end{figure}

In the model by
Aoki-Kanemura-Seto (AKS) \cite{Aoki:2008av},
two Higgs doublets $\Phi_1$ and $\Phi_2$ to have a physical charged Higgs scalar ($H$) in the loop,
and one real scalar singlet $\eta^0$ are introduced.
The neutrino mass is generated by three-loop diagrams such as shown in Fig.\ref{fig:AKS}:
\begin{equation}
(M_\nu)_{\alpha\beta} = \frac{\kappa^2 \tan^2\beta}{(16\pi^2)^3}  m_\alpha m_\beta h_{\alpha i} h_{\beta i} 
\frac{F_{\rm AKS} (M_{Ni}, M_\eta, M_S, M_{H^+}  ) }{M_{Ni}},
\end{equation}
where $\kappa$ is $\Phi_1 \Phi_2 \eta S$ coupling.

Let us consider if the observed parameters of neutrino oscillations can be reproduced
with satisfying the $\mu \to e\gamma$ constraint,
and see if the Mu-to-$\mubar$ transition can be induced by the $h_{\alpha i}$ coupling.
To do that, let us consider
\begin{equation}
h_{\alpha i} = \left(
 \begin{array}{ccc}
  h_{e 1} & 0 & 0 \\
  0 & h_{\mu 2} & 0 \\
  h_{\tau 1} & h_{\tau 2} & h_{\tau 3} 
 \end{array}
\right),
\end{equation}
similarly to Eq.(\ref{DiracY-me}),
in order to eliminate the one-loop $\mu \to e \gamma$ amplitude via $h_{\alpha i}$ coupling.
Then the neutrino mass matrix is
\begin{equation}
M_\nu = \frac{\kappa^2 \tan^2\beta}{(16\pi^2)^3}
 \left(
 \begin{array}{ccc}
  \frac{m_e^2 h_{e 1}^2}{\tilde M_1} & 0 &  \frac{m_e m_\tau h_{e 1} h_{\tau 1}}{\tilde M_1} \\
  0 &  \frac{m_\mu^2 h_{\mu 2}^2}{\tilde M_2} & \frac{m_\mu m_\tau h_{\mu 2} h_{\tau 2}}{\tilde M_2}  \\
\frac{m_e m_\tau h_{e 1} h_{\tau 1}}{\tilde M_1} & \frac{m_\mu m_\tau h_{\mu 2} h_{\tau 2}}{\tilde M_2}  & 
m_\tau^2 \left( \frac{h_{\tau 1}^2}{\tilde M_1} + \frac{h_{\tau 2}^2}{\tilde M_2} + \frac{h_{\tau 3}^2}{\tilde M_3} \right)
 \end{array}
\right),
\end{equation}
where $\tilde M_i = M_{Ni}/F_{\rm AKS} (M_{Ni})$.
%
We obtain the mass scale of $\tilde M_2$ as
\begin{equation}
\tilde M_2 \simeq 100 \ {\rm GeV} \times h_{\mu 2}^2 \kappa^2 \tan^2\beta .
\end{equation}

There are two types of solutions:
\begin{enumerate}
\item
Naive solution: $h_{\mu 2}  \ll h_{e1} \sim 1$.

If there is no cancellation in the $\tau \tau$ element of $M_\nu$,
one needs $h_{\mu 2}  \ll h_{e1}$ to realize the neutrino mixings.
This is due to $m_e \ll m_\mu$.
Since $h_{\mu 2}$ is small, a large value of $\kappa \tan\beta$ is needed
to obtain the proper size of the neutrino mass in this solution.
Any observed neutrino mixings (within errors) and a PMNS phase $\delta$ can be realized.
The Mu-to-$\mubar$ transition is estimated as
$|G_2|/G_F \alt O(10^{-7})$.
(If one allows a nearly non-perturbative value of $h_{e1} \sim 10$, the Mu-to-$\mubar$ transition can be enlarged, though.)

\item
$h_{e1} \sim h_{\mu 2}$.

If we allow a tuning of the $\tau\tau$ element of $M_\nu$
($h_{\tau 1}^2/\tilde M_1 + h_{\tau 3}^2/\tilde M_3 \ll h_{\tau 2}^2/\tilde M_2$)
to obtain the atmospheric mixing properly, 
$h_{e1} \sim h_{\mu 2}$ can be allowed.
%
%
Both $ee$ and $e\mu$ elements of $M_\nu$ are much smaller than $\mu\mu$ element in this solution,
and therefore, the neutrino mixings and the Dirac phase $\delta$ are related as shown in
Fig.\ref{fig-tex2}.
To realize the solar neutrino mixing, one needs to enlarge $h_{\tau 1}$ compared to the naive solution.
Since $h_{\tau 1}$ becomes $O(1)$, one needs to care about $\tau \to e \gamma$ process.
Our estimation of the Mu-to-$\mubar$ transition is
$|G_2|/G_F \alt O(10^{-6})$.

\end{enumerate}

\section{Neutral scalar exchange}
\label{sec8}

\begin{figure}
\center
\includegraphics[width=7cm]{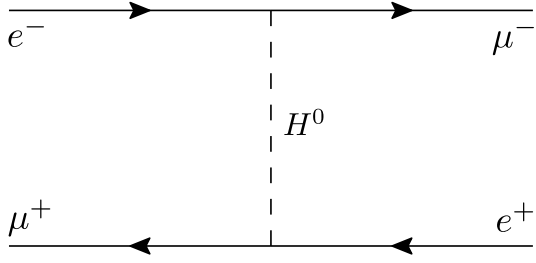}
\caption{
The tree-level exchange of a neutral scalar boson ($H^0$, here) to induce the Mu-to-$\mubar$ transition.
}
\label{fig:neutral-scalar}
\end{figure}

The Mu-to-$\mubar$ transition induced by the neutral scalar exchange 
shown in Fig.\ref{fig:neutral-scalar} is considered~\cite{Hou:1995dg}.

In general two-Higgs-doublets model, so-called type-III,
the Yukawa couplings can be written as
\begin{equation}
-{\cal L} = (Y_1)_{ij} \overline{ \ell_{iL}} e_{jR} \Phi_1'+  (Y_2)_{ij} \overline{\ell_{iL}} e_{jR} \Phi_2' + h.c.
\end{equation}
and the vevs of the neutral components are $\langle \Phi_1^{\prime 0} \rangle= v \cos \beta'$, $\langle \Phi_2^{\prime 0} \rangle= v \sin \beta'$.
Redefining the Higgs fields so that $\Phi_1$ does not acquire a vev,
\begin{equation}
\left(
 \begin{array}{c}
  \Phi_2 \\
  \Phi_1
 \end{array}
\right)
= \left(
\begin{array}{cc}
 \sin\beta' & \cos\beta' \\
  -\cos\beta' & \sin \beta'
 \end{array}
\right)
\left(
 \begin{array}{c}
  \Phi_2' \\
  \Phi_1'
 \end{array}
\right),
\end{equation}
we can rewrite the Yukawa interaction as
\begin{equation}
-{\cal L} = (Y_e)_{ij} \overline{\ell_{iL} }e_{jR} \Phi_2 + \rho_{ij} \overline{\ell_{iL}} e_{jR} \Phi_1 + h.c.,
\end{equation}
where
\begin{equation}
Y_e = Y_1 \sin\beta' + Y_2 \cos\beta',\qquad
\rho = -Y_1 \cos\beta' + Y_2 \sin\beta'. 
\end{equation}
We can redefine $\sin\beta = 1$.
The Yukawa coupling $Y_e$ generates the charged lepton masses,
and thus, we work on the flavor basis where $Y_e$ is diagonal, $Y_e = {\rm diag} (y_e, y_\mu, y_\tau)$.
The neutral physical Higgs interaction can be written as
\begin{equation}
-{\cal L} = \frac1{\sqrt2} (Y_e c_\alpha + \rho s_\alpha)_{ij} \overline{e_{iL}} e_{jR} h
+ \frac1{\sqrt2} (Y_e s_\alpha - \rho c_\alpha)_{ij} \overline{e_{iL}} e_{jR} H 
+ i \frac1{\sqrt2} \rho_{ij}  \overline{e_{iL}} e_{jR} A + h.c.,
\end{equation}
where $s_\alpha = \sin \alpha$ and $c_\alpha = \cos\alpha$
are the mixings of CP even Higgs bosons ($h$ and $H$).
Integrating out the neutral Higgs bosons, $h$, $H$, and $A$,
we extract the terms which can be the transition operators:
\begin{eqnarray}
-{\cal L} = -\frac14 \left(\rho_{21}^2 (\overline{\mu_L} e_R)^2 + \rho_{12}^{*2} (\overline{\mu_R} e_L)^2\right)
\left( \frac{s_\alpha^2}{m_h^2} + \frac{c_\alpha^2}{m_H^2} - \frac{1}{m_A^2} \right) \nonumber \\
- \frac12 \rho_{21} \rho_{12}^* (\overline{\mu_L} e_R)(\overline{\mu_R} e_L) 
\left( \frac{s_\alpha^2}{m_h^2} + \frac{c_\alpha^2}{m_H^2} + \frac{1}{m_A^2} \right) + h.c. ,
\end{eqnarray}
and 
\begin{eqnarray}
\frac{G_3}{\sqrt2} &=& \frac{1}{16} \rho_{21} \rho_{12}^* \left( \frac{s_\alpha^2}{m_h^2} + \frac{c_\alpha^2}{m_H^2} + \frac{1}{m_A^2} \right), \\
\frac{G_4}{\sqrt2} &=& -\frac{1}{16}  \rho_{12}^{*2} \left( \frac{s_\alpha^2}{m_h^2} + \frac{c_\alpha^2}{m_H^2} - \frac{1}{m_A^2} \right), \\
\frac{G_5}{\sqrt2} &=& -\frac{1}{16}  \rho_{21}^{2} \left( \frac{s_\alpha^2}{m_h^2} + \frac{c_\alpha^2}{m_H^2} - \frac{1}{m_A^2} \right).
\end{eqnarray}

If we suppose that $\rho_{21}$ and $\rho_{12}$ are not zero with the other $\rho_{ij} = 0$
and $\sin \alpha (= \cos(\beta - \alpha)) \to 0$ (alignment limit),
LFV processes such as $\mu \to e \gamma$ and $\mu \to 3 e$ do not occur.
In the limit, one obtains $G_4$, $G_5 \to 0$.
Actually, in that case, there is a global discrete $Z_4$ symmetry, and the charge assignments are
\begin{equation}
e_L : 1, \quad e_R : 1, \quad
\mu_L : 3, \quad \mu_R : 3, \quad \Phi_1 : 2,
\end{equation}
and the charges of the others are 0.
Therefore, once the selection of $\rho$ with the alignment limit is given,
LFV is not generated perturbatively.
We note that
the electron mass can have a loop correction from $\rho_{12} \rho_{21}$.
The $\rho$ terms can also induce the electron $g-2$ and electric dipole moment 
(therefore, $\rho_{12} \rho_{21}$ should be real).
The coefficient $G_3$ can be generated satisfying the LFV constraints in the alignment limit,
and the Mu-to-$\mubar$ transition can be around the current experimental bound.

\section{SUSY}
\label{sec9}

Similarly to the meson mixings,
the box diagram in which superparticles propagate can generate the transition operators.
Indeed, the minimally extended SUSY standard model (MSSM)
contains Majorana particles known as gauginos: Bino $\tilde B$ and Wino $\tilde W^0$.
Though the Majorana property of gauginos can be utilized to generate the Mu-to-$\mubar$ transition,
the transition probability is bounded by the $\mu \to e\gamma$ constraint
as we have seen in various models.

In the MSSM, the left-handed slepton doublet 
and the down-type Higgs doublet have the same quantum numbers,
and thus, so-called $R$-parity is introduced to distinguish them.
The $R$-parity is also needed to avoid rapid proton decays.
If the $R$-parity is broken in the lepton sector, 
the Mu-to-$\mubar$ transition can be induced at the tree-level sneutrino exchange \cite{Halprin:1993zv},
as a simple corollary
of the neutral Higgs exchange (in the alignment limit) in the previous section.

In this section, we first briefly describe the previously well-known Mu-to-$\mubar$ transition 
in the $R$-parity violating SUSY model,
and then study the box contribution of the neutralinos and charginos in
extended models.

\subsection{R-parity violation}
\label{sec9.1}

If we consider $R$-parity violating terms, the transition operators can be induced at the tree level \cite{Halprin:1993zv}.
The superpotential is
\begin{equation}
W  = \frac12 \lambda_{ijk} \ell_i \cdot \ell_j e^c_k,
\end{equation}
where $\lambda_{jik} = - \lambda_{ijk}$.
We introduce only $\lambda_{312}$ and $\lambda_{321}$:
\begin{equation}
W = \lambda_{312} (\nu_3 e_1 - e_3 \nu_1) e^c_2 + \lambda_{321} (\nu_3 e_2 - e_3 \nu_2) e^c_1,
\end{equation}
and Lagrangian contains
\begin{equation}
-{\cal L} \supset (\lambda_{312} \tilde \nu_\tau \overline \mu P_L e + \lambda_{321} \tilde \nu_\tau \overline e P_L \mu 
+ h.c.)
+ m_{\tilde \nu_\tau}^2 |\tilde \nu_\tau|^2,
\end{equation}
where $m_{\tilde \nu_\tau}$ is a SUSY breaking tau-sneutrino mass.
By the tau-sneutrino exchange, one obtains
\begin{equation}
\frac{G_3}{\sqrt2} = \frac{ \lambda_{321}^* \lambda_{312}}{8m^2_{\tilde \nu_\tau}}.
\end{equation}

The tau-neutrino mass can be generated by loop:
\begin{equation}
m_{\nu_\tau} \sim \frac{\lambda_{231} \lambda_{132}}{16\pi^2} \frac{A m_e m_\mu}{m_{\tilde \ell}^2},
\end{equation}
where $A$ is a SUSY breaking trilinear scalar coupling.

\subsection{Gaugino contribution}
\label{sec9.2}

In this subsection, we describe how large Mu-to-$\mubar$ transition can be induced by box contribution
in the MSSM with $R$-parity.

Though gauginos are new Majorana particles in the MSSM,
the Mu-to-$\mubar$ transition is strongly restricted by $\mu \to e\gamma$
constraints as we have seen several times
since the gaugino interactions are 
$(\Delta L_e, \Delta L_\mu) = (\pm 1, 0)$, $(0,\pm 1)$ processes.
Therefore, it is not very worth describing it in detail, and
we here note the rough estimation of the bound of the Mu-to-$\mubar$ transition 
from the $\mu \to e\gamma$ bound. 

Using mass insertion approximation \cite{Gabbiani:1996hi,Hisano:1995nq},
the $\mu \to e\gamma$ constraints of the off-diagonal elements of slepton mass matrix can be written as
\begin{equation}
 \frac{g_2^2 \tan\beta |\delta_{12}^{LL}|}{m_{\tilde \ell}^2 G_F^2}  \alt O(10^{-5})
\end{equation}
with\begin{equation}
\delta_{12}^{LL} = (M_{\tilde \ell}^2)_{12} / m_{\tilde \ell}^2.
\end{equation}
Here, $M_{\tilde \ell}^2$ is a $3\times 3$ SUSY breaking left-handed slepton squared mass matrix,
and $m_{\tilde \ell}$ is an averaged left-handed slepton mass.
This $LL$ constraint mainly comes from the chargino loop diagram,
and
the neutralino loop can also constrain $\delta_{12}^{RR,LR,RL}$ similarly.
The coefficient $G_1$ of the transition operator from Wino loop
can be roughly written as
\begin{equation}
\frac{G_1}{\sqrt2} \sim
\frac{ g_2^4 (\delta^{LL}_{12})^2}{512\pi^2 m_{\tilde \ell}^2 } I_2(x,x)
+ 
\frac{ g_2^4 (\delta^{LL}_{12})^2}{256\pi^2 m_{\tilde \ell}^2 } x I_1(x,x),
\end{equation}
where $x \simeq m_{\tilde W}^2/m_{\tilde \ell}^2$.
One should notice that the Mu-to-$\mubar$ transition needs the mass insertion twice 
irrespective of whether the Majorana property of Wino is used or not.
Because $G_2^b (x,x)\simeq x G_1^b(x,x) \alt 1$,
it is estimated as
\begin{equation}
\frac{|G_1|}{G_F} \alt O(10^{-8}) \frac{g^2_2 |\delta_{12}^{LL}|}{\tan\beta}.
\end{equation}
Therefore, the Mu-to-$\mubar$ transition becomes the largest if $\delta_{12}^{LL} \sim 1$ is allowed
for the slepton mass to be several times 10 TeV.
The Bino and charged Wino box loop contribution with $RR$, $LR$ mass insertions can be also estimated similarly.
The coefficients can depend on the detail of the superparticle spectrum,
but we do not survey the numerical detail since the Mu-to-$\mubar$ transition is tiny.
The chargino and neutralino contribution of $\mu_R \to e_L \gamma$ amplitude can be canceled,
but the $\mu$\dash$e$ conversion is not simultaneously canceled, and thus, it constrains the Mu-to-$\mubar$ transition.

\subsection{Charged Higgsino contribution}
\label{sec9.3}

If $\Delta L = \pm 2$ neutrino mass can be utilized,
the Mu-to-$\mubar$ transition can be potentially generated avoiding the $\mu \to e\gamma$ constraint.
However, as we have seen,
the size of neutrino mass restricts the Mu-to-$\mubar$ transition. 
This is because
Dirac neutrino masses
induce light--heavy neutrino mixings,
and large active neutrino masses can be generated
at the loop level, even if we adjust the tree-level neutrino masses to be zero.
In the situation that the loop-induced mass is suppressed by a flavor symmetry,
the box contribution of the Mu-to-$\mubar$ transition is also suppressed.
In the SUSY limit, the loop-induced masses are zero due to the non-renormalization theorem.
By soft terms of SUSY breaking, the neutrino masses are induced 
but the size of the loop-induced mass can be suppressed compared to the non-SUSY case \cite{daSilva:2020adr}.
As a result, the Dirac mass or the light--heavy neutrino mixing can be larger than the one in the non-SUSY case,
and the Mu-to-$\mubar$ transition can be larger.
Although it can be larger, it is not large enough to observe in near-future experiments.
Therefore, similarly to the non-SUSY model, let us consider a model
with inert Higgs doublets and Dirac mass is forbidden by discrete symmetry.

The model we now consider is the SUSY version of the model in Section \ref{sec7.2}.
We introduce additional Higgs doublets $\eta$ and $\eta^\prime$
(because of gauge anomaly, we need to introduce a pair of Higgs doubles in the SUSY model),
and the Yukawa coupling to $\eta$ with hypercharge $Y=1/2$:
\begin{equation}
W = y_{\alpha i}\, \eta \cdot \ell_\alpha  N_i = y_{\alpha i} (\eta^+ e_\alpha  - \eta^0 \nu_\alpha) N_i.
\end{equation}
We suppose that $\eta^0$ does not acquire a vev and the Dirac neutrino mass is absent.
In order to avoid too large neutrino masses, we introduce a discrete symmetry to forbid 
a $\eta \cdot H_d$ mass term.
In the SUSY model, the so-called $\lambda_5$ term $(\eta H_d)^2$ is absent
in the $F$-term and $D$-term scalar potentials in the SUSY limit,
and the neutrino mass shown in Section \ref{sec7.2} is not generated. 
Introducing the $(\eta H_d)^2$ term in the superpotential, one obtains the $\lambda_5$ term
by Bino and Wino dressing, and the tiny neutrino masses can be generated. 
Alternatively, one can apply the type-II seesaw to generate the neutrino mass.

The box loop via the charged scalar $\eta^+$ is the same as in Section \ref{sec7.2}.
We here consider the box loop via the charged Higgsino $\tilde \eta^+$
and sneutrino $\tilde N$.
The sneutrino masses are given as
\begin{equation}
-{\cal L} \supset (m_{\tilde N}^2)_{ij} \tilde N_i \tilde N_j^\dagger + (\frac12 B_{ij} \tilde N_i \tilde N_j + h.c.),
\end{equation}
where $m_{\tilde N}^2$ contains the SUSY breaking squared mass and 
 Majorana mass of $N$, $1/2 M_i N_i N_i$,
 and $B_{ij}$'s are the coefficients of the SUSY breaking bilinear terms.
For simplicity, we assume that $(m_{\tilde N}^2)_{ij}$ and $B_{ij}$ are flavor diagonal.
By a field redefinition, $B_{ii}$ can be made to be real and positive.
Then, masses of the sneutrinos for $\sqrt2\, {\rm Re}\, \tilde N$ and $\sqrt2\, {\rm Im}\, \tilde N$
are $(m_{\tilde N}^2)_{ii} \pm B_{ii}$.
%
The coefficient of the Mu-to-$\mubar$ transition operator from the sneutrino box loop diagram is obtained as
\begin{eqnarray}
\frac{G_1}{\sqrt2} &=& \frac1{2048 \pi^2 m_{\tilde \eta}^2}  
\left(
y_{ei} y_{\mu i}^* y_{ej} y_{\mu j}^*
\left(I_2 (x_{Ri}, x_{Rj}) + I_2 (x_{Ri}, x_{Ij}) + I_2 (x_{Ii}, x_{Rj}) + I_2 (x_{Ii}, x_{Ij}) \right)
\right. \nonumber \\
&&
\left.
+ (y_{ei})^2 (y_{\mu j}^*)^2 
\left(I_2 (x_{Ri}, x_{Rj}) - I_2 (x_{Ri}, x_{Ij}) - I_2 (x_{Ii}, x_{Rj}) + I_2 (x_{Ii}, x_{Ij}) \right)
\right), \label{G1-sneutrino}
\end{eqnarray}
where
\begin{equation}
x_{Ri} = \frac{(m^2_{\tilde N} + B)_{ii}}{m^2_{\tilde \eta}},
\qquad
x_{Ii} = \frac{(m^2_{\tilde N} - B)_{ii}}{m^2_{\tilde \eta}}.
\end{equation}
We suppose that the Yukawa coupling is given in 
Eq.(\ref{DiracY-me})
to eliminate $\mu \to e\gamma$.
Then, the first term of $G_1$ in Eq.(\ref{G1-sneutrino}) becomes zero,
and the second term remains if $B_{ii} \neq 0$
(which reminds us the $B$-term violates lepton number conservation).
The Yukawa coupling can generate $\tau \to \mu (e) \gamma$:
\begin{equation}
{\rm Br} (\tau \to l_\alpha \gamma) =
 \frac{3 \alpha}{16 \pi} \left| y_{\alpha i}^* y_{\tau i}
\left( \frac{ F_N (x_{Ri})+ F_N(x_{Ii})}{2 G_F m_{\tilde \eta}^2} -
 \frac{ F_N (x_{i})}{G_F m_{\eta}^2}  \right) \right|^2
{\rm Br} (\tau \to l_\alpha \nu\bar\nu), 
\end{equation}
where $x_i = M_i^2/m_\eta^2$, and the loop function 
$F_N(x)$ is given in Eq.(\ref{FN_function}).
%
The Mu-to-$\mubar$ transition can be the largest 
when $x_{Ii} \ll 1 \ll x_{Ri}$,
and our estimation is $|G_1|/G_F \alt O(10^{-5})$.

\section{Dilepton gauge bosons}
\label{sec10}

\begin{figure}
\center
\includegraphics[width=7cm]{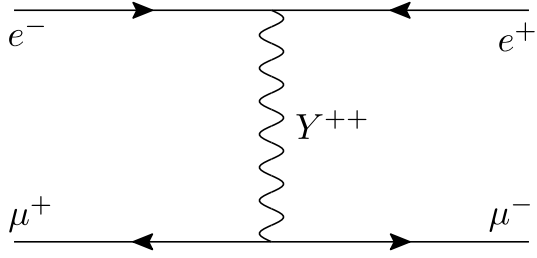}
\caption{
The tree-level exchange of the doubly charged gauge boson to induce the Mu-to-$\mubar$ transition.
}
\label{fig:dilepton}
\end{figure}

In a gauge extension model where the left- and right-handed charged leptons 
are in one multiplet, a doubly-charged dilepton gauge boson can be introduced
\cite{Frampton:1989fu,Frampton:1992wt,Frampton:1991mt,Frampton:1997in,Fujii:1992np,Fujii:1993su}:
\begin{equation}
{\cal L } = \frac{g_{3l}}{\sqrt2}(Y^{++}_\mu \overline{\sf e^c}_i \bar \sigma^\mu {\sf e}_i 
+ Y^{--}_\mu {{\sf e^c}_i}  \sigma^\mu \overline{\sf e}_i ) + M_Y^2 Y_\mu^{++} Y^{\mu --}.
\end{equation}
The gauge coupling $g_{3l}$ is the same as the $SU(2)_L$ gauge coupling $g_2$ (up to a renormalization correction)
in the $SU(3)_c \times SU(3)_L \times U(1)_X$ model.
The tree-level exchange of the doubly-charged dilepton gauge boson in Fig.\ref{fig:dilepton}
can induce the Mu-to-$\mubar$ transition.
Integrating out the gauge boson,
one obtains
\begin{equation}
{\cal L} = - \frac{g_{3l}^2}{2 M_Y^2} (\overline{\sf e^c}_i \bar \sigma^\mu {\sf e}_i ) ({{\sf e^c}_j}  \sigma^\mu \overline{\sf e}_j ) 
=- \frac{g_{3l}^2}{M_Y^2}  (\overline{\sf e^c}_i \overline{\sf  e}_j)  ({\sf e^c}_j {\sf e}_i) .
\end{equation}
In four-component convention, it is written as
\begin{equation}
{\cal L} = - \frac{g_{3l}^2}{M_Y^2} (\overline{e}_j P_R e_i) (\overline{e}_j P_L e_i)
= + \frac{g_{3l}^2}{2M_Y^2} (\overline{e}_j \gamma^\mu P_R e_i) (\overline{e}_j \gamma_\mu P_L e_i).
\end{equation}
Assuming that the current eigenstates  ${\sf e}_i$, ${\sf e^c}_i$ correspond to the mass eigenstates of charged leptons,
one finds that the transition operator $Q_3$ is generated and the coefficient is
\begin{equation}
\frac{G_3}{\sqrt2} = - \frac{g_{3l}^2}{8 M_Y^2}.
\end{equation}
The transition bound, Eq.(\ref{G3_bound}),
implies 
\begin{equation}
M_{Y} > 1.8 \times \frac{g_{3l}}{g_2}\ {\rm TeV}.
\end{equation}

In general, the 
current eigenstates are not the same as the mass eigenstates.
Writing the four-component charged leptons of the mass eigenstates as
\begin{equation}
e_i = \left(
 \begin{array}{c}
  {\sf e}_i \\
  (V^\dagger \overline{\sf e^c})_i
 \end{array}
\right),
\end{equation}
we obtain the gauge interaction as
\begin{equation}
{\cal L} \supset  -\frac{g_{3l}}{2\sqrt2} \left(  \frac{V_{ij}- V_{ji}}{2} e^T_i C \gamma^\mu  e_j
 +  \frac{V_{ij}+ V_{ji}}{2} e^T_i C \gamma^\mu \gamma_5 e_j
 \right) Y^{++}_\mu + h.c.
\end{equation}
Due to the anti-commutation of the fermions, 
the vector term is anti-symmetric 
and the axial-vector term is symmetric under the flavor indices.
Therefore, LFV decays such as $\mu \to 3e$ are induced by the gauge boson exchange in general.
If the charged lepton mass matrix $M_{ij}$ is symmetric (where $M_{ij} {\sf e}_i {\sf e^c}_j$ is the mass term)
and no extra fields are mixed with the charged leptons,
one finds 
that the current eigenstates correspond to the mass eigenstates
and $V_{ij} = \delta_{ij}$.
Then, LFV decays are not induced by this term.

%

\section{Flavored gauge bosons}
\label{sec11}

\begin{figure}
\center
\includegraphics[width=7cm]{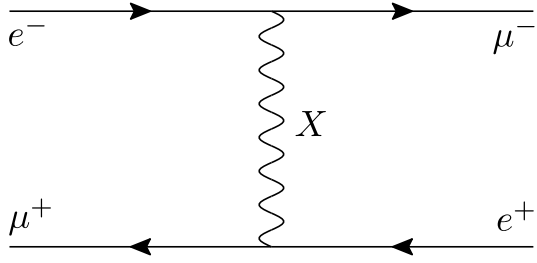}
\caption{
The tree-level exchange of a neutral gauge boson to induce the Mu-to-$\mubar$ transition.
}
\label{fig:flavor-gauge-boson}
\end{figure}

If there is a neutral gauge boson whose interaction is
$\Delta L_e = - \Delta L_\mu = \pm 1$,
the Mu-to-$\mubar$ transition can be induced by its exchange at the tree level as in Fig.\ref{fig:flavor-gauge-boson}.
The gauge boson should have a flavor-dependent charge.
For example, the Mu-to-$\mubar$ transition in non-Abelian flavor gauge symmetry
 is discussed in Ref.\cite{Koide:2010qc}.
We here consider an Abelian flavor gauge symmetry to induce the Mu-to-$\mubar$ transition.

As an example, the extra $U(1)$ charges of the lepton doublets $\ell = (\nu_L,e_L)$ and scalars $\phi_i$ are assigned as
in Table \ref{table1},
\begin{table}
\caption{An example of the extra $U(1)$ charge assignments for the flavored gauge model.}
\begin{center}
\begin{tabular}{cccc}
\\\hline 
$\ell_1$ & $\ell_2$ & $\phi_1$ & $\phi_2$ \\ 
\hline
1 & $-1$ & $-1$ & 1 \\
\hline
\end{tabular}
\end{center}
\label{table1}
\end{table}
and the charges of the other fields are zero.
We introduce vector-like lepton doublets $L$, $\bar L$.
The mass terms for the first and second generations can be written as
\begin{equation}
-{\cal L} = y_1 \phi_1 \overline{ L_R} \ell_{1} + y_2 \phi_2  \overline{L_R} \ell_{2} + M_L \bar L L  
+  \bar y_1 \overline{e_{1R}} L_L H_d + \bar y_2 \overline{e_{2R}} L_L H_d.
\end{equation}
Integrating out the vector-like lepton doublets, 
we obtain the Yukawa interactions of the first and second generations of charged leptons.
The electron and muon mass eigenstates are given by
\begin{equation}
e_L = \cos \theta\, e_1 + \sin \theta\, e_2, \qquad \mu_L = -\sin \theta\, e_1 + \cos\theta\, e_2,
\end{equation}
where $\tan\theta = -y_1 \phi_1/(y_2 \phi_2)$.
Denoting the extra $U(1)$ gauge field as $X_\mu$,
we can write the gauge interaction
as
\begin{equation}
{\cal L} = g_X X_\mu (\bar \ell_1 \gamma^\mu \ell_1 - \bar \ell_2 \gamma^\mu \ell_2) + \frac12 M_X^2 X_\mu X^\mu.
\end{equation}
The gauge interaction of the electron and muon is
\begin{equation}
g_X X_\mu (\bar e_1 \gamma^\mu e_1 - \bar e_2 \gamma^\mu e_2)
= g_X X_\mu ((\bar e_L \gamma^\mu \mu_L + \bar \mu_L \gamma^\mu e_L)\sin2\theta
+ (\bar e_L \gamma^\mu e_L - \bar\mu_L \gamma^\mu \mu_L) \cos2\theta).
\end{equation}
The $\sin2\theta$ term can induce the Mu-to-$\mubar$ transition,
\begin{equation}
{\cal L} \supset - \frac{g_X^2 \sin^2 2\theta}{2M_X^2} (\bar \mu_L \gamma_\mu e_L)(\bar \mu_L \gamma^\mu e_L),
\end{equation}
and
\begin{equation}
\frac{G_1}{\sqrt2} = \frac{g_X^2 \sin^2 2\theta}{8 M_X^2}.
\end{equation}

We need to assume $|\cos2\theta| \ll 1$ to suppress $\mu \to 3e$.
(Surely, we assume that the mixings of $X$ to photon and $Z$ boson are negligibly small).
Naively, Br($\mu \to 3e) < 10^{-12}$ provides a bound:
\begin{equation}
\frac{G_1}{G_F}  |\cot 2\theta|  \alt \frac1{2\sqrt2} \times 10^{-6}.
\end{equation}
Since the signs of $ee$ and $\mu\mu$ couplings to $X_\mu$ in the above are opposite,
the constraint of $\cos2\theta$ from $\mu \to e\gamma$ is loose for $m_e, m_\mu \ll M_X$.
Theoretically,
if there is an exchange symmetry
$\phi_1 \leftrightarrow \phi_2$, $\ell_1 \leftrightarrow \ell_2$
in the Lagrangian,
one finds that $|\tan\theta| = 1$
and the unwanted LFV is absent.


One can similarly consider a model where
$X_\mu$ can couple to both left- and right-handed electron and muons.
In that model,
the bounds of the diagonal couplings to electrons and muons
are stronger.
The right-handed operators for $\mu \to 3e$
are simply added at the tree level,
while the bound becomes stronger due to the chirality flipping in the internal line
for $\mu \to e\gamma$.
%
%
%
In any cases, the models are free from LFV constraints by a choice that the diagonal couplings are absent.
%
Even in the choice, the electron and muon $g-2$ can be induced.
If $X_\mu$ couples only to the left-handed leptons (as the example above),
the ratio of the induced anomalous magnetic moments\footnote{
The current muon $g-2$ measurement \cite{Abi:2021gix} implies 
\begin{equation}
\Delta a_\mu = (2.51 \pm 0.59) \times 10^{-9}.
\end{equation}
The values of the anomalous electron $g-2$ are reported by 
Berkeley \cite{Parker:2018vye} and Laboratoire Kastler Brossel (LKB) \cite{Morel:2020dww} groups:
\begin{equation}
\Delta a_e ({\rm Berkeley})= (-8.8 \pm 3.6) \times 10^{-13}, \qquad \Delta a_e ({\rm LKB}) = (4.8 \pm 3.0) \times 10^{-13},
\end{equation}
which seem to depend on the fine structure constant extracted from the Rydberg constant.
} is
\begin{equation}
\frac{\Delta a_e}{\Delta a_\mu} = \frac{m_e^2}{m_\mu^2}.
\end{equation}
If $X_\mu$ couples to both left- and right-handed ones,
the muon and electron mass can be hit at internal lines,
and thus, the above flavor relation is violated.
One can find that the magnitude of $\Delta a_e/m_e^2$ can be much larger than $\Delta a_\mu/m_\mu^2$.

\begin{figure}
\center
\includegraphics[width=8cm]{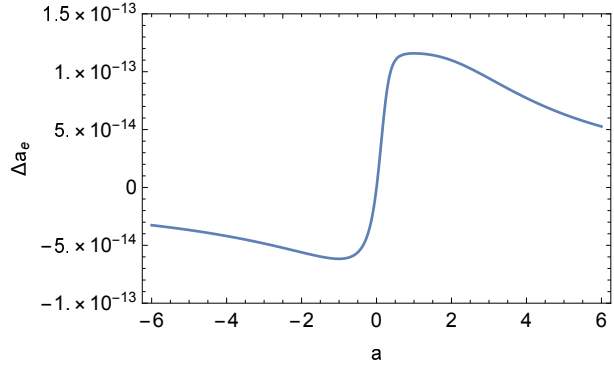}
\caption{
The contribution to the electron $g-2$ ($\Delta a_e$) is shown
as a function of the $U(1)$ charge $a$ for the right-handed charged leptons
when the Mu-to-$\mubar$ transition is assumed to be just the same as the experimental upper bound.
}
\label{fig-ae}
\end{figure}

Let us consider the model where both left- and right-handed leptons couple to the extra gauge boson
in the absence of the diagonal couplings:
\begin{equation}
{\cal L} = g_X X_\mu (\overline{e_L} \gamma^\mu \mu_L + a \overline{e_R} \gamma^\mu \mu_R + (\mu \leftrightarrow e)),
\end{equation}
where $a$ is a $U(1)$ charge for the right-handed charged leptons.
In this case,
we obtain
%
\begin{equation}
\frac{G_1}{\sqrt2} = \frac{g_X^2}{8M_X^2}, \qquad
\frac{G_2}{\sqrt2} = \frac{a^2 g_X^2}{8M_X^2}, \qquad
\frac{G_3}{\sqrt2} = \frac{2a g_X^2}{8M_X^2}.
\end{equation}
The contributions to $(g-2)/2$ of the muon and electron can be calculated as 
\begin{eqnarray}
\Delta a_\mu &=& 
-m_\mu^2 \frac{(1+a^2) g_X^2}{32 \pi^2 M_X^2} \times \frac83, \\
\Delta a_e &=& 
m_e m_\mu \frac{2a g_X^2}{32 \pi^2 M_X^2} G_X\left( \frac{m_\mu^2 }{M_X^2}\right), 
\end{eqnarray}
where the loop function is
\begin{eqnarray}
G_X(x) =  \frac{4-3x-x^3 + 6x \ln x}{(1-x)^3}.
\end{eqnarray}
%
The contribution of the neutral gauge boson to the muon $g-2$ is negative.
The experimental bound of the Mu-to-$\mubar$ transition implies
\begin{eqnarray}
-\Delta a_\mu \alt 2 \times 10^{-11}.
\end{eqnarray}
Therefore, the muon $g-2$ anomaly cannot be explained.
In Fig.\ref{fig-ae},
we show the electron $g-2$ as a function of $a$ when the Mu-to-$\mubar$ transition is just on the current experimental bound.
The magnitude of the electron $g-2$ is one digit smaller than the current central value.
%
Though the electron $g-2$ can be larger than the naive expectation
without flavor violation,
 the Mu-to-$\mubar$ transition constrains the magnitude of it.

\section{Other models}
\label{sec12}

\subsection{Leptoquark}
\label{sec12.1}

If there are $\bar F P_L \ell_i  S$ and/or $\bar F P_R e_i S$ types of interactions ($F$: fermion, $S$: boson),
the Mu-to-$\mubar$ transition can be induced by a box diagram.
As an example, we consider a ``leptoquark'' scalar,
\begin{equation}
D : ({\bf 3}, {\bf 1}, -\frac13),
\end{equation}
and
we allow the following interactions\footnote{In this section, we use two-component fermion convention to avoid complicated expressions.}:
\begin{equation}
- {\cal L} = \bar y_{ij}  \ell_i q_j D^* + y_{ij} e^c_i u^c_j D + h.c.
\end{equation}
We must forbid ``diquark'' interactions ($qq D$ and $u^c d^c D^*$ terms) to avoid a rapid proton decay.
There is a fairly merit to introduce the leptoquark coupling: 
it induces the $b \to s e^+ e^-$, $b \to s \mu^+ \mu^-$ for lepton non-universality at the tree level \cite{Aaij:2021vac,Bauer:2015knc}.

Let us consider $\mu \to e\gamma$ constraints.
We suppose $y_{ij} \to 0$.
Then, the bound is roughly
\begin{equation}
\frac{\bar y_{1i}^* \bar y_{2i} }{M_D^2 G_F} F_1 \alt O(10^{-5}),
\end{equation}
where $F_1$ is a form factor.
If both $y_{ij}$ and $\bar y_{ij}$ are switched on,
the chirality can flip at an internal line,
and the bound can be roughly
\begin{equation}
\frac{m_{u_i}}{m_\mu}\frac{ y_{1i} \bar y_{2i} }{M_D^2 G_F} F_2 \alt O(10^{-5}),
\end{equation}
where $F_2$ is a form factor.
Therefore,
for $y, \bar y < O(1)$, 
the $\mu \to e\gamma$ constraints roughly give the bound,
\begin{equation}
\frac{|G_{1,2,3}|}{G_F} \alt O(10^{-8}).
\end{equation}
This is a consequence when only $(\Delta L_e, \Delta L_\mu) = (\pm1,0),(0,\pm1)$ interactions
are introduced.

\subsection{Vector-like fermions}
\label{sec12.2}

Vector-like fermions are often introduced in flavor models.
Via the box diagram in which the vector-like fermions propagate,
the Mu-to-$\mubar$ transition can be generated.
Here, we consider a minimal version to illustrate the essence. 
We introduce $SU(2)_L$ singlet fermions:
\begin{equation}
\bar E^c : ({\bf 1}, {\bf 1}, -1), \qquad E^c : ({\bf 1}, {\bf 1}, 1),
\end{equation}
and SM singlet flavon complex scalars $\phi_e$ and $\phi_\mu$.
The global $U(1)$ charges are assigned as in Table \ref{table2},
\begin{table}
\caption{The global $U(1)$ charge assignments for the model with vector-like fermions discussed in Section \ref{sec12.2}.}
\begin{center}
\begin{tabular}{cccccc}
\hline 
$e^c$ & $\mu^c$ & $\phi_e$ & $\phi_\mu$ & $\bar E^c$ & $E^c$ \\ 
\hline
$a$ & $b$ & $-a$ & $-b$ & 0 & 0 \\
\hline
\end{tabular}
\end{center}
\label{table2}
\end{table}
and 
Lagrangian is
\begin{equation}
-{\cal L} = h_e \phi_e e^c \bar E^c + h_\mu \phi_\mu \mu^c \bar E^c + \sum_{i=1,2} y_i \ell_i H E^c + M_E \bar E^c E^c.
\end{equation}
The mass matrix is given as
\begin{equation}
-{\cal L} \supset \left(
 \begin{array}{ccc}
  e & \mu & \bar E^c
 \end{array}
\right)
\left(
\begin{array}{ccc}
 0 & 0 & y_1 v \\
 0 & 0 & y_2 v \\
 h_e \phi_e & h_\mu \phi_\mu & M_E
\end{array}
\right)
\left(
\begin{array}{c}
 e^c \\ \mu^c \\ E^c
\end{array}
\right).
\end{equation}
The muon mass is generated as
\begin{equation}
m_\mu \simeq \frac{\sqrt{y_1^2+y_2^2} \sqrt{(h_e \phi_e)^2 + (h_\mu \phi_\mu)^2}}{M_E} v.
\end{equation}
Even if the Yukawa couplings are $O(1)$, the smallness of muon mass is explained
by the vev hierarchy $\phi_e \ll \phi_\mu \ll M_E$.
Surely, the electron is massless at this stage, unless we introduce additional heavier vector-like fermions.
To suppress $\mu \to e\gamma$ via the vector-like fermions, $\phi_e \ll \phi_\mu$ is needed.

The $\phi_e$ field has partial lepton numbers as $(L_e,L_\mu) = (-1,1)$.
Once $\phi_e$ and $\phi_\mu$ acquire vevs,
partial lepton number symmetry is broken (while total lepton number symmetry is kept).
Then, the Mu-to-$\mubar$ transition can be generated as
\begin{eqnarray}
\frac{G_2}{\sqrt2} = 
\frac{ h_e^{*2} h_\mu^{2}}{2048 \pi^2 M_{E}^2}  
\left(I_2 (x_{Re}, x_{R\mu}) - I_2 (x_{Re}, x_{I\mu}) - I_2 (x_{Ie}, x_{R\mu}) + I_2 (x_{Ie}, x_{I\mu}) \right), 
\end{eqnarray}
where
\begin{equation}
x_{R\alpha} = \frac{m_{{\rm Re}\,\phi_\alpha}^2}{M^2_{E}},
\qquad
x_{I\alpha} = \frac{m_{{\rm Im}\,\phi_\alpha}^2}{M^2_{E}}.
\end{equation}
If soft breaking terms of the global $U(1)$ symmetry are absent,
the imaginary parts are massless.
Because the masses of real and imaginary parts split, $G_2$ becomes non-zero.

\subsection{Axion-like particle}

If a global flavor symmetry is broken spontaneously,
the following flavored axion-like coupling can be considered \cite{Calibbi:2020jvd,Bauer:2019gfk,Endo:2020mev}:
\begin{equation}
{\cal L} = \frac{\partial_\mu a}{2 f_a } \bar f_i \gamma^\mu (v_{ij} + \gamma_5 a_{ij} ) f_j.
\end{equation}
By a field redefinition $f$, 
one can consider the axion-like coupling as
\begin{equation}
a \bar f ( y^S_{ij} + \gamma_5 y^P_{ij} ) f.
\end{equation}
Therefore, one can think that the coupling can be a global symmetry version 
of the model in Section 11, 
or the Higgs scalar has a Peccei-Quinn-like global charge in the model in Section~\ref{sec8}.
The Mu-to-$\mubar$ transition can be induced by the exchange of the axion-like particle at the tree level.
The Mu-to-$\mubar$ transition 
induced by the axion-like particle is intensively
discussed in Ref.\cite{Endo:2020mev}.

\section{Conclusion}

The Mu-to-$\mubar$ transition is one of the interesting probes for physics beyond the SM.
In near future, the search experiments will be performed in some facilities, such as J-PARC in Japan and CSNS in China.
In anticipation of those upcoming experiments, we have evaluated their impact on models and the connection to other experiments.
Assuming appropriate mediators, we are allowed to consider five independent effective operators to induce the transition.
In terms of the effective couplings, we have shown the general formula of the transition probability in the zero or nonzero magnetic field.
We have pointed out that the magnetic-field dependence helps us to identify the type of dominant effective operators.

For each plausible new physics model, we have estimated the maximum size of the induced effective couplings $G_i$ ($i=1$\dash$5$) given in Eq.(\ref{eq:Gi}), taking into account current experimental constraints.
The result shows that
 the Mu-to-$\mubar$ transition is most sensitive to the cases 
 where the mediator has $\Delta L_e - \Delta L_\mu = \pm 2$ interactions in the model: 
 such as a doubly charged scalar, doubly charged gauge boson, 
 neutral scalar, 
 and neutral gauge boson.
Since the transition is generated at the tree level by the exchange of the mediator, 
the size of the effective couplings can be $|G_i|/G_F \alt O(10^{-3})$, 
which is as large as the current experimental bound (See Eqs.(\ref{transition-probability})-(\ref{G3_bound})
for the bound).
The severe experimental constraints from LFV decays such as $\mu \to e\gamma$ and $\mu \to 3e$
 can be avoided for the mediators with the $\Delta L_e - \Delta L_\mu = \pm 2$ interactions.
The left--right model with $SU(2)_R$ triplet is typically the case, for example.
To generate the sizable Mu-to-$\mubar$ transition avoiding the LFV decays, 
one needs some ideas such as introducing a discrete flavor symmetry.
Thus, the possible observation of the Mu-to-$\mubar$ transition at the near-future experiment can give us the drastic paradigm change of the understanding of the lepton flavors.

In the models with $\Delta L_e - \Delta L_\mu = \pm 1$ interactions, the Mu-to-$\mubar$ transition can be generated by box loop diagrams.
However, the $\mu \to e\gamma$ constraint severely bounds the transition, and we have obtained $|G_i|/G_F \alt O(10^{-8})$ at most.

If mass terms violate the (partial) lepton numbers, the Mu-to-$\mubar$ transition can be generated by box loop diagrams
even without the LFV interactions.
The light--heavy neutrino mixing needs to be large
in order to enlarge the Mu-to-$\mubar$ transition induced by the lepton number violation.
However, 
the large light--heavy neutrino mixing can induce sizable active neutrino masses radiatively,
which spoils the realization of the sub-eV neutrino mass. 
In the sense, 
the Mu-to-$\mubar$ transition induced by the lepton number violation
is conceptually constrained to realize the sub-eV neutrino masses.
We need to assume the neutrino masses to be zero by a discrete symmetry at the tree level, 
and the sub-eV active neutrino masses are generated radiatively
to obtain a large size of the Mu-to-$\mubar$ transition.
We have found $|G_i|/G_F \alt O(10^{-5})$ in such models.

\begin{table}[t]
\caption{
The models to generate the Mu-to-$\mubar$ transition at the tree level and the corresponding coefficients $G_i$.
For each model, a checkmark $\checkmark$ is placed in the column where
the coefficient can be obtained around the current experimental bound ($|G_i|/G_F \sim O(10^{-3})$).
In the column where the coefficient that is generated but becomes smaller due to LFV bounds,
a triangle mark $\triangle$ is placed.
The section column contains the section number in which the model is described.
}
\center
\begin{tabular}{|c|c|c|c|c|c|c|}
\hline
 Model & $G_1$ & $G_2$ & $G_3$ & $G_4$ & $G_5$ & Section \\
 \hline 
 Type I + II hybrid seesaw & \checkmark & --&-- &-- &-- & \ref{sec5}
 \\
  \hline
 Left-right model with $SU(2)_R$ triplet &-- & \checkmark &-- &-- &-- & \ref{sec6.2}
 \\
 \hline
 Inert Higgs doublet &-- &-- & \checkmark & $\triangle$ & $\triangle$ & \ref{sec8}
 \\
 \hline
  $R$-parity violating SUSY &-- &-- & \checkmark & -- & -- & \ref{sec9.1}
 \\
 \hline
 Dilepton gauge boson &-- &-- & \checkmark & -- & -- & \ref{sec10}
 \\
 \hline
 Neutral flavor gauge boson &\checkmark &\checkmark & \checkmark & -- & -- & \ref{sec11}
 \\
 \hline
\end{tabular}
\label{Table3}
\end{table}

\begin{table}[t]
\caption{
The models in which the Mu-to-$\mubar$ transitions are induced at the one-loop level 
by $\Delta L_e - \Delta L_\mu = \pm 1$ interactions,
and the estimation of the corresponding coefficients.
Due to the LFV constraints, the induced size is tiny in this category of the interaction.
}
\center
\begin{tabular}{|c|c|c|c|c|}
\hline
 Model & $|G_1|/G_F$ & $|G_2|/G_F$ & $|G_3|/G_F$  & Section \\
 \hline 
 Heavy singlet neutrino & $\alt O(10^{-8})$ & --&--  & \ref{sec4}
 \\
  \hline
 Left-right model without $SU(2)_R$ triplet & $\alt O(10^{-8})$ & $\alt O(10^{-8})$ & $\alt O(10^{-10})$& \ref{sec6.1}
 \\
 \hline
 SUSY (Gaugino loop) & $\alt O(10^{-8})$  &-- & -- &  \ref{sec9.2}
 \\
 \hline 
 Leptoquark & $\alt O(10^{-8})$  & $\alt O(10^{-8})$  & $\alt O(10^{-8})$  &  \ref{sec12.1}
 \\
 \hline 
 \end{tabular}
\label{Table4}
\end{table}

\begin{table}[t]
\caption{
The radiative neutrino mass models in which the Mu-to-$\mubar$ transitions are induced
at the one-loop level by using the lepton number violation.
The constraints from LFV decays via $\Delta L_e - \Delta L_\mu = \pm 1$ interactions can be avoided.
$^{(*)}$In order to realize the size of the active neutrino masses in the KNT model, 
one needs to consider a non-perturbative region of the couplings.
Then, the induced size of the coefficient can be larger, but the quality of the estimate is different from others.
}
\center
\begin{tabular}{|c|c|c|c|c|}
\hline
 Model & $|G_1|/G_F$ & $|G_2|/G_F$  & Section \\
 \hline
 Charged Higgs(ino) & $\alt O(10^{-5})$ &--  & \ref{sec7.2}, \ref{sec9.3}
 \\
 \hline
 KNT model &  --  & $\alt O(10^{-5}) {}^{(*)}$   & \ref{sec7.3.1}
 \\
 \hline 
 AKS model &  --  & $\alt O(10^{-6}) $ &   \ref{sec7.3.2}
 \\
 \hline 
\end{tabular}
\label{Table5}
\end{table}

\begin{table}[t]
\caption{
The neutrino mass models in which the Mu-to-$\mubar$ transitions 
are induced at the tree level 
via $\Delta L_e - \Delta L_\mu = \pm 2$ interactions with doubly charged scalars.
$^{(\#)}$In the type-II seesaw, 
the induced size of the coefficient is bounded by the absolute neutrino mass 
from the cosmological measurements.
If neutrinos are not stable in the cosmological time, the bound is not applied
and the coefficient of $|G_1|$ can be larger.
}
\center
\begin{tabular}{|c|c|c|c|}
\hline
 Model & $|G_1|/G_F$ & $|G_2|/G_F$  & Section \\
 \hline 
 Type-II seesaw & $\alt O(10^{-5}) {}^{(\#)}$ & --& \ref{sec5}
 \\
  \hline
 Zee-Babu model &  --  & $\alt O(10^{-3})$ &  \ref{sec7.1.1}
 \\
 \hline
 Cocktail model &  --  & $\alt O(10^{-5})$ &  \ref{sec7.1.2}
 \\
 \hline 
\end{tabular}
\label{Table6}
\end{table}

The existence of the doubly charged scalar can be related to the neutrino mass generation.
As examples of such models, we have investigated the type-II seesaw, Zee-Babu, and cocktail models.
The Zee-Babu model especially leaves the possibility of $|G_2|/G_F \sim O(10^{-3})$, which is related to $\tau^- \to \mu^+ \mu^- \mu^-$ and $\tau^- \to \mu^+ e^- e^-$.
In the type-II seesaw model, the large Mu-to-$\mubar$ transition favors degenerate neutrino masses, which results in $|G_1|/G_F \alt O(10^{-5})$ if the cosmological neutrino mass bound is applied.
The cocktail model predicts $|G_2|/G_F \alt O(10^{-5})$ due to the constraints of $\tau^- \to e^+ e^- e^-$ and $\tau^- \to e^+ \mu^- \mu^-$.

The Mu-to-$\mubar$ transition is model-independently related to the measurements of the Mu HFS interval
 by the effective couplings.
Although the current sensitivity of the HFS interval is less than that of the Mu-to-$\mubar$ transition, future excellent upgrades will 
have the potential to provide us with a way to check the Mu-to-$\mubar$ transition.
Furthermore, through specified models, the Mu-to-$\mubar$ transition is connected to the other observables in flavor physics (e.g., neutrino masses, muon/electron $g-2$, LFV searches, and so on) and direct particle searches by collider experiments.
The near-future experiments for the Mu-to-$\mubar$ transition will give us complementary information to investigate the detailed structure of high-energy physics.

\medskip

We conclude with model-by-model tables of the Mu-to-$\mubar$ transition
for the convenience of readers,
though there are some overlaps with what we have already described in this section.
In Table \ref{Table3},
we list the models
in which the Mu-to-$\mubar$ transition operators 
are generated at the tree-level 
and they are tested by the near-future Mu-to-$\mubar$ transition experiments.
In Table~\ref{Table4},
we list the models
in which the transition operators are generated
by box diagrams via $\Delta L_e - \Delta L_\mu = \pm 1$ interactions,
and the size of the transitions is suppressed by the constraints from LFV decays.
If the active neutrino masses are generated radiatively,
the lepton number violation is utilized to
generate the Mu-to-$\mubar$ transitions by box diagrams avoiding the
constraints from LFV decays.
We list those models in Table \ref{Table5}.
In Table \ref{Table6},
we list the predictive neutrino mass models with doubly charged scalars
which can induce the Mu-to-$\mubar$ transitions at the tree level.

\section*{Acknowledgements}

We would like to thank N.~Kawamura and K.~Shimomura for their useful comments.
This work was supported by JSPS KAKENHI Grant Numbers JP18H01210 and JP21H00081 (Y.U.).

\appendix

\section{Transition amplitude} \label{app:transition_amplitude}

Here, we show a calculation for the amplitude of the Mu-to-$\mubar$ transition~\cite{Conlin:2020veq}.

The charged lepton fields $l=\mu,e$ included in the operators, $Q_i$ in Eqs.\eqref{eq:Q1}-\eqref{eq:Q5}, are explicitly written as
\begin{align}
l\left(x\right)=\int\frac{d^3p}{(2\pi)^3}\frac{1}{\sqrt{2E_p}}\sum_{s}\left\{a_l^{\bm{p},s}u_l^s\left(p\right)\exp\left(-ipx\right)+b_l^{\bm{p},s\dagger}v_l^s\left(p\right)\exp\left(ipx\right)\right\},
\end{align}
where $u_l^s\left(p\right)$ and $v_l^s\left(p\right)$ are the four-component Dirac spinors.
In the nonrelativistic limit, the spinors are given as
\begin{align}
u_l^s=\sqrt{m_l}
\begin{pmatrix}
\xi^s \\
\xi^s
\end{pmatrix}, \hspace{5mm}
v_l^s=\sqrt{m_l}
\begin{pmatrix}
\eta^s \\
-\eta^s
\end{pmatrix},
\end{align}
where $\xi^{+1/2}=\left(1,0\right)^T$, $\xi^{-1/2}=\left(0,1\right)^T$, $\eta^{+1/2}=\left(0,1\right)^T$, and $\eta^{-1/2}=\left(-1,0\right)^T$.
The creation and annihilation operators satisfy the anticommutation relation,
\begin{align}
\left\{a_l^{\bm{p},s},a_{l'}^{\bm{p}',s'\dagger}\right\}=\left\{b_l^{\bm{p},s},b_{l'}^{\bm{p}',s'\dagger}\right\}=(2\pi)^3\delta^{(3)}\left(\bm{p}-\bm{p}'\right)\delta_{s,s'}\delta_{l,l'}.
\end{align}
Using the creation operators, we define the Mu state as
\begin{align}
\ket{\textrm{Mu}\left(\bm{P}\right);F,m}=\sum_{s_e,s_\mu}\left(1/2,s_\mu,1/2,s_e|F,m\right)\int\frac{d^3q}{(2\pi)^3}\tilde{\varphi}\left(\bm{q}\right)a_e^{\bm{p}_e,s_e\dagger}b_\mu^{\bm{p}_\mu,s_\mu\dagger}\ket{0},
\label{eq:muonium_state}
\end{align}
where $\bm{P}$ and $\bm{q}$ are the total and relative momentums of the muon-electron system, respectively, so that $\bm{p}_e=m_e/\left(m_e+m_\mu\right)\bm{P}+\bm{q}$ and $\bm{p}_\mu=m_\mu/\left(m_e+m_\mu\right)\bm{P}-\bm{q}$.
Here, $\left(1/2,s_\mu,1/2,s_e|F,m\right)$ is the Clebsch-Gordan coefficient.
The wave function $\tilde{\varphi}\left(\bm{q}\right)$ of the ground state in the momentum space is normalized to be
\begin{align}
\int\frac{d^3q}{\left(2\pi\right)^3}\left|\tilde{\varphi}\left(\bm{q}\right)\right|^2=1.
\end{align}
With the above definition, the Mu state obeys
\begin{align}
\braket{\textrm{Mu}(\bm{P});F,m|\textrm{Mu}(\bm{P}');F',m'}=(2\pi)^3\delta^{(3)}\left(\bm{P}-\bm{P}'\right)\delta_{F,F'}\delta_{m,m'}.
\end{align}
On the other hand, the $\mubar$ state is defined by the charge conjugation of Eq.\eqref{eq:muonium_state} \footnote{
We note that the definition of the charge conjugations of the creation and annihilation operators can have unphysical sign freedom.
We define the $\mubar$ state by Eq.(\ref{eq:antimuonium_state}).
},
\begin{align}
\ket{\overline{\textrm{Mu}}\left(\bm{P}\right);F,m}=\sum_{s_e,s_\mu}\left(1/2,s_\mu,1/2,s_e|F,m\right)\int\frac{d^3q}{(2\pi)^3}\tilde{\varphi}\left(\bm{q}\right)b_e^{\bm{p}_e,s_e\dagger}a_\mu^{\bm{p}_\mu,s_\mu\dagger}\ket{0}.
\label{eq:antimuonium_state}
\end{align}

We show the calculation for the transition amplitude via the operator $Q_1$,
\begin{align}
\braket{\overline{\mathrm{Mu}};F,m|Q_1|\mathrm{Mu};F,m}=\braket{\overline{\mathrm{Mu}};F,m|(\bar \mu \gamma_\mu (1- \gamma_5) e ) ( \bar \mu \gamma^\mu (1- \gamma_5) e )|\mathrm{Mu};F,m},
\end{align}
as an example.
By substituting the explicit form of the field into the equation and using the anticommutation relation, we obtain
\begin{align}
\braket{\overline{\mathrm{Mu}};F,m|Q_1|\mathrm{Mu};F,m}=& \ \frac{\left|\varphi\left(0\right)\right|^2}{m_\mu m_e}\sum_{\overline{s}_e,\overline{s}_\mu}\left(1/2,\overline{s}_\mu,1/2,\overline{s}_e|F,m\right)\sum_{s_e,s_\mu}\left(1/2,s_\mu,1/2,s_e|F,m\right) \nonumber\\
&\!\!\!\!\!\! \times\left[-\left(\overline{u}_\mu^{\overline{s}_\mu}\gamma_\alpha P_Lv_e^{\overline{s}_e}\right)\left(\overline{v}_\mu^{s_\mu}\gamma^\alpha P_Lu_e^{s_e}\right)-\left(\overline{v}_\mu^{s_\mu}\gamma_\alpha P_Lu_e^{\overline{s}_e}\right)\left(\overline{u}_\mu^{\overline{s}_\mu}\gamma^\alpha P_Lv_e^{s_e}\right)\right. \nonumber\\
&\left.+\left(\overline{v}_\mu^{s_\mu}\gamma_\alpha P_Lv_e^{\overline{s}_e}\right)\left(\overline{u}_\mu^{\overline{s}_\mu}\gamma^\alpha P_Lu_e^{s_e}\right)+\left(\overline{u}_\mu^{\overline{s}_\mu}\gamma_\alpha P_Lu_e^{s_e}\right)\left(\overline{v}_\mu^{s_\mu}\gamma^\alpha P_Lv_e^{\overline{s}_e}\right)\right],
\label{eq:amplitude_Q1}
\end{align}
where $\varphi$ is the wave function in the coordinate space,
\begin{align}
\varphi(\bm{r})=&\int\frac{d^3q}{\left(2\pi\right)^3}\tilde{\varphi}\left(\bm{q}\right)\exp\left(i\bm{q}\cdot\bm{r}\right).
\end{align}
The first term in the square bracket of Eq.~\eqref{eq:amplitude_Q1} reduces to
\begin{align}
\left(\overline{u}_\mu^{\overline{s}_\mu}\gamma_\alpha P_Lv_e^{\overline{s}_e}\right)\left(\overline{v}_\mu^{s_\mu}\gamma^\alpha P_Lu_e^{s_e}\right)=m_\mu m_e\mathrm{Tr}\left[\eta^{\overline{s}_e}\xi^{\overline{s}_\mu\dagger}\overline{\sigma}_\alpha\right]\mathrm{Tr}\left[\xi^{s_e}\eta^{s_\mu\dagger}\overline{\sigma}^\alpha\right],
\end{align}
where $\overline{\sigma}^\alpha=\left(\sigma_0^0,-\sigma_1^i\right)$ and $\sigma_0^0$ is the $2\times 2$ identity matrix, and $\sigma_1^i$ ($i=1,2,3$) is the Pauli matrix.
The products of spinors, $\eta$ and $\xi$, can be replaced with the Pauli matrices by
\begin{align}
\sum_{s_e,s_\mu}\left(1/2,s_\mu,1/2,s_e|F,m\right)\xi^{s_e}\eta^{s_\mu\dagger}=&\frac{\overline{\sigma}_F^m}{\sqrt{2}}, \label{eq:xieta} \\
\sum_{\overline{s}_e,\overline{s}_\mu}\left(1/2,\overline{s}_\mu,1/2,\overline{s}_e|F,m\right)\eta^{\overline{s}_e}\xi^{\overline{s}_\mu\dagger}=&-\frac{\sigma_F^m}{\sqrt{2}},
\label{eq:etaxi}
\end{align}
where $\sigma_0^0=\overline{\sigma}_0^0$ is the 2$\times$2 identity matrix, and $\sigma_1^m$ ($m=0,\pm 1$) in those equations represents the Pauli matrix in the spherical basis: $\sigma_1^{\pm 1}=-\overline{\sigma}_1^{\pm 1}=\left(\mp \sigma_1-i\sigma_2\right)/\sqrt{2}$, and $\sigma_1^{0}=-\overline{\sigma}_1^{0}=\sigma_3$.
After calculating the traces, we obtain
\begin{align}
\sum_{\overline{s}_e,\overline{s}_\mu}\left(1/2,\overline{s}_\mu,1/2,\overline{s}_e|F,m\right)\sum_{s_e,s_\mu}\left(1/2,s_\mu,1/2,s_e|F,m\right)\left(\overline{u}_\mu^{\overline{s}_\mu}\gamma_\alpha P_Lv_e^{\overline{s}_e}\right)\left(\overline{v}_\mu^{s_\mu}\gamma^\alpha P_Lu_e^{s_e}\right)=& 2m_\mu m_e.
\end{align}
By the Fierz transformation, the third term is changed into
\begin{align}
\left(\overline{v}_\mu^{s_\mu}\gamma_\alpha P_Lv_e^{\overline{s}_e}\right)\left(\overline{u}_\mu^{\overline{s}_\mu}\gamma^\alpha P_Lu_e^{s_e}\right)=&-\left(\overline{u}_\mu^{\overline{s}_\mu}\gamma_\alpha P_Lv_e^{\overline{s}_e}\right)\left(\overline{v}_\mu^{s_\mu}\gamma^\alpha P_Lu_e^{s_e}\right).
\end{align}
The second and fourth terms are the same as the first and third ones in the case of the operator $Q_1$.
Thus, we obtain
\begin{align}
\braket{\overline{\mathrm{Mu}};F,m|Q_1|\mathrm{Mu};F,m}=-8\left|\varphi(0)\right|^2,
\end{align}
for $(F,m)=(1,\pm 1)$, $(1,0)$, and $(0,0)$.

By applying the similar procedure, we also find the amplitudes for $Q_2$\dash $Q_5$,
\begin{align}
&\braket{\overline{\mathrm{Mu}};F,m|Q_2|\mathrm{Mu};F,m}=-8\left|\varphi(0)\right|^2, \\
&\braket{\overline{\mathrm{Mu}};0,0|Q_3|\mathrm{Mu};0,0}= 12\left|\varphi(0)\right|^2, \hspace{3mm} \braket{\overline{\mathrm{Mu}};1,m|Q_3|\mathrm{Mu};1,m}=-4\left|\varphi(0)\right|^2, \\
&\braket{\overline{\mathrm{Mu}};F,m|Q_4|\mathrm{Mu};F,m}=\braket{\overline{\mathrm{Mu}};F,m|Q_5|\mathrm{Mu};F,m}=2\left|\varphi(0)\right|^2.
\end{align}
We now easily obtain Eqs.\eqref{eq:amplitude00}-\eqref{eq:amplitude10} \footnote{Ref.\cite{Conlin:2020veq} includes some typos in the signs of amplitudes.}.

\section{Muonium ground-state spectroscopy}  \label{app:spectroscopy}

The Zeeman effect splits the energy levels of the Mu states.
The spin Hamiltonian of Mu is given as
\begin{equation}
{\cal H} = a \, {\bm S}_\mu \cdot {\bm S}_e - {\bm \mu}_{e^-} \cdot {\bm B} - {\bm \mu}_{\mu^+} \cdot {\bm B}.
\end{equation}
Here, $a = h \nu_{\rm HFS}= 2\pi \nu_{\rm HFS}$ is the HFS coupling constant,
and ${\bm \mu}_{e^-} , {\bm \mu}_{\mu^+}$ are the magnetic moments of the electron and anti-muon:
\begin{equation}
{\bm \mu}_{e^-} = - g_e \mu_B {\bm S}_e, \qquad {\bm \mu}_{\mu^+} = g_\mu \frac{m_e}{m_\mu} \mu_B {\bm S}_\mu,
\end{equation}
where $\mu_B$ is the Bohr magneton, and $g_{e,\mu}$ are $g$-factors of the electron and muon.

We first ignore the Mu\dash$\mubar$ mixing interactions.
The energy eigenvalues of Mu states $|{\rm Mu}; F,m \rangle_B$ are given as
\begin{align}
E_B\left(\mathrm{Mu};1,\pm 1\right)=&\, E_0+\frac{a}{2}\left(\frac{1}{2}\pm Y\right), \label{E_B_pm1}\\
E_B\left(\mathrm{Mu};1,0\right)=&\, E_0+\frac{a}{2}\left(-\frac{1}{2}+\sqrt{1+X^2}\right), \\
E_B\left(\mathrm{Mu};0,0\right)=&\, E_0+\frac{a}{2}\left(-\frac{1}{2}-\sqrt{1+X^2}\right).
\end{align}
Here, $E_0\simeq -m_\textrm{red}\alpha^2/2$ is the $1S$ binding energy and
\begin{align}
\label{eq:X} X =& \, \frac{\mu_B B}{a} \left(g_e + \frac{m_e}{m_\mu} g_\mu\right) \simeq 6.31 \,  \frac{B}{\rm Tesla}, \\
Y =& \, \frac{\mu_B B}{a} \left(g_e - \frac{m_e}{m_\mu} g_\mu\right) \simeq 6.25 \, \frac{B}{\rm Tesla},
\end{align}
where we have used 
$\mu_B\simeq 5.788\times 10^{-5}$ eV/Tesla, $a\simeq 1.846\times 10^{-5}$ eV, and $g_e\simeq g_\mu\simeq 2.002$.
Since the response of $\mubar$ to the magnetic field is opposite to that of Mu, we obtain the formulae for $\mubar$ by replacing $X$ and $Y$ with $-X$ and $-Y$, respectively.
It is found that the energy values for $\mubar$ are
\begin{align}
E_B\left(\overline{\mathrm{Mu}};1,\pm 1\right)=&\, E_0+\frac{a}{2}\left(\frac{1}{2}\mp Y\right), \\
E_B\left(\overline{\mathrm{Mu}};1,0\right)=&\, E_0+\frac{a}{2}\left(-\frac{1}{2}+\sqrt{1+X^2}\right), \\
E_B\left(\overline{\mathrm{Mu}};0,0\right)=&\, E_0+\frac{a}{2}\left(-\frac{1}{2}-\sqrt{1+X^2}\right).
\end{align}
We note that the $(0,0)$ and $(1,0)$ states for the zero magnetic field are mixed as
\begin{align}
\begin{pmatrix}
\ket{\mathrm{Mu};1,0}_B \\
\ket{\mathrm{Mu};0,0}_B
\end{pmatrix}=&
\begin{pmatrix}
C & -S \\
S & C
\end{pmatrix}
\begin{pmatrix}
\ket{\mathrm{Mu};1,0} \\
\ket{\mathrm{Mu};0,0}
\end{pmatrix}, \\
\begin{pmatrix}
\ket{\overline{\mathrm{Mu}};1,0}_B \\
\ket{\overline{\mathrm{Mu}};0,0}_B
\end{pmatrix}=&
\begin{pmatrix}
C & S \\
-S & C
\end{pmatrix}
\begin{pmatrix}
\ket{\overline{\mathrm{Mu}};1,0} \\
\ket{\overline{\mathrm{Mu}};0,0}
\end{pmatrix},
\end{align}
where the mixing is given as 
\begin{align}
C=\frac{1}{\sqrt2} \left(1+\frac{1}{\sqrt{1+X^2}}\right)^{\frac12}, \qquad
S=\frac{1}{\sqrt2} \left(1-\frac{1}{\sqrt{1+X^2}}\right)^{\frac12}.
\end{align}
%

The experimental measurement of the HFS interval in the strong magnetic field is obtained as
\begin{equation}
a  = h\nu_{12} + h\nu_{34},
\end{equation}
where
\begin{eqnarray}
h\nu_{12} &\equiv& E_{B} ({\rm Mu}; 1, 1) - E_B ({\rm Mu}; 1,0) = \frac{a}2\left(1 + Y - \sqrt{1+X^2}\right), \\
h\nu_{34} &\equiv& E_{B} ({\rm Mu}; 1, -1) - E_B ({\rm Mu}; 0,0) = \frac{a}2\left(1 - Y + \sqrt{1+X^2}\right).
\end{eqnarray}

Now we switch on the Mu\dash$\mubar$ mixing interactions
and ${\cal M}_{F,m} \neq 0$.
To calculate the energy states, we drop the $\Gamma_{12}$ part of ${\cal M}_{12} = M_{12} - i \Gamma_{12}/2$
and we denote the $M_{12}$ parts of $F=0$ and $F=1$ states
as $M_0$ and $M_1$, which can be made to be real by rephasing of the states.

If the external magnetic field is zero, 
Mu and $\mubar$ are maximally mixed 
and the energy eigenstates correspond to the CP eigenstates:
\begin{equation}
| \pm ; F, m \rangle \equiv \frac1{\sqrt2} ( | {\rm Mu} ; F, m \rangle \pm  | \overline{\rm Mu} ; F, m \rangle ).
\end{equation}

The energy matrix of $( | {\rm Mu}, 1, 1 \rangle, |\mubar, 1, 1\rangle )$
can be written as
\begin{equation}
E_0 {\bf 1}+ \left(
 \begin{array}{cc}
   \frac{a}4 + \frac{a}2 Y & M_1 \\
   M_1 & \frac{a}4 - \frac{a}2 Y
 \end{array}
\right) .
\end{equation}
The energy eigenvalues are
\begin{equation}
E_{(1,1),\pm} = E_0 + \frac{a}4 \pm \sqrt{ \left(\frac{a}2 Y\right)^2 + M_1^2 },
\end{equation}
and $\Delta E \simeq a Y$.
The Mu\dash$\mubar$ mixing becomes small in the magnetic field due to $\Delta E \gg M_1$.
The $m= -1$ state is similar to the $m=1$ state.

The energy matrix of $m=0$ states
$( | {\rm Mu};1,0 \rangle , | {\rm Mu}; 0, 0 \rangle, 
 | \overline{\rm Mu}; 1,0 \rangle, | \overline{\rm Mu}; 0,0 \rangle )$
in an appropriate sign convention can be written as
\begin{equation}
E_0 {\bf 1} + 
\left(
 \begin{array}{cccc}
  \frac{a}4 & -\frac{a}2 X & M_1 & 0 \\
  -\frac{a}2 X & -\frac{3}4 a & 0 & M_0 \\
  M_1 & 0 & \frac{a}4 & \frac{a}2 X \\
  0 & M_0 & \frac{a}2 X & -\frac34 a
 \end{array}
\right).
\end{equation}
The energy eigenstates in the magnetic field are
\begin{eqnarray}
 | + ; 1,0 \rangle \cos\theta_M -  | - ; 0,0 \rangle \sin\theta_M 
 &\simeq& 
 (|{\rm Mu}; 1,0 \rangle_B +  |\mubar; 1,0 \rangle_B)/\sqrt2 ,
 \\
 | + ; 1,0 \rangle \sin\theta_M +  | - ; 0,0 \rangle \cos\theta_M
& \simeq& 
 (|{\rm Mu}; 0,0 \rangle_B -  |\mubar; 0,0 \rangle_B)/\sqrt2, 
\\
 | - ; 1,0 \rangle \cos \theta_{\bar M} -  | + ; 0,0 \rangle \sin \theta_{\bar M}
& \simeq &
 (|{\rm Mu}; 1,0 \rangle_B -  |\mubar; 1,0 \rangle_B)/\sqrt2, 
 \\
| - ; 1,0 \rangle \sin\theta_{\bar M} + | + ; 0,0 \rangle \cos\theta_{\bar M} 
&\simeq & 
 (|{\rm Mu}; 0,0 \rangle_B +  |\mubar; 0,0 \rangle_B)/\sqrt2, 
\end{eqnarray}
where
\begin{equation}
\tan2\theta_M = \frac{a X}{a + M_1 + M_0},
\qquad
\tan2\theta_{\bar M} = \frac{a X}{a - M_1 - M_0}.
\end{equation}
The energy eigenvalues are
\begin{eqnarray}
E_{(1,0), \pm} &=& E_0 - \frac{a}4 \pm \frac{M_1 - M_0}{2} + \frac12 \sqrt{ (a \pm M_1 \pm M_0)^2 + (aX)^2} \nonumber \\
&\simeq& E_0 - \frac{a}4 + \frac{a}2 \sqrt{1+X^2} \pm \left(\frac{M_1 - M_0}{2} + \frac{M_1+M_0}{2\sqrt{1+X^2}}\right), \\
E_{(0,0), \pm} &=& E_0 - \frac{a}4 \pm \frac{M_1 - M_0}{2} - \frac12 \sqrt{ (a \pm M_1 \pm M_0)^2 + (aX)^2} \nonumber \\
&\simeq& E_0 - \frac{a}4 - \frac{a}2 \sqrt{1+X^2} \mp \left(- \frac{M_1 - M_0}{2} + \frac{M_1+M_0}{2\sqrt{1+X^2}}\right). 
\end{eqnarray}
The precise development is described by the $4\times 4$ matrix,
but the transitions can approximately happen only 
$| {\rm Mu} ; 1,0 \rangle_B \to | \mubar ; 1,0 \rangle_B$ and
$| {\rm Mu} ; 0,0 \rangle_B \to | \mubar ; 0,0 \rangle_B$,
and the mass differences are
\begin{eqnarray}
\frac{\Delta M^B_{1,0}}2 \,(= M_{1}^B)& =& C^2 M_1 - S^2 M_0 = \frac{M_1-M_0}{2}+ \frac{M_1+M_0}{2\sqrt{1+X^2}} ,\\
\frac{\Delta M^B_{0,0}}2 \,(= M_{0}^B)& =& C^2 M_0 - S^2 M_1 =- \frac{M_1-M_0}{2}+ \frac{M_1+M_0}{2\sqrt{1+X^2}} .
\end{eqnarray}

\section{Diagonalization of neutrino mass matrix}
\label{app:diagonalization}


We work on the basis where the charged-lepton mass matrix is diagonal.
The $6\times 6$ neutrino mass matrix ${\cal M}$ is written as
\begin{equation}
-{\cal L}_m = \frac12 \left( 
 \begin{array}{cc}
   \overline{(\nu^c)_R} & \overline{N_R}
 \end{array}
 \right)
{\cal M}
\left( 
 \begin{array}{c}
   \nu_L \\ (N^c)_L 
 \end{array}
 \right) + h.c.,
\end{equation}
where $\nu$ and $N$ are
 current-basis left- and right-handed neutrinos, and
 \begin{equation}
{\cal M} = 
\left( 
 \begin{array}{cc}
   0 & m_D \\
   m_D^T & M_N 
 \end{array}
 \right).
\end{equation}
 The mass eigenstates $\nu', N'$ are given as
\begin{equation}
\left( 
 \begin{array}{c}
   \nu_L \\ (N^c)_L 
 \end{array}
 \right) = {\cal U} 
 \left( 
 \begin{array}{c}
   \nu'_L \\ N'_L
 \end{array}
 \right),
\end{equation}
and 
\begin{equation}
{\cal U}^T {\cal M} \, {\cal U} = {\rm diag} (M_{\cal I}) =  {\rm diag} (m_i, M_{I}) .
\end{equation}
We choose phases in $\cal U$ so that $M_{\cal I}$'s are real.
We use index 
$i$ 
 for the light neutrino mass eigenstates,
index $I$ for the ``heavy'' neutrino mass eigenstates,
and index ${\cal I}$ for both states.
We use the Greek characters $\alpha, \beta$ for the generation index in the current basis.
%
For convenience, we define
\begin{equation}
{\cal U} = \left( 
 \begin{array}{cc}
   U & X \\
   V & Y
 \end{array}
 \right).
\end{equation}
Namely,
\begin{eqnarray}
(\nu_L)_\alpha &=& U_{\alpha i} \nu'_i + X_{\alpha I} N'_I , \\
(N^c_L)_\alpha &=& V_{\alpha i} \nu'_i + Y_{\alpha I} N'_I .
\end{eqnarray}
%
In the following, the mass eigenstates $\nu_i$ and $N_I$ are defined as Majorana fermions,
e.g. $\nu_i = \nu'_i + (\nu'_i)^c$.

The interactions to the $W_L$ and $W_R$ gauge bosons 
are written as
\begin{eqnarray}
%
\frac{g_L}{\sqrt2 }W_L^\mu \, \bar l_L^\alpha \gamma_\mu (U_{\alpha i} \nu_i + X_{\alpha I} N_I)_L
+
\frac{g_R}{\sqrt2} W_R^\mu \, \bar l_R^\alpha \gamma_\mu (V^*_{\alpha i} \nu_i + Y^*_{\alpha I} N_I)_R + h.c..
\end{eqnarray}

If one adds three gauge singlets $S$, the expressions can be easily extended.
We define so that the $9\times 9$ mass matrix ${\cal M}$ for 
${\cal N} = ( \nu_L, (N^c)_L, S)^T$
is diagonalized by 
$9\times 9$ diagonalization unitary matrix 
\begin{equation}
{\cal U} = \left( 
 \begin{array}{cc}
     U & X \\
     V & Y \\
     W & Z
 \end{array}
 \right),
 \label{9x9}
\end{equation}
where $U,V,W$ are $3\times 3$ matrices, and $X,Y, Z$ are $3\times 6$ matrices.
Then, the expression of the gauge interaction is unchanged under this convention, but the indices of the `heavy' neutrinos 
are summed by $I = 1,2,..., 6$.

\section{Box loop function}
\label{app:loop_function}

In the box loop calculation,
one encounters an integral such as
\begin{eqnarray}
I_n (x,y,z; \xi) &=& 
\int_0^\infty dt \frac{(-t)^n}{(t+x)(t+y)(t+z)(t+1)(t+\xi_1)^{a_1} (t+\xi_2 z)^{a_2}},
\end{eqnarray}
where $a_i = 0$ or $1$, and $\xi_i$ is a gauge parameter for $R_\xi$ gauge, e.g., $\xi_1=\xi_2= \xi$ and $z=1$ in the 
box diagram with $W_L$\dash $W_L$ exchanges. 
Since 
\begin{equation}
\frac{t^2}{(t+x)(t+y)} = 1 - \frac{x}{t+x} - \frac{y}{t+y} + \frac{xy}{(t+x)(t+y)},
\end{equation}
one finds
\begin{equation}
\sum_{i,j} \lambda_i \lambda_j I_{n+2} (x_i, x_j, z; \xi) = \sum_{i,j} \lambda_i \lambda_j x_i x_j I_n(x_i,x_j,z;\xi),
\label{xi-independent}
\end{equation}
if $\sum_i \lambda_i = 0$.
Due to this equation, the $\xi$ dependence in the box loop calculation can vanish.
For example, for
\begin{equation}
\lambda_i = V_{\alpha i} V^*_{\beta i}, \quad x_i = \frac{m_i^2}{M_W^2},
\end{equation}
$\sum_i \lambda_i = 0$ is satisfied due the unitarity of the mixing matrix $V$,
and thus, the box loop contribution of the meson mixings is gauge independent \cite{Inami:1980fz}.

We define
\begin{eqnarray}
I_n (x,y,z) &=& \int_0^\infty dt \frac{(-t)^n}{(t+x)(t+y)(t+z)(t+1)} \\ 
&=& \frac{x^n \ln x}{(x-1)(x-y)(x-z)} + \frac{y^n \ln y}{(y-1)(y-x)(y-z)} + \frac{z^n \ln z}{(z-1)(z-x)(z-y)}, \nonumber
\end{eqnarray}
for $n= 0,1,2$,
and the loop functions of the box diagrams are given as
\begin{eqnarray}
E_0(x,y,z) &=& x y \left(  I_0(x,y,z) - \left(1+\frac1{z}\right) I_1(x,y,z) + \frac1{4z}I_2 (x,y,z) \right), \\
E_1(x,y,z) &=& 2\sqrt{xy}  \left(  I_1(x,y,z) + \frac{xy}{4z}  I_1(x,y,z) - \frac14\left(1+\frac1{z}\right) I_2 (x,y,z) \right),
\end{eqnarray}
where $E_0$ is a function for the contributions where momentums ($k\!\!\!/$) are picked in the fermion propagators,
and $E_1$ is a function for the ones where masses ($M_{\cal I}$) are picked.
The coefficients of the transition operators can be written by $E_0$ and $E_1$ terms. 
The gauge invariance of the $E_0$ term is assured by the unitarity of the neutrino mixing matrix ${\cal U}$.
On the other hand, $E_1$ term is not necessarily gauge invariant.
For the $W_L$\dash $W_L$ exchange diagram in the type-I seesaw case,
the function $E_1$ is gauge invariant due to
\begin{equation}
{\cal U}^* {\cal M}^{\rm diag} {\cal U}^\dagger = {\cal M},
\end{equation}
and
\begin{equation}
({\cal U}^* {\cal M}^{\rm diag} {\cal U}^\dagger)_{e_L, e_L} = 0 ,
\quad 
({\cal U}^* {\cal M}^{\rm diag} {\cal U}^\dagger)_{\mu_L, \mu_L} = 0 .
\end{equation}
Similarly, 
for the $W_R$\dash $W_R$ exchange diagram in the left--right model without $SU(2)_R$ triplet,
the $E_1$ term is gauge invariant.
However, for the $W_R$\dash $W_R$ exchange with $SU(2)_R$ triplet, and $W_L$\dash $W_R$ exchange box diagrams,
the $E_1$ term is not gauge invariant.
In those cases, adding the loop corrections of the triplet and bi-doublet Higgs scalar exchange diagram,
the gauge dependence is canceled.
See Refs.\cite{Hou:1985ur,Basecq:1985cr} for the gauge invariance in the case of the $K$\dash$\overline K$ mixing.
(Strictly speaking, in the case of the type-II seesaw with $SU(2)_L$ triplet, the box loop is not gauge invariant similarly
unless the light neutrino masses are ignored.)
The loop function $E_1$ above is given in the 't Hooft-Feynman gauge.

The loop function $E_0$ is usually redefined as given by the Inami-Lim function \cite{Inami:1980fz},
\begin{equation}
\tilde E_0 (x,y,z) = E_0(x,y,z) - E_0 (x,0,z) - E_0 (0,y,z) + E_0 (0,0,z).
\end{equation}
The function $E_0$ above is already redefined using 
Eq.(\ref{xi-independent}) by the unitarity of the mixing matrix,
and thus, $\tilde E_0 (x,y,z) = E_0 (x,y,z)$.

To express the loop functions for $W_L$\dash $W_L$ and $W_R$\dash $W_R$ box diagrams shortly, we define
\begin{equation}
E_0 (x,y) \equiv E_0 (x,y,1), \quad E_1 (x,y) \equiv E_1 (x,y,1).
\end{equation}
We note
\begin{eqnarray}
I_n (x,y) \equiv I_n (x,y,1) &=& 
\frac{x^n \ln x}{(1-x)^2(x-y)} + \frac{y^n \ln y}{(1-y)^2(y-x)} + \frac1{(1-x)(1-y)} \nonumber \\
&=& \frac{1}{x-y} \left( \frac{x^n \ln x}{(1-x)^2} -\frac{y^n \ln y}{(1-y)^2}
+ \frac1{1-x} - \frac1{1-y}
\right),
\end{eqnarray}
and 
\begin{eqnarray}
I_n(x,x,z) &=& \frac{z^n \ln z}{(z-1)(x-z)^2} + \frac{x^{n-1} (1+ n \ln x)}{(x-1)(x-z)} 
-  \frac{(2x-1-z)x^n \ln x}{(x-1)^2(x-z)^2},
\\
I_n(x,x) &=&\frac{1+ x^{n-1}(1+n \ln x)}{(1-x)^2} + \frac{2x^n \ln x}{(1-x)^3}.
\end{eqnarray}

\end{document}